\documentclass{lmcs}
\pdfoutput=1

\usepackage{lastpage}
\lmcsdoi{22}{3}{9}
\lmcsheading{}{\pageref{LastPage}}{}{}%
{Feb.~04,~2025}{Sep.~22,~2026}{}


\usepackage{amsmath,amssymb,amsfonts}
\usepackage[utf8]{inputenc}
\usepackage[T1]{fontenc} 
\usepackage{microtype}
\usepackage{graphicx}
\usepackage{color, xcolor}
\usepackage{algorithmic}
\usepackage{xspace}
\usepackage{todonotes}
\usepackage{mathtools}
\usepackage{comment}
\usepackage{multicol}
\usepackage{listings, listings-rust}
\PassOptionsToPackage{hyphens}{url}\usepackage{hyperref}
\usepackage[inline]{enumitem}
\usepackage{cleveref}
\usepackage{tikz-cd} \usetikzlibrary{decorations.pathmorphing}
\usepackage{framed}

\newcommand{\dlneg}{Datalog$^{\neg}$\xspace}
\newcommand{\slick}{Slick\xspace}
\newcommand{\brane}{\textsc{brane}\xspace}

\newcommand{\mlst}[1]{\text{\lstinline!#1!}}
\newcommand{\slickalign}[2]{\makebox[3em][#1]{\mlst{#2}}}

\newcommand{\dplus}{\mathbin{+\!\!+}}

\newcommand{\funarrow}[1]{\raisebox{-2pt}{\hspace{1pt}$\xrightarrow{\text{\scriptsize\smash{\raisebox{-2pt}{\hspace{1pt}#1\hspace{1pt}}}}}$\hspace{1pt}}}

\input{dasharrows}

\makeatletter
\let\lmcs@makecaption\@makecaption
\newcommand{\raggedcaption}{%
  \long\def\@makecaption##1##2{\@rightskip\@flushglue\lmcs@makecaption{##1}{##2}}}
\makeatother

\usepackage{listings}
\definecolor{green1}{HTML}{00BB00}
\definecolor{lstmaincol}{HTML}{0A0600}
\definecolor{specialFactCol}{HTML}{A03080}
\lstdefinelanguage{slick}{ 
    sensitive=true,    
    keywordstyle={[1]\color{blue}},
    keywordstyle={[2]\color{green1}},
    keywordstyle={[3]\color{blue}},
    keywordstyle={[4]\color{magenta}},
    keywordstyle={[5]\color{specialFactCol}},
    stringstyle=\color{specialFactCol},
    keywords=[1]{QQQQ},
    keywords=[2]{Author,Truth,Relates,Event,Var,L,Checker,Uses,Trusted,Patient,V,Accesses,Fact,List,MsgDiscriminator,Name,Msg1,Msg2,Variable,T,O,Driver,Label,R,S1,S2,C1,C2,Driver,Source,Contents,F,M,A,Dependency,Msg,Agent,Task1,Task2,Worker,Task,Data,Anyone,Relation,Sayer,X,Y,Z,Name1,Name2,Verb,Reason},  
    keywords=[3]{and,if},
    morecomment=[l][\color{gray}]{//},
    keywords=[4]{not,diff,same},
    keywords=[5]{error,within},
    morestring=[b]"
}

\lstdefinelanguage{rustagentscript}{
  sensitive=true,
  morecomment=[l]{//},
  commentstyle=\color{green1},
  morestring=[b]",
  stringstyle=\color{brown},
}

\lstdefinelanguage{coq}{ 
    sensitive=true,    
    keywordstyle={[1]\color{blue}},
    keywordstyle={[2]\color{green1}},
    keywordstyle={[3]\color{blue}},
    keywordstyle={[4]\color{magenta}},
    stringstyle=\color{specialFactCol},
    keywords=[1]{QQQQ},
    keywords=[2]{Definition,Parameter,Parameters},  
    keywords=[3]{Prop,Module,End,Type},
      literate=
        {->}{{\textcolor{purple}{$\rightarrow$}}}1
        {:}{{\textcolor{purple}{:}}}1
        {.}{{\textcolor{purple}{.}}}1,
    morecomment=[s]{(*}{*)},
    commentstyle=\color{gray}\itshape,
    keywords=[4]{->},
    keywords=[5]{:},
    morestring=[b]"
}
\definecolor{slickBackgroundColor}{rgb}{0.98,0.98,0.92}
\definecolor{rustBackgroundColor}{rgb}{0.96,0.98,1.00}
\definecolor{coqBackgroundColor}{rgb}{1.00,0.97,0.99}
\usepackage{microtype} 

\newcommand{\Prop}{\mathsf{Prop}}

\newtheorem{parry}[thm]{Parameter}
\Crefname{lstlisting}{Listing}{Listings}
\Crefname{parry}{Parameter}{Parameters}
\Crefname{exa}{Example}{Examples}
\Crefname{defi}{Definition}{Definitions}
\Crefname{lem}{Lemma}{Lemmas}
\Crefname{thm}{Theorem}{Theorems}
\Crefname{cor}{Corollary}{Corollaries}

\def\eg{{\em e.g.}}
\def\Eg{{\em E.g.}}

\def\ie{{\em i.e.}}

\begin{document}

\title[JustAct]{
JustAct: A Framework for Auditable Multi-Agent Systems Regulated by Inter-Organisational Policies\texorpdfstring{\hyperlink{extendedbabyyy}{\rsuper*}}{}}
\titlecomment{\hypertarget{extendedbabyyy}{\lsuper*}This article is a significantly extended version of \cite{DBLP:conf/forte/EsterhuyseMB24}.
    We define the same framework more precisely and extensively.
    The prior implementation discussion is replaced: we present our new Rust implementation and Rocq formalisation of our framework, policy language, its interpreter, and runtime system.
    The prior case study is replaced by a new more extensive one: recreating the \brane medical data processing system.
    }

\author[C.~A.~Esterhuyse]{Christopher A.\ Esterhuyse\lmcsorcid{0000-0002-9124-9092}}
\author[T.~M\"uller]{Tim M\"uller\lmcsorcid{0000-0002-9759-5973}}
\author[L.~T.~van~Binsbergen]{L.\ Thomas van Binsbergen\lmcsorcid{0000-0001-8113-2221}}

\address{Informatics Institute, University of Amsterdam, Amsterdam, The Netherlands}	
\email{c.a.esterhuyse@uva.nl, t.muller@uva.nl, ltvanbinsbergen@acm.org}  


\begin{abstract}
\noindent In open multi-agent systems that cross organisational boundaries, agent actions must be regulated by complex policies.
Consider medical data processing systems, which must observe generic laws (\eg, EU data protection regulations) and also specific participants' resource conditions (\eg, Bob consents to sharing his X-Rays with EU hospitals).
Presently, we address the implementation of these systems as distributed software.
Solutions to key sub-problems are available:
existing policy languages capture the necessary normative concepts and formalise the computational representation and reasoning about policies, and
existing distributed algorithms and protocols coordinate agents' changing actions and policies.
But which policies and protocols are useful in application?


With the \textit{JustAct} framework, we characterise a class of multi-agent systems where actors justify their actions with sufficient policy information collected from dynamic policy statements and agreements.
We prove key properties of these systems, \eg, any decision that an action is permitted now cannot be refuted later, regardless of any added statements or updated agreements.
We study a particular instance of the framework by specifying (in Rocq) and implementing (in Rust) a particular policy language and runtime system for mediating agent communications.
We demonstrate and assess JustAct via a case study of this implementation: we reproduce the usage scenarios of Brane, an existing policy-regulated, inter-domain, medical data processing system.

\end{abstract}

\maketitle

\section{Introduction}
\label{sec:introduction}

In this article, we consider multi-agent software systems that are \textit{open} (\ie, agents may join and leave at runtime) and subject to regulation by \textit{policies}: specifications of how agents are permitted to act.
These systems are needed in practice whenever software systems cross organisational boundaries, as physical distance separates institutional entities from the users or software components that they aim to control.
Such users, components, and entities as unified in \textit{agents} able to act, and their dependencies are captured as policies.
For example, to enable medical research on the rare DIPG disease, hospitals must share their medical records.
But, to maintain control, participants are required to accept the data sharing agreement of the DIPG registry ecosystem~\cite{veldhuijzen2017development}.
In general, participants rely on the \textit{enforcement} the policy, which aligns agents' actions with what is permitted.
For example, \textit{auditor} agents trigger compensatory actions to punish actors of prohibited actions after the fact, or automated monitors prevent prohibited actions just in time.
The terms in which policies define permission are dependent on the case.
For example, when agents' access to data is modelled as instantaneous events, policies are formulated in terms of \textit{access control}~\cite{DBLP:journals/iotj/QiuTDZSF20,DBLP:conf/fosad/SamaratiV00}, and when data access is a continuous process, policies are formulated in terms of usage control~\cite{Jung2022,DBLP:journals/tissec/ZhangPSP05,MUNOZARCENTALES2019590}.
In any case, real systems reflect complex governance models~\cite{Torre-Bastida2022}, and because they integrate with society, their policies capture (inter)national laws, whose enforcement is legally required~\cite{Curry2022}.
For example, controllers of data-processing systems within the European Union must enforce the General Data Protection Regulation~(GDPR)~\cite{european_commission_regulation_2016}, \eg, by always ensuring `the controller shall be able to demonstrate that the data subject has consented to processing of his or her personal data'~(\S{}7.4).
The policies are only useful if their meaning is clear, \eg, so that disputes between agents can be reliably resolved as per the agents' agreements~\cite{shakeri2019}.

\Cref{sec:background} overviews some of the extensive literature on the notions of \textit{policy} that are relevant to our work.
To participants in these systems, it is crucial that policies can faithfully capture the social relationships that comprise laws and contracts.
For our purposes, it is also crucial that policies are expressed in formal languages, in the mathematical sense.
Thus, computational agents can reason unambiguously about the meaning of policies that they observe, \eg, by applying inference rules and interrogating relational databases.

Our contributions address the problems that arise when agents cannot maintain complete knowledge of the policies being enforced.
We are particularly motivated by cases of distributed processing of (sensitive) medical data.
Here, firstly, data processing is sometimes regulated by policies that are also sensitive.
For example, the fact that data processor Dan is obligated to demonstrate Pam's consent to the processing of \mlst{cancer_patient_55.csv} reveals sensitive information about Pam!
In general, is it \textit{undesirable} for policy enforcement process to expose policy information to more agents than is necessary.
This rules out some typical solutions.
For example, updates to policies cannot be shared via a distributed ledger that every agent replicates.
Moreover, as the systems and its policies become more complex and dynamic, it becomes \textit{infeasible} to synchronise all agents' views of all (changes to) the policy, \eg, as laws are amended each week, as users join the network each day, and as metadata is updated each second.
Of course, synchronisation and consensus is necessary in our distributed, multi-agent software systems, to some extent.
But what exactly must be synchronised to preserve the universal meaning of permission with respect to policies in general, on the one hand, with the local knowledge of particular policy details, on the other hand?

\begin{figure}
\centering
\begin{minipage}[b]{.43\textwidth}
  \centering
  \includegraphics[width=1\linewidth]{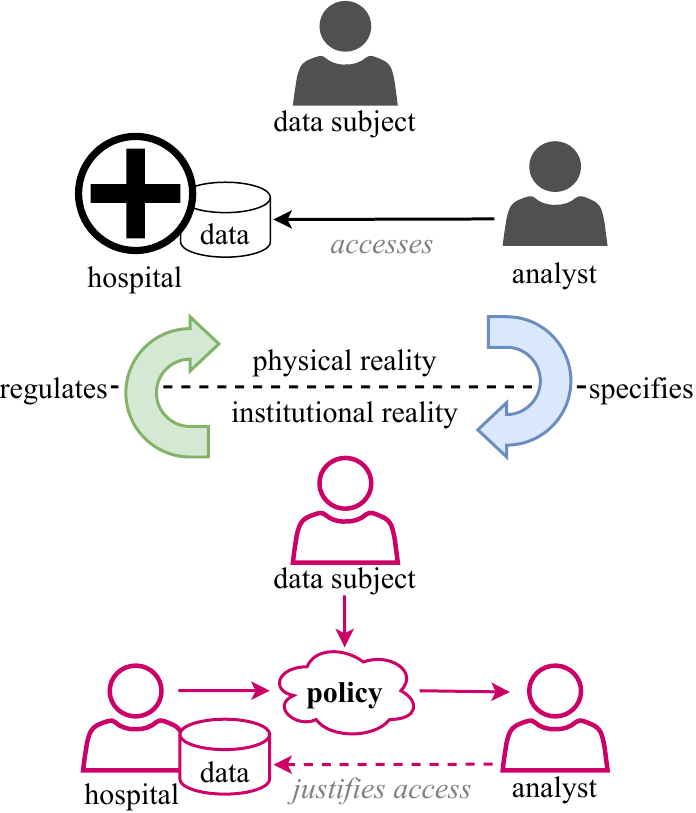}
  \raggedcaption
  \caption{User experience in systems implementing JustAct: agents physically update shared policies which specify the agents' institutional relationships.
  In turn, these relationships regulate how agents physically act.}
  \label{fig:overview}
\end{minipage}%
\hfill
\begin{minipage}[b]{.57\textwidth}
  \centering
  \vfill
  \includegraphics[width=1\linewidth]{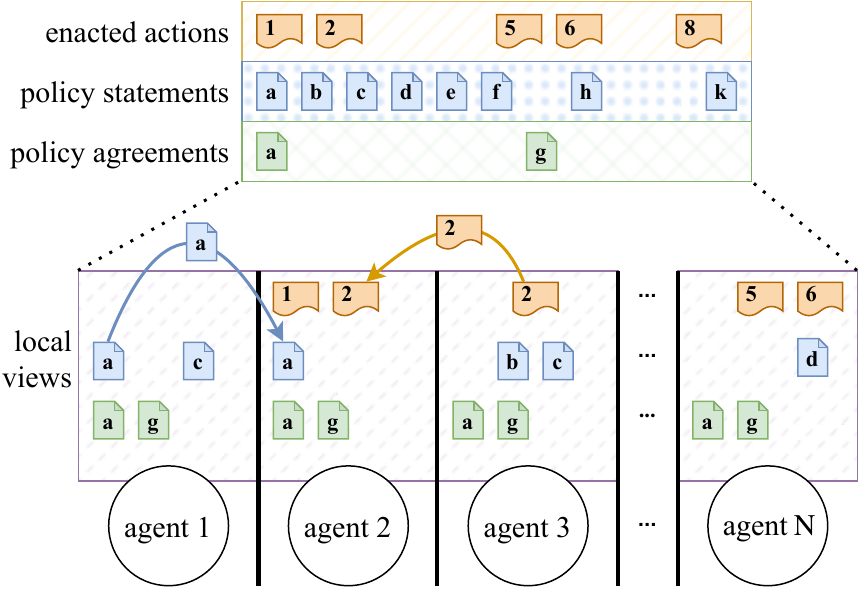}
  \raggedcaption
  \caption{Communication in systems implementing JustAct: agents update action permission by sharing updates to policy as statements and agreements.
  JustAct fixes how core concepts relate (\eg, action vs.\ agreement) such that any agent can audit the permission and effects of enacted actions, despite their partial local view of the system.}
  \label{fig:views}
\end{minipage}
\end{figure}

Our approach is the JustAct \textbf{framework}, which characterises systems that are used as is visualised in \Cref{fig:overview}: agents interleavedly share policies and take policy-regulated actions.
The framework formalises the dependency between these activities: actors must \textit{justify} their actions with sufficient policy information, such that observers necessarily agree \textit{whether} (and obligatorily agree \textit{that}) the information suffices to prove that the action is permitted.
\Cref{fig:views} visualises how agents create and communicate policy information.
\textit{Agreements} are crucial, as they are the bases of justifications.
We expect agreements to be carefully designed by committee a priori (and revised infrequently) to shape the space of permissible actions.
But the bulk of policy information is expressed by agents' subjective policy \textit{statements}, which agents create autonomously and share with their peers asynchronously, at their own pace, for their own reasons.
\Eg, statements are sent to actors on a needs-to-know basis.
The utility of JustAct is that it gives complex and vague problems (\eg, policy regulation and agent communication) a precise framing, and provides a path to formal specification and implementation that gives participants the guarantees that they need.
For example, the relations between policies, actions, and messages becomes clear, the decidability of key agent activities (\eg, auditing the permission of an action) is assured, and actors at runtime can rely on their proofs of permission being verified by any auditors in the future.

Many different implementations of JustAct are conceivable.
Throughout the article, we characterise this space by discussing various design considerations.
But we concretise our ideas by studying a particular implementation of the framework, which we define in two parts.
Firstly, 
\Cref{sec:slick} defines the \slick \textbf{policy language}, which is taken from \cite{chris_thesis}.
\slick inherits the essence of the eFLINT normative specification language~\cite{DBLP:conf/gpce/BinsbergenLDE20}, but \slick is adapted for incremental specification by cooperating agents~\cite{esterhuyse2024cooperative}.
Secondly,
\Cref{sec:impl} defines a prototype \textbf{data exchange runtime system}, which mediates the agents' communication of policy information and regulates their access to a shared dataspace.

In \Cref{sec:case_study}, we evaluate our framework by conducting a \textbf{case study}: we apply our implementation to new and existing usage scenarios of the \textit{\brane} system~\cite{DBLP:conf/eScience/ValkeringCB21}, which is the component of the \textit{EPI framework} which coordinates and regulates the execution of medical data processing workflows~\cite{DBLP:conf/eScience/KassemVBG21,DBLP:series/oasics/KassemAAKMTBBGHGLK24}.
This reflects our ongoing developments of data exchange systems in general, and the EPI framework in particular, where dynamic organisational policies, resource conditions, and legal regulations are systematically enforced.

In summary, after we overview some background literature (\Cref{sec:background}), we contribute:
\begin{enumerate}
    \item the definition of a \textbf{framework} for multi-agent runtime systems in which agent actions are regulated by policies that agents autonomously create and share (\Cref{sec:framework}), 
    \item the definition of the \slick \textbf{policy language} and interpreter, whose policies are assembled into specifications of permitted behaviour (\Cref{sec:slick}), 
    \item the implementation of a case- and policy language-generic \textbf{data exchange runtime system} (\Cref{sec:impl}), which regulates agents' access to sensitive data and mediates their communication of meta data carrying policy information, and
    \item a \textbf{case study}, evaluating our framework, instantiated for data exchange, regulated by \slick policies, in application to \brane's medical workflow processing scenarios (\Cref{sec:case_study}).
\end{enumerate}
We discuss our contributions by their own merits (\Cref{sec:discussion}) and in comparison to related work (\Cref{sec:related_work}) before we conclude with a summary (\Cref{sec:conclusion}).

The definitions and claims of the framework, policy language, and runtime system are backed by our \textbf{machine-checked specification} in the Rocq theorem prover.
These analytical results mirror our \textbf{executable implementation} in Rust, which produced the experimental results of our case study.
Both of these artefacts are included in
the ancillary files~\cite{esterhuyse_2026_21240014} that supplement the article, such that the reader can reproduce our results.

\section{Background}
\label{sec:background}

This section contextualises our contributions by overviewing the conventions and terminology that we inherit from various fields of study.
Our presentation of specific works that complement or compete with our contributions are postponed the related work section: \Cref{sec:related_work}. 

\subsection{Distributed and Multi-Agent Systems}
\textit{Distributed systems} model the distribution of a stateful \textit{configuration} over a set of \textit{processes}; each process has its own local \textit{state}.
Our article inherits this meaning of `configuration', which the reader should not confuse with `configuration' to mean to specialise the initialisation of a software system. 
\textit{Distributed algorithms}, when implemented by each process, give systems useful emergent properties.
Often, these algorithms are defined in terms of only basic message-passing primitives, comparable to the IP and UDP protocols in networking: messages are asynchronous, and message delivery is unreliable.
Distributed algorithms solve distributed problems and build abstractions atop the network.
For example, Awerbuch's synchroniser adapts synchronous distributed algorithms for use in asynchronous networks~\cite{DBLP:journals/jacm/Awerbuch85}.

In this work, we refer to two classes of algorithms.
Firstly, \textit{gossip} protocols disseminate information by replicating and forwarding messages from peer (process) to peer, resulting in decentralisation and robustness, by imposing minimal requirements on the network topology and process behaviour~\cite{bakhshi2008meanfield}.
We will show peers gossiping \textit{policy statements} in systems implementing JustAct.
Secondly, \textit{consensus algorithms} establish fundamental agreement on the selection of a particular value, consistently among processes.
Consensus has been well-studied for decades~\cite{ren2005survey}, and newer algorithms such as \textit{Raft}~\cite{DBLP:conf/usenix/OngaroO14} are still actively developed and implemented as part of modern production systems including CockroachDB, Docker Swarm, and Kubernetes.
Consensus has seen renewed interest even more recently in application to blockchain technologies in, for example \cite{DBLP:journals/sensors/KhanHH22,DBLP:conf/blockchain/KhobragadeT22}.
We will show peers maintaining consensus on \textit{policy agreements} in systems implementing JustAct.

The field of \textit{multi-agent systems} studies processes called \textit{agents} that exhibit \textit{autonomy}: agents are motivated to act on shared resources and interact with other agents.
Limitations on multi-agent systems are shaped by the limitations on their implementations as distributed systems.
This is one reason why agents have access to only partial and stale knowledge of their environment.
Multi-agent systems are used to model and study social phenomena, \eg, the incentives to cooperate in consortia~\cite{castro2023data} and to compete in markets~\cite{ZHOU2023128172}.
Work in this field on incentivising and policing agent behaviour overlaps with the study of normative policy specification languages (see \Cref{sec:policy_overview}, to follow).
We will characterise constituents of JustAct systems as agents who autonomously communicate, reason, and act.

\textit{Agent-oriented programming} studies the programming of agents, balancing the concerns of software (language) engineering and object-oriented programming, but with a unique emphasis on agent autonomy.
For example, agent autonomy also tends to improve system scalability and robustness.
These ideas are present in seminal agent-oriented programming works such as \cite{DBLP:journals/ai/Shoham93} and persist into more recent works like the survey \cite{DBLP:journals/wias/MaoWY17}.
To a large extent, our contributions are orthogonal to agent-oriented programming, because we generally leave unspecified what each agent chooses to do.
But we rely on agent-oriented programming in \Cref{sec:agent_scripting} to fix agent behaviour for our experiments.

\subsection{Policy Specification Languages}
\label{sec:policy_overview}

We give a brief overview of the various notions of `policy' which were developed by various disciplines and which influenced our work.

\textit{Access control} is a mainstay in cyber-physical systems (e.g., from databases to IoT networks) that revolves around the regulation of agents accessing resources.
For example, a policy specifies in which case a given agent is \textit{authorised} to read or write a given data asset.
Policies often take the form of conditional rules \cite{sandhu1998role,DBLP:journals/thms/FragkosJT22}, sometimes applied in the context of metadata attributes~\cite{servos2017current}.
\textit{Usage control} generalises access control to the control of usage events that occur for the durations of time the resource is being used.
Usage authorisation must be maintained and can be interrupted~\cite{DBLP:journals/corr/abs-2203-04800}, which may be realised as runtime monitoring~\cite{Hariri2023} and bookkeeping of agent attributes~\cite{sandhu2003}.
XACML~\cite{anderson2003extensible} and ODRL~\cite{ianella2007open} are popular, standardised policy languages for expressing access and usage conditions as policies.
For example, \cite{DBLP:conf/cisim/Um-e-GhaziaMSB12} implements usage control in XACML.

\textit{Normative specifications} define fundamental normative positions (relations) such as powers, duties, rights, obligations, and permissions between actors~\cite{DBLP:conf/dagstuhl/2013dfu4}.
Normative specifications can formalise (and make machine-readable) the norms described in normative documents such as laws and regulations, organisational policies, and contracts.
Laws and regulations are specified to apply to activities of particular kinds within particular jurisdictions. 
For example, the GDPR~\cite{european_commission_regulation_2016} regulates the processing of personal data within the European Union, but its impact makes it influential even outside the EU.

The study of norms reflects its long history in its rich nomenclature, for example, \cite{DBLP:journals/japll/BoellaT08} clarifies the relationship between \textit{substantive} and \textit{procedural} norms, and \cite{DBLP:journals/ail/GovernatoriIMRS18} characterises \textit{open-texture terms} in norms as intentionally opaque, leaving room for interpretation later, \eg, by a judge during the resolution of a particular dispute. 
Normative specification languages attempt to capture these concepts, thus enabling their systematic enforcement within software systems, \eg, semi-automated by trusted enforcement services.

A wealth of literature connects all these notions of norms and policies, with the goal of relating (abstract) social relations to (concrete) software events, and connecting unpredictable human behaviour to programmable software processes.
For example, the eFLINT language~\cite{DBLP:conf/gpce/BinsbergenLDE20} formalises norms using the Hohfeldian framework of legal proceedings~\cite{wesley1913some} and has been used for access control~\cite{DBLP:conf/euspn/BinsbergenKBEV21}.
Symboleo~\cite{DBLP:conf/re/SharifiPALM20} is also based on Hohfeld's framework but, contrary to eFLINT, is focused on contracts, \eg, tenancy agreements.
Fievel~\cite{vigano2007symbolic} is a language used to verify properties of social institutions. 
These languages afford the application of various tools and techniques to policies.
For example, model-driven development~\cite{schmidt2006model} and model-checking for high-level properties in policies of both Fievel~\cite{vigano2007symbolic} and Symboleo~\cite{DBLP:conf/models/ParvizimosaedRR22}.

Recently, \cite{esterhuyse2024cooperative} studied the \textit{cooperative specification} of policies, where multiple agents develop a shared policy together, by taking turns contributing parts to the policy.
Crucially, the ongoing cooperation is regulated by a meta-level agreement between the agents to preserve the \textit{validity} of the shared policy.
Suitable policy languages must fix a semantic definition of validity which gives the agents meaningful control over how the policy is developed.
For example, after Amy's contribution to the specification, it expresses `Only Amy can access Amy's data', and would  be invalidated if Bob contributes `Bob accesses Amy's data'.
Thus, the policy itself captures Amy's specification of how Bob may contribute to the policy.

\subsection{Logic Programming for Modelling and Specification}
\label{sec:dlneg}
\textit{Logic programming} languages are designed to operationalise various logics: logic programs encode logical theories.
Logic programming has been used to automate logical reasoning embedded in social policies.
For example the constitutive norm `hybrid cars are eligible for the subsidy' is encoded in a logic program as an inference rule, predicating `is eligible for the subsidy' to every car with `hybrid'.
\Eg, the eFLINT language uses these kinds of rules to capture these kinds of norms.
The \slick language which we detail in \Cref{sec:slick} and apply in \Cref{sec:case_study} is similar;
in fact, \slick policies boil down to Datalog-style inference rules.
Here, we give a brief account of Datalog-style logic programming, so readers can anticipate how the related \slick language is ultimately used by agents to express policies that regulate their shared system.

Datalog, overviewed in \cite{DBLP:journals/tkde/CeriGT89}, is a simple logic programming language, which has become the common ancestor of many logic programming languages since.
Each Datalog program consists of a set of \textit{rules}, which assert the \textit{truth} of \textit{facts} on the condition that other facts are true.
The semantics of Datalog is desirable for its proof-theoretic interpretation: every truth is the root of a proof tree comprised of finite applications of program rules, representing an inductive reasoning process.
Users appreciate that every truth has an underlying explanation in terms of program rules they presumably can understand.
At the same time, this inductive semantics lends itself to implementation.
Datalog engines automate the evaluation of a program's \textit{model}, which incrementally builds the (finite) set of true facts, implicitly assigning falsity to all omitted facts.
Users can interrogate this model, \eg, using it to query the truth of any fact.
Importantly, evaluation is deterministic: evaluating the same program yields the same model.
Multiple agents can thus use explicit agreement on (a~few) program rules to reach implicit agreement on the (many) answers to queries.

Datalog programs can model and specify systems by denoting \textit{relational databases}, upon which normative meaning may be additionally layered.
Each fact is some $p(c_1, c_2, c_3, \dots, c_n)$, interpreted as the assertion that \textit{constants} $c_{1 \dots n}$ are related by the $n$-ary relation identified by~$p$, \ie, asserting the presence of row $c_1~c_2~c_3~\dots~c_n$ in a table called~$p$.
For flexibility, Datalog rules admit \textit{variables} in the position of constants (inside facts), such that each such rule represents the set of \textit{ground} (variable free) rules resulting from any consistent substitution of its variables.
For example, the rule \mlst{knows(amy, Person) :- knows(Person, amy)} is applicable to any person (\eg, Bob) meeting the condition of knowing Amy.

Various dialects of Datalog have been studied in the literature, exploring combinations of extra features.
\dlneg~\cite{DBLP:journals/jcst/ShiZ94} is a useful generalisation:
rules may be conditioned on the \textit{falsity} of facts.
This improves expressiveness~\cite{DBLP:conf/icdt/Ketsman020}, because it affords \textit{non-monotonic reasoning}: sometimes, adding a rule to a program \textit{removes} truths from its model~\cite{sep-logic-nonmonotonic}.
For example, fact \lstinline{sun} is true in program \lstinline{sun :- not clouds}, but false after rule \lstinline{clouds} is added.
Unfortunately, not all \dlneg programs have unique logical interpretations.
Accordingly, different semantics exist (\eg, stable model \cite{DBLP:conf/iclp/GelfondL88} and well-founded~\cite{van_gelder_well-founded_1991}) that attribute different models to these \textit{unstratified} (defined in \cite{DBLP:conf/pods/Ross90}) programs, whose rules express non-deterministic or contradictory reasoning.
\Eg, what should be the truth of fact~$p$ given a program asserting `$p$~unless~$p$'?
\Cref{sec:slick} details how \slick handles these cases.

From another perspective, logic programming in general, and Datalog in particular, are formal specification languages that go all-in on automation by sacrificing expressivity.
In contrast, Rocq specifications can express more than Datalog-style inductive rules can, but users cannot rely on the Rocq prover to automatically prove logical propositions by itself.

\subsection{Mathematical Preliminaries}
\Cref{sec:framework,sec:slick,sec:impl} presents our framework, policy language, and runtime system (respectively), approximating our formalisation in Rocq~\cite{esterhuyse_2026_21240014} with the following conventional mathematical terminology and notations.

\begin{itemize}
    \item \textbf{Definition} $d:T \coloneqq v$ lets $d$ identify the value~$v$ of a term with defined type~$T$.
    In general, we express the assertion that $d$ \textbf{has type} $T$ with the notation $d:T$.

    \item
    Each term~$f$ of any type matching $A \rightarrow B$ is \textbf{a pure function} from $A$-type to $B$-type values.
    We denote the application of $f$ to some $a:A$ as $f(a)$.
    We always consider the evaluation of any such function application $f(a)$ to some $B$-type value as \textbf{computable}, \eg, by any agent inside the system, and that its result is always \textbf{deterministic}.

    \item
    We \textbf{chain $\rightarrow$} to represent functions with \textbf{multiple arguments}.
    We read $\rightarrow$ as right-associative to avoid some parenthesis.
    For example, $+: \mathbb{N} \rightarrow \mathbb{N} \rightarrow \mathbb{N}$ is a binary \textit{add} function over the \textbf{natural numbers} ($\mathbb{N}$), and so $(+1): \mathbb{N} \rightarrow \mathbb{N}$ increments a given number.
    
    \item
    \textbf{Proposition} (which we shorten to $\Prop$, as Rocq calls it) is the type of logical propositions.
    \Eg, if $P: \Prop$, $P$ is a proposition, proving $P$ true means finding any~$h: P$.
    We adopt the usual logical operators
    $\rightarrow$, $\leftrightarrow$, $\wedge$, $\vee$, and $\neg$.
    For example, $\wedge$ is read as conjunction and $\rightarrow$ is read as logical implication.
    Mechanically, they are proposition combinator functions. 
    For example, $(\neg): \Prop \rightarrow \Prop$, so whenever $P: \Prop$, then $\neg P: \Prop$.
    \\
    Note that this appears to introduce ambiguity in the reading of $\rightarrow$, which was already defined as the function arrow.
    But there is no ambiguity, because we adopt the \textbf{Curry-Howard isomorphism} by conflating both readings. For example, take $v: A \rightarrow B$:
    \begin{enumerate}
        \item $v$ is a function that maps any given $A$-type term into some $B$-type term, and
        \item $v$ is the proof of the proposition that the truth of $A$ implies the truth of $B$.
    \end{enumerate}
    
    \item A \textbf{relation} over $T_1$, $T_2$, ... , $T_n$ is a function from $T_1$ to $T_2$ to ... to $T_n$ to $\Prop$.
    For example, $<: \mathbb{N} \rightarrow \mathbb{N} \rightarrow \Prop$ denotes the usual \textit{less-than} total ordering of natural numbers, where $<(1,2): \Prop$, more conventionally notated as $1 < 2$, proposes that $1$ is less than~$2$.

    \item 
    We use \textbf{predicate} and \textbf{property} (interchangeably) of any $T$ to refer to unary relations over some type~$T$.
    for example, $\mathit{even}: \mathbb{N} \rightarrow \Prop$ is a property of natural numbers.
    Then $\mathit{even}(n)$ is the proposition that $n$ in particular is even, and any $h: \mathit{even}(n)$ proves it.

    \item Any new type $T$ is \textbf{defined inductively} given a set of \textbf{constructor} terms with prescribed types of (functions to functions to ...) $T$.
    For example, $\mathit{even}: \mathbb{N} \rightarrow \Prop \coloneqq \mathit{E0}: \mathit{even}(0) \mid \mathit{En: \forall n, \mathit{even}(n) \rightarrow \mathit{even}(n+2)}$ gives an inductive definition of the \textit{even} property over natural numbers: $E0$ is the (only) constructor that trivially proves that $0$ is even, and $En$ transforms a proof that given $n$ is even to a proof that $n+2$ is even.
    Intuitively, the proof that any~$n+2$ is even is a linear proof tree with leaf~$E0$, rooted in the application of $En(n)$.
    For example, $En(4,En(2,En(0,E0))): \mathit{even(6)}$.
    Our mathematical foundations mean there is no significant difference between inductively-defined \textit{propositions} and inductively-defined data of another sort.
    For example, with $\mathbb{N} \coloneqq 0 \mid \mathit{Succ}: \mathbb{N} \rightarrow \mathbb{N}$, we adopt Peano's pervasive encoding of the natural numbers.
    For larger, more complex cases, we prefer to notate inductive definitions as \textbf{Gentzen-style rule schemas}, as is shown for \textit{even}, below:
    \begin{align*}
        \dfrac{}{\mathit{even(0)}}
        \quad
        \dfrac{\mathit{even}(n)\hfill}{\mathit{even}(n+2)}
    \end{align*}

    \item \textbf{Finite lists} are the typical polymorphic inductive data type: $\mathit{list}(t) \coloneqq \mathit{Nil} \mid \mathit{Cons}(t, \mathit{list}(t))$.
    $X \dplus Y$ concatenates lists $X$ and~$Y$.
    We reason about list elements in set-like fashion: $x \in L$ and $L \subseteq L'$ are decidable propositions over lists, and $L \setminus x$ removes every $x$ from~$L$.
    As is typical, $[t_1, t_2, ..., t_n]$ notates list $\mathit{Cons}(t_1, \mathit{Cons}(t_2, ... \mathit{Cons}(t_n,\mathit{Nil}) ... )))$.
    For brevity, our article uses \textit{list comprehensions}.
    \Eg, $[f(a) \mid \forall a \in L]$ results from mapping $f$ over~$L$. 

    \item  \textbf{Sets} are conceptually ubiquitous, but they are not explicitly present in our formalisation.
    Instead, each $T$-type set is encoded as some predicate $\mathit{p}: T \rightarrow \Prop$.
    Thus, subset ($P~\subseteq~Q$) is simply notation for ($\forall x, P(x) \rightarrow Q(x)$).
    The largest set of $T$ elements is $T$ itself.

    \item
    Sets are generally not finite, so we represent \textbf{finite and enumerable} collections as lists.
    \Eg, $\mathit{payload}: \mathit{action} \rightarrow \mathit{list}(\mathit{message})$ in \Cref{fig:framework} relates each action to finite messages.

    \item $\forall (a : A), B$ generalises $A \rightarrow B$, binding the input to $a$ so the \textbf{output type can depend} on~$a$.
    For example, if $P \coloneqq \forall (n:\mathit{N}), g(n)$, then $f: P$ supports two isomorphic readings:
    \begin{enumerate}
        \item $f$ proves the proposition $P$, asserting that each natural number has property~$g$.
        \item function $f$ maps any~$n$ to some term of type $g(n)$, a proof that $n$ has property~$g$.
    \end{enumerate}

    \item Term $a = b$ denotes \textbf{Leibniz equality} between terms $a$ and $b$, meaning that $a$ and $b$ are indistinguishable and interchangeable under evaluation. 
    But like any other proposition, $a = b$ is not necessarily true until some proof $H : a = b$ is available. 
    Take care not to confuse $\coloneqq$ (definitional equality) with $=$, although $a \coloneqq b$ implies $a = b$.

    \item \textbf{Parameter} $p: T$ fixes an arbitrary term $p$ of type $T$ without yet providing it a definition.
    In the meanwhile, all results in terms of $p$ treat it as arbitrary, necessarily holding regardless of how $p$ is ultimately defined.
    A helpful approximation is to read definitions using $p$ as preceded by an implicit quantification $\forall(p: T)$.
    We call JustAct a \textit{specification} because it contains parameters which impede the application of functions and normalisation of values.
    But as \Cref{sec:slick,sec:impl} fill its parameters, JustAct is refined toward some \textit{implementation}, an executable Rocq program which satisfies the specification by construction.

    \item \textbf{Theorems} are propositions whose proofs we have defined, but we have omitted them from the article for brevity.
    They are included in the Rocq supplement for scrutiny.

    Note that this makes theorems and parameters appear similar in our article; their terms are treated as defined, but not shown.
    But bear the key difference in mind:
    theorems have unique definitions that are just omitted, while parameters have no unique definitions (yet).

    Finally, \textbf{lemmas} are theorems that signal intermediate results as usual (also in Rocq).
\end{itemize}

\section{The JustAct Framework for Multi-Agent Runtime Systems}
\label{sec:framework}

In this section, we define and explain the JustAct framework and its properties.

\subsection{Overview}

Informally, the JustAct framework is a scaffolding for a class of designs, formalisations and software implementations of multi-agent systems.
Readers benefit from keeping in mind the fundamental activities of the agents participating in these systems:
\begin{quote}{
\textit{%
Agents model their system and specify its desired attributes via policies.
\\
Agents create, share, and collect messages that carry policies.
\\
Agent actions carry messages, determining the action effects and permission.
}}\end{quote}

Formally, we define the framework as a collection of definitions and theorems atop a collection of \textit{parameters}: sets, functions, and relations with specified relationships, but which are otherwise arbitrary.
On its own, the framework thus prescribes a highly abstract relational ontology for multi-agent systems: it names and structures the key concepts.
However, as users fill in the parameters, the framework is gradually refined to something concrete:
a software architecture, its executable implementation, and finally, a particular system configuration
at runtime.
The idea is that the framework thus offers a means for users to acquire these valuable artefacts.
The utility is that, by working within the parameters, the framework-generic definitions and theorems apply to all its (partial) instances.
The premise of our work is that users are able to fill in the parameters and that the resulting artefacts are desirable.

\Cref{fig:framework} visualises the parameters, whose structural constraints can be understood as the requirements for using the framework.
For example, \Cref{fig:framework} shows that the $\mathit{payload}$ parameter must be defined as a total, pure function from \textit{actions} to \textit{message lists}.
\Cref{lst:coq_ontology} shows the same ontology as it is formalised as in Rocq as parametrised data types, applicable functions over types, and relations over types, \ie, as functions to propositions.

\begin{figure}[bth]
{
\setlength{\abovecaptionskip}{0pt}
\centering
\begin{tikzcd}[cramped,column sep=5em,row sep=3em,math mode=false,cells={font=\normalsize\itshape},labels={inner sep=2pt,font=\footnotesize,pos=0.5}]
&[-2em]
fact
    \arrow[squiggly,dash,start anchor={[yshift=-16ex,xshift=5.2ex]center},end anchor={[yshift=12ex,xshift=5.2ex]center},blue]{d}[pos=0.05,inner sep=1em]{policy language}
&[-0.2em]
agent
    \arrow[squiggly,dash,start anchor={[yshift=-16ex,xshift=-7.5ex]center},end anchor={[yshift=12ex,xshift=-7.5ex]center},blue]{d}[pos=0.05,swap,inner sep=4.4em]{communication statics}
    \arrow[dashed]{l}[pos=0.55,swap]{affects}
&[+2em]
action
    \arrow[squiggly,dash,start anchor={[yshift=-16ex,xshift=6.6ex]center},end anchor={[yshift=12ex,xshift=6.6ex]center},blue]{d}[pos=0.05,swap,inner sep=1em]{dynamics}
    \arrow[]{d}[swap]{basis}
    \arrow[bend right=15]{dl}[sloped,near start,swap]{payload}
    \arrow[]{l}[swap]{actor}
&[+1em]
config
    \arrow[dashed]{l}[swap,pos=0.4]{enacted}
    \arrow[dashed,bend right=15]{ld}[sloped,swap,pos=0.3]{stated}
    \arrow[dashed,out=270,in=0]{ld}[pos=0.8]{agreed}
\\
$\Prop$
&
policy
    \arrow[dashed]{u}[]{truth}
    \arrow[]{l}[]{valid}
&
message list
    \arrow[]{l}[pos=0.46]{extract}
&
message
    \arrow[dashed]{l}[]{$\in$}
    \arrow[bend left=15]{ul}[sloped,near start]{author}
\end{tikzcd}
\vspace{-3.4em}
\caption{
Graphical depiction of the \textit{framework ontology} or \textit{signature} and its constituent collection of \textit{parameters}: sets, functions, and relations over sets to be user-defined.
Vertices (italicized text) depict the sets.
Each solid arrow $X \rightarrow Y$ depicts a pure, total, unary function from $X$ to~$Y$.
\Eg, each action has a unique \textit{basis} message.
Each dashed arrow $X \xdashrightarrow{R} Y$ depicts a binary relation $R \subseteq X \times Y$, \ie, a function $R: X \times Y \rightarrow \Prop$.
The arrows orient how relations are read. \Eg, agents \textit{affect} facts.
Blue squiggly lines group parameters into suggestively named clusters.
}
\label{fig:framework}
}
\end{figure}
\begin{figure}[tbh]
\begin{lstlisting}[language=Coq, backgroundcolor=\color{coqBackgroundColor}, multicols=2]
Module Type Ontology.

  (* policy language statics *)
  Parameters
    (policy fact: Type)
    (truth  : policy -> fact -> Prop)
    (valid  : policy -> Prop).


  (* communication statics *)
  Parameters
    (agent action message: Type)
    (author : message -> agent)
    (actor  : action  -> agent)
    (basis  : action  -> message)
    (payload: action  -> list message).
    
  (* the policy-communication bridge *)
  Parameters
    (affects: agent -> fact -> Prop)
    (extract: list message -> policy).

  (* dynamics *)
  Parameters
    (config : Type)
    (enacted: config -> action  -> Prop)
    (stated : config -> message -> Prop)
    (agreed : config -> message -> Prop).
End Ontology.
\end{lstlisting}
\refstepcounter{lstlisting}\label{lst:coq_ontology}
\addcontentsline{lol}{lstlisting}{\protect\numberline{\thelstlisting}Rocq formalisation of the JustAct ontology}
\makeatletter
\@makecaption{\lstlistingname~\thelstlisting}{Rocq formalisation of the parameters comprising the JustAct ontology shown graphically in \Cref{fig:framework}.}
\makeatother
\end{figure}

We expect system designers and implementers to fix all the JustAct parameters prior to the system's deployment and usage.
Then at runtime, the distributed configuration of the system corresponds to one configuration value (of the type $\mathit{config}$), and the evolution of the system over time traces a path through the configuration space.

\subsection{Separating the Study of the Framework from its Implementation}

The remainder of the section builds definitions and presents theorems atop the framework parameters, and remarks and demonstrates which definitions of parameters are intended and useful.
Roughly, the parameters are addressed in the order shown in \Cref{fig:framework} from left to right.

In \Cref{lst:coq_ontology} and henceforth, (only) listings in the Rocq language have this pink colour.

To aid in the presentation of framework concepts, we use illustrative concrete examples, \eg, of a particular message data structure. 
To avoid confusion now, we isolate these examples to \textbf{Example} environments (which our article uses in no other way), avoid presently dwelling on all their details, and remark on which details are presently relevant.
To avoid confusion later, these examples reflect the same instantiation of JustAct that we detail in \Cref{sec:slick,sec:impl} to come, \eg, wherein \Cref{def:message} precisely defines the message data type.

\subsection{The Policy Language}
\label{sec:framework_policy_lang}
The leftmost cluster of parameters in \Cref{fig:framework} comprise the \textit{policy language}: how agents model and reason about their system.
Intuitively, \textit{policies} are the syntactic objects encoding what agents require, believe, desire, intend, expect, etc.

The primary facet of a policy's semantics is the subset of facts it predicates as \textit{true}.
Intuitively, facts represent the evaluable Boolean policy-queries.
\Cref{eg:policies_facts_truth} shows some policies, facts, and truths after \textit{Slick} is fixed as the policy language, as is detailed in \Cref{sec:slick}, where \lstinline{bob reads data1} represents the Boolean query: does Bob read Data1?
Here, it suffices to observe that agents reason about the truth of facts and that truth is relative to a policy.

The secondary facet of a policy's semantics is validity: a subset of policies are \textit{valid}, characterising those with some fundamentally desirable characteristic such as `sensible' or `internally consistent'.
In \Cref{fig:framework}, we express this subset as a predicate over policies, such that any $\mathit{valid}(p)$ is a proposition.
\Cref{eg:policies_facts_truth} demonstrates how validity is like the truth of any fact: a semantic valuation purely determined by the given policy.
But the agents agree that truth has special significance; 
roughly, invalid policies are useless on their own.
Specifically, the validity of policies is a key component of permitted actions.
We focus on this connection and how agents reason about (and decide) validity and permission in \Cref{sec:permission_of_enacted_actions}.
However, policies (valid and invalid) are ubiquitous in the framework and in agent activities;
agents reason about and manipulate policies in various ways.
For example, note in \Cref{eg:policies_facts_truth} how larger policies are clearly (syntactically) composed of smaller policies.
As such, we have chosen to distinguish policies from valid policies.

\begin{figure}[htb]
\begin{exa}
\label{eg:policies_facts_truth}
Illustrative Slick policies (left) and all the facts (right) they are related to by \textit{truth} (lines).
Policies are valid unless \mlst{error} is true, which is an otherwise ordinary fact.

\noindent
\begin{tikzcd}[math mode=false,cells={font=\normalsize\itshape},labels={inner sep=2pt,font=\footnotesize,pos=0.5},column sep=5em,row sep=0.8ex,start anchor=east,end anchor=west]
&\phantom{thisishacky}
&
\slickalign{r}{error if bob reads data1.}
\\
\normalfont{(invalid)}&&
\slickalign{r}{error if not bob reads data1.}
    \arrow[dash]{r}[]{}
&
\slickalign{l}{error}
\\
&&
\slickalign{r}{error and amy says error if bob reads data1.}
&
\slickalign{l}{amy says error}
\\[+0.3ex]
&&
\slickalign{r}{bob reads data1 and bob says (bob reads data1).}
    \arrow[dash]{r}[]{}
    \arrow[dash]{rd}[]{}
&
\slickalign{l}{bob reads data1}
\\
\normalfont{(invalid)}&&
\makebox[3em][r]{$
{
{\renewcommand{\arraystretch}{0.7}
\begin{array}{l}
     \textnormal{\lstinline|bob reads data1 and bob says (bob reads data1).|}
     \\
     \textnormal{\lstinline|error and amy says error if bob reads data1.|}
\end{array}}}$}
    \arrow[dash]{uuur}[]{}
    \arrow[dash]{uur}[]{}
    \arrow[dash]{ur}[]{}
    \arrow[dash]{r}[]{}
&
\slickalign{l}{bob says (bob reads data1)}
\end{tikzcd}
\end{exa}
\end{figure}

\noindent 
We say that (all parameters comprising) the policy language are entirely \textit{static}, because they are not contextualised by any dynamic runtime configuration (\textit{config} in \Cref{fig:framework}).
Static terms are advantageous in practice, because they do not need to be recorded or communicated between agents at runtime.
For example, once an agent has (dynamically) sent a policy to their peers, all recipients necessarily agree on its (static) truth and validity. 

\subsection{Agent Messages Carrying Policies}
\label{sec:framework_messages}
The set of autonomous \textit{agents} are the entities comprising the multi-agent system.
We generally let these agents coincide with the networked processes that comprise the underlying distributed system.

Agents express themselves by \textit{authoring} \textit{messages} from which policies can be \textit{extracted}.
The intention is that each message is understood as the subjective assertion of its author.
This subjective context is evident in the message, by applying $\mathit{author}: \mathit{message} \rightarrow \mathit{agent}$.

\Cref{eg:message} 
uses the instance of JustAct that we define in \Cref{sec:slick,sec:impl} to show some message alongside its author and extracted policy.
This example demonstrates a useful pattern, where \textit{extract} seems to `inject' some contextual information into the resulting policy.
In this case, \textit{extract} injected policy syntax \mlst{and amy says error}, which was not originally inside the message;
$\mathit{extract}$ forced the policy to reflect its author.
Later, we demonstrate repeatedly how this pattern lays the groundwork for formalising power dynamics between agents.

\begin{exa}
\label{eg:message}
Pair $(\mlst{amy}, \ \mlst{error if bob reads data1})$ is a message whose author is \mlst{amy} and whose extracted policy is (\mlst{error and amy says error if bob reads data1}).
\end{exa}

\noindent
As with policies, messages are an entirely static data type.
Consequently, given the same message, all agents agree on which policy is extracted from it, and who is its author.
But on their own, messages have no effect.
Instead, they have effects indirectly, via actions.

\subsection{Agent Actions Carrying Messages}
Like messages, \textit{actions} are a static data type that is associated with a particular agent, who is called its \textit{actor}, in this case.
Also like messages, actions carry policies, in this case, via
\textit{payload} messages\footnote{
    In \cite{DBLP:conf/forte/EsterhuyseMB24}, the first version of this article, we called payloads \textit{justifications} because the extracted policies were \textit{only} used for permission.
    We have introduced \textit{payload} as a more accurately neutral term to reflect its roles in both permission \textit{and} effects.
    \Cref{sec:justification} gives our new formulation of `justification'.
}
from which a unique policy is extracted.
But unlike messages, actions represent the concerted effort of several agents: their payloads may include messages authored by several agents.
\Cref{eg:action} shows such a case: the payload of Bob's action includes $m_b$, a message authored by Amy.

\begin{exa}
\label{eg:action}
Action $(\mlst{bob}, m_b, [m])$ has 
actor \mlst{bob},
 basis $m_b$, and
 payload $[m_b, m_r, m]$, with
\begin{align*}
    m_b \coloneqq (\mlst{amy}, \ \mlst{error if bob reads data})
&&
    m \coloneqq (\mlst{bob}, \ \mlst{bob reads data1})
&&
    m_r \coloneqq (\mlst{bob}, \ \mlst{actor bob})
&&
\end{align*}
\end{exa}

\noindent 
Until \Cref{def:action} precisely defines the action type and payload function underlying \Cref{eg:action}, it suffices to understand what is says about Bob.
In determining the payload, Bob had extensive choice in including extra messages $[m]$ (from \textit{stated} messages), had little choice in including $m_b$ (from the limited \textit{agreed} messages), but no choice in including $m_r$, which was derived and injected by the $\mathit{payload}$ function, to reflect Bob as the actor.

Like messages, actions are statically defined, and do very little on their own.
But unlike messages, actions have direct effects when they are instantiated at runtime.

\subsection{Dynamic Configurations}
\label{def:dynamic_configurations}

The \textit{configurations} model the snapshots of the (distributed) system state that can be encountered at runtime.
Agents work to reason about and update their shared, \textit{current} configuration.
Intuitively, the configuration prescribes a subset of actions that are \textit{enacted} and subsets of messages that are \textit{stated} and \textit{agreed}.
At the outset, the configuration space is arbitrary, and these sets may be entirely unrelated.

However, our most significant results apply to configuration updates which are \textit{growing}, as per \Cref{def:growing}: 
while agreed messages are unconstrained, existing stated messages (`statements') and enacted actions are preserved.
Intuitively, statements and enacted actions represent \textit{historical} data: messages which have \textit{ever} been stated and actions which have \textit{ever} been enacted.
Later, we will discuss how these conceptual, infinitely growing collections do not imply that agents must really store or search through infinite collections.

\begin{defi}
\label{def:growing}
$\mathit{growing}(c,c': \mathit{config}) \coloneqq \mathit{stated}(c) \!\subseteq\! \mathit{stated}(c') \ \wedge \ \mathit{enacted}(c) \!\subseteq\! \mathit{enacted}(c')$.
\end{defi}
\begin{lem}
$\forall (c: \mathit{config}), \ \mathit{growing}(c,c)$.
\end{lem}
\begin{lem}
$\forall(c_1,c_2,c_3: \mathit{config}), \ \mathit{growing}(c_1,c_2) \wedge \mathit{growing}(c_2,c_3) \ \rightarrow \ \mathit{growing}(c_1,c_3)$.
\end{lem}

\subsection{Effects of Actions}

Effects are just what we call the mappings of the \textit{affects} function: data structures identifying \textit{affector}-fact pairs.
\Cref{def:encated_effect} lifts the dynamic \textit{enacted} property of actions to their associated effects:
effects of enacted actions are enacted.
\begin{defi}
\label{def:effect_of}
$\mathit{effect}\text{-}\mathit{of}(a, (\alpha,f)) \coloneqq \mathit{truth}(\mathit{extract}(\mathit{payload}(a)),f) \wedge \mathit{affects}(\alpha,f)$.
\end{defi}

\begin{defi}
\label{def:encated_effect}
$\mathit{enacted}\text{-}\mathit{effect}(c,e) \coloneqq  \exists a, \mathit{enacted}(c,a) \wedge \mathit{effect}\text{-}\mathit{of}(a,e)$.
\end{defi}

\noindent 
Effects are intended as `hooks' to external events.
For example, our case study models data processors reading and writing of datasets as effects.
As such, \Cref{thm:enacted_effect_pres_while_growing} characterises how enacted effects are preserved no matter how the configuration grows in the future.
Every effect has a determined \textit{affector} capable of enacting it, which must be determined and computable (by function $\mathit{affector}\text{-}\mathit{of}$).
Our case study demonstrates the value of letting auditors use this function, to trace enacted effects to the agents to be held accountable.

\begin{thm}
\label{thm:enacted_effect_pres_while_growing}
$\forall (c, c', e), \mathit{enacted}\text{-}\mathit{effect}(c,e) \wedge \mathit{growing}(c,c') \ \rightarrow \ \mathit{enacted}\text{-}\mathit{effect}(c',e)$.
\end{thm}

\noindent 
\Cref{eg:effect} demonstrates a suggestive case.
Statically, $(\mlst{bob}, \ \mlst{bob reads data1})$ is an effect of this action, so it is enacted once the action is enacted.
Then affector Bob works to externalize the effect by reading the data identified by \mlst{data1} in some external database.

\begin{exa}
\label{eg:effect}
Action $(\mlst{bob}, (\mlst{amy}, \mlst{error if bob reads data1}), [(\mlst{bob}, \mlst{bob reads data1})])$ has the effect $(\mlst{bob}, \ \mlst{bob reads data1})$, because $\mlst{affects}(a~\mlst{reads}~f) \coloneqq (a, \ a~\mlst{reads}~f)$ and because the truths extracted from its payload include $\mlst{bob reads data1}$.
\end{exa}

\noindent
We frame the desires of our autonomous agents as extrinsic desires to affect the world by enacting effects.
By \Cref{def:effect_of}, this requires solving a search problem: which actions which have the desired effects?
We expect prospective actors themselves to take on the burden of solving this search problem, because they are incentivised to solve it.
At the same time, they must balance these extrinsic motivations to act with the intrinsic obligation that JustAct prescribes: to act only as \textit{permitted}.

\subsection{Permission of Enacted Actions}
\label{sec:permission_of_enacted_actions}
We characterise \textit{well-behaved} agents as only enacting permitted actions, where \textit{permission} is defined by \Cref{def:permitted} as a relation over actions and configurations, as satisfying three criteria: \textit{valid}, \textit{sourced}, and \textit{based}.
To follow, we define each criterion in turn, and explain its role in giving agents control of which actions are permitted, and otherwise \textit{prohibited}.
From the perspective of the (prospective) actor, their obligation to stay well-behaved imposes a burden of proof, that each criterion is met.

\begin{defi}
\label{def:permitted}
$\mathit{permitted}(c: \mathit{config}, a: \mathit{action}) \coloneqq \mathit{valid}\text{-}\mathit{act}(a) \wedge \mathit{sourced}(c,a) \wedge \mathit{based}(c,a)$.
\end{defi}
\begin{defi}
\label{def:prohibited}
$\mathit{prohibited}(c,a) \coloneqq \neg \ \mathit{permitted}(c,a)$.
\end{defi}

\noindent
Firstly, \Cref{def:valid_act} characterises \textit{valid} actions, which simply lifts the validity of policy $\mathit{extract}(\mathit{payload}(a))$ to action~$a$.
Thus, in selecting the static definition of validity, the permissible actions are statically constrained.
For instance, once agents have agreed on the policy semantics sampled in \Cref{eg:policies_facts_truth}, without further communication, all agents agree that \Cref{eg:action} is invalid, and that enacting it would witness the misbehaviour of Bob.

\begin{defi}
\label{def:valid_act}
$\mathit{valid}\text{-}\mathit{act}(a: \mathit{action}) \coloneqq \mathit{valid}(\mathit{extract}(\mathit{payload}(a)))$.
\end{defi}

\noindent
Secondly, \Cref{def:sourced} characterises \textit{sourced} actions, whose payloads contain only statements.
The intention is that $\mathit{author}(m)$ is uniquely able to \textit{state}~$m$, \ie, grow the stated messages to include~$m$.
Thus, restricting statements restricts permitted actions: until~$m$ is stated, $\mathit{author}(m)$ prohibits all actions with payloads including~$m$.
Thus, actions generally require authors to cooperate in permitting a given action, by sourcing its payload.

While the \textit{sourced} property is not static, it is preserved no matter how the configuration grows (\Cref{lem:grows_pres_sourced}).
In practice, this lets agents reason meaningfully about the past and future configurations, even if the agents have only have partial information.

\begin{defi}
\label{def:sourced}
$\mathit{sourced}(c: \mathit{config},a:\mathit{action}) \coloneqq \forall m \in \mathit{payload}(a), \ \mathit{stated}(c,m)$.
\end{defi}

\begin{thm}
\label{lem:grows_pres_sourced}
$\forall (c,c',a), \
    \mathit{growing}(c,c') \wedge
    \mathit{sourced}(c,a) \ \rightarrow \
    \mathit{sourced}(c',a)$.
\end{thm}

\noindent
Finally, permitted actions are required to satisfy \Cref{def:weak_based}: they include \textit{some} agreed message (`agreement') in their payload.
This can be understood as the dual to \textit{sourced}, because it forces (agreed) messages \textit{into} payloads, where \textit{sourced} forces (un-stated) messages \textit{out of} payloads.
By guaranteeing that some agreement is included, control over agreements controls action payloads.
Later, we will see how our choice to name these \textit{agreements} is suggestive of their intended usage: infrequently, agents `reconfigure' their system by adjusting the shared agreements, such that undesirable policies are invalid, and hence, prohibited. 

Actually, we strengthen this criterion to \Cref{def:based} and call it \textit{based}, such that each action~$a$ itself identifies the agreement in question as $\mathit{basis}(a)$.
Conceptually, the difference between \Cref{def:weak_based} and \Cref{def:based} is small (\Cref{lem:based_impl_weak_based}), but this choice reflects our effort to move most of the burden of reasoning about actions from auditors (who check whether actions are permitted) to actors (who choose how they act). 

\begin{defi}
\label{def:weak_based}
$\mathit{weak}\text{-}\mathit{based}(c: \mathit{config},a:\mathit{action}) \coloneqq \exists m \in \mathit{payload}(a), \ \mathit{agreed}(c,m)$.
\end{defi}

\begin{defi}
\label{def:based}
$\mathit{based}(c,a) \coloneqq m \in \mathit{payload}(a) \wedge \mathit{agreed}(c,m)$ where $m \coloneqq \mathit{basis}(a)$.
\end{defi}

\begin{thm}
\label{lem:based_impl_weak_based}
$\forall (c: \mathit{config},a:\mathit{action}), \ \mathit{based}(c,a) \rightarrow \mathit{weak}\text{-}\mathit{based}(c,a).$
\end{thm}

\noindent 
Again, consider \Cref{eg:action}.
Bob may search for different actions which produce the same effect, but which are permitted, by considering different bases.
However, Bob has no chance if $(\mlst{bob},\ \mlst{error if bob reads data1})$ is the only agreement.
Necessarily, choosing any other basis preserves the action's prohibition, because it falsifies \textit{based}.

Note \Cref{lem:grow_nimpl_based_pres}, which observes that \textit{based} is generally \textit{not} preserved as configurations grow, because growing can remove existing agreements.
Consequently, growing a configuration sometimes changes which actions are permitted in the future.
Later, we shall see how agents use changes to the agreements to dynamically restrict which actions are permitted in the future.
But note a subtlety: changing agreements preserves each actor's well-behavedness; what matters is whether each of their actions was permitted \textit{when it was enacted}.

\begin{thm}
\label{lem:grow_nimpl_based_pres}
$\quad \mathbf{\neg} \  \forall (c,c',a), \ \mathit{growing}(c,c') \wedge \mathit{based}(c,a) \ \rightarrow \ \mathit{based}(c',a)$.    
\end{thm}

\noindent 
Nevertheless, results like \Cref{lem:grows_pres_sourced} let agents reason about permission of actions, up and down the timeline, despite having only partial dynamic information.
\Cref{lem:prospection} is particularly useful to agents: the permission of any based actions is preserved no matter how the configuration grows.
This reduces the burden on auditors to reason retroactively.
Knowledge that an action is currently not sourced suffices to prove that it was also not sourced in the past; hence it witnesses its actor's misbehaviour.
Lenient auditors can choose to assume the inverse: actions currently sourced were presumably sourced when they were enacted.\footnote{In \Cref{sec:impl}, we \textit{enforce} this inverse: enacted actions are \textit{sourced} by construction.
Consequently, auditors can enforce well-behavedness in retrospect without knowing which (other) messages were stated at the time.}
\Cref{lem:prospection} also gives actors peace of mind, because they can more accurately predict how auditors in the future will reason about permission in retrospect.

\begin{thm}
\label{lem:prospection}
$\forall (c,c',a), \
    \mathit{growing}(c,c') \wedge
    \mathit{permitted}(c,a) \wedge
    \mathit{based}(c',a)  \ \rightarrow \
    \mathit{permitted}(c',a).
$
\end{thm}

\subsection{Operations Handling Actions}
\label{sec:computable_operations}

For many purposes, it is sufficient -- even beneficial -- for the framework to remain abstract by characterising sets and relations in a declarative manner, and not specifying how agents represent, remember, and reason about their system.

However, atop the parameters depicted in \Cref{fig:framework}, we define two more parameters which refine the specifications of \textit{permitted} and \textit{effect-of} to computable functions, operationalising reasoning about these relations, to support key agent activities.
Precisely, implementers must define \Cref{par:dec_enacted_permitted,par:enum_effect_of}, which we call the \textit{operators}.

Firstly, \textit{enum-effects-of} maps any action~$a$ to an exhaustive list the effects of~$a$.
Note that this operator is inherently static, and does not depend on any particular runtime configuration, or the perspective of any particular agent.
This operator is essential to \textit{affector}~$\alpha$, who externalizes enacted effect~$(\alpha,f)$, \eg, of some action~$a$ after learning that $a$ is enacted.

\begin{parry}
\label{par:enum_effect_of}
$\mathit{enum}\text{-}\mathit{effects}\text{-}\mathit{of}: \forall(a: \mathit{action}), \ [e \ \mid \ \forall e, \ \mathit{effect}\text{-}of(a,e)] : \mathit{list~effect}$.
\end{parry}

\noindent
Secondly, the \textit{dec-permitted} function \textit{decides} whether the given action~$a$, enacted in configuration~$c$, was permitted or prohibited.
This formulation as a pure and total function necessitates that extra context (\eg, which agent evaluates the application of $\mathit{dec}\text{-}\mathit{permitted}$) does not change the result; consequently, all observers across time and space necessarily observe the same results.
As our formalisation is based on \textit{constructive logic}, the reader should understand that \textit{decisions}, \ie, terms of the form $P \mid \neg P$, are far from trivial: their evaluation produces one of two values: a \textit{proof} of proposition~$P$, or a proof of its inverse.

\begin{defi}[decision of proposition~$P$]
$P? \coloneqq \mathit{Prove}: P \rightarrow P? \mid \mathit{Disprove}: \neg P \rightarrow P?$
\\
and we coerce any decision to a Boolean value with $\{\mathit{Prove}(\_) \mapsto \top, \mathit{Disprove}(\_) \mapsto \bot \}$.
\end{defi}

\begin{parry}
\label{par:dec_enacted_permitted}
$\mathit{dec}\text{-}\mathit{permitted}: \forall(c,a), \ \mathit{enacted}(c,a) \ \rightarrow \ \mathit{permitted}(c,a)?$.
\end{parry}

\noindent 
At first glance, \Cref{par:dec_enacted_permitted} seems to imply the decidability validity for arbitrary policies, which we expect to be costly, because validity it is entangled with the semantics of policies.
But actually filling \Cref{par:dec_enacted_permitted} requires something weaker.
For auditors, it suffices that the policies extracted from enacted actions is decidable.
Notably, these are only the policies in the image of $\mathit{extract}$, which may have additional structure and properties.
Actors have an even weaker requirement,
because they are free to restrict their attention to actions of their choice.
For example, Bob may choose to avoid enacting \Cref{eg:action} after failing to prove its validity quickly enough.
However, to afford the tasks of auditors and well-behaved actors, system designers should consider the cost of reasoning about validity when selecting the policy language.
For example, we consider \slick to be desirable, in part because of \Cref{def:valid_dec}: the validity of any \slick policy is decidable.

Our Rocq formalisation includes theorems over various decisions of various properties.
From the logical perspective, these theorems reveal emergent relationships between parameters: decidability of some implies decidability of others.
But, owing to Rocq's foundations in constructivism, these same theorems can be seen from the computational perspective:
they define \textit{decision combinators}, which can help users to define the operators.
\Cref{lem:decide_via,lem_dec_stated_impl_dec_sourced} are noteworthy combinators which can help users to define $\mathit{dec}\text{-}\mathit{permitted}$.

\begin{lem}
\label{lem:decide_via}
$\forall (c,a), \
    \mathit{sourced}(c,a)? \wedge
    \mathit{based}(c,a)? \wedge
    \mathit{valid}\text{-}\mathit{act}(a)? \ \rightarrow \
    \mathit{permitted}(c,a)?$.
\end{lem}

\begin{lem}
\label{lem_dec_stated_impl_dec_sourced}
$\forall (c,a), \
    (\forall m: \mathit{message}, \ \mathit{stated}(c,m)?) \ \rightarrow \
    \mathit{sourced}(c,a)?$.
\end{lem}

\noindent
As the framework is instantiated, it becomes equally important to consider which properties are \textit{not} decidable, and which sets are \textit{not} enumerable, such that agents can work despite partial resources and knowledge.
Notably, filling in \Cref{par:dec_enacted_permitted,par:enum_effect_of} does \textit{not} require agents to know all the stated messages and enacted actions.
Thus, this information can be unfolded by autonomous authors and actors, and communicated asynchronously.

In contrast, agents must know more about the \textit{agreements} to support \Cref{par:dec_enacted_permitted}.
In \Cref{sec:impl}, we take a simple approach: 
agents' views of the agreements are \textit{synchronised}.\footnote{
    Here, we briefly illustrate that other approaches exist which reduce the burden on agents to synchronise.
    For example, $\forall a, \mathit{sourced}(a) \rightarrow \mathit{based}(a)?$ holds if $\mathit{agreed}(c,m) \coloneqq stated(c,m) \wedge \mathit{author}(m)=\alpha$ for some fixed~$\alpha$.
    But such agreements cannot be removed.
    We can bound their applicability if each action $a$ enacted at \textit{timestamp}~$t$ carries proof that $\alpha$ approved $\mathit{basis}(a)$ for~$t$, although these approvals are irrevocable.
    Then the agents must only synchronise on the current time, \eg, via the Global Positioning System (GPS).
}

\subsection{The Justification of Desired Actions and Effects}
\label{sec:justification}

\Cref{sec:computable_operations} reflected our primary concern of how actors handle actions already enacted:
permission is decidable, and affectors can enumerate enacted effects.
But the system only comes to life when agents are motivated to act.
Generally, we assume that agents are ultimately extrinsically motivated to enact desired effects, \eg, to read a particular medical dataset.
Necessarily, agents must explore hypothetical configurations, and realise them via the cooperation with the prospective authors and actors.
We characterise this pursuit as \textit{justification}: the search to identify configuration~$c$ and action~$a$ which \textit{justifies} a desired effect, and whatever steps are taken to reach~$c$ or to enact~$a$: realising the desired effects while preserving well-behavedness.
For example, actor~$\alpha$ may reason in terms of fixed agreement~$m$ and desired effect~$e$, and ask: which $\alpha$-authored statements are needed to build an $m$-based payload in order to justify~$e$?

\begin{defi}
\label{def:justifies}
$
\mathit{justifies}(c, a, e) \coloneqq\mathit{effect}\text{-}\mathit{of}(a, e) \ \wedge \ \mathit{permitted}(c,a).
$
\end{defi}

\noindent
In many JustAct instantiations that we imagine, and in that we define in \Cref{sec:slick,sec:impl} to follow, we accept that agents' pursuit to justify effects and actions is not defined as a deterministic function.
We prefer not to assume that agents maintain the necessary overview of the configuration, or the power to explore the necessary combinations of messages to be stated and agreed, and actions to be enacted.
Instead, we design our system implementations to make the expected justification processes tractable, and we embrace agent cooperation via their use of reasoning and communication channels outside the framework.

\section{The \slick Policy Language \& Interpreter}
\label{sec:slick} 

This section instantiates the JustAct framework parameters comprising the \textit{policy language}, discussed in \Cref{sec:framework_policy_lang}, concerning the syntax of policies and facts, the semantic \textit{truth} relation between them, and the semantic \textit{valid} predicate over policies.

In particular, this section fixes the policy language as \slick, a language presented in \cite{chris_thesis} by building on \cite{esterhuyse2024cooperative}, where policies are composed of parts from multiple agents.

\subsection{Design Considerations}

While many languages exist that can capture agents' norms of permitted behaviour in some sense or another, the particular role of policies in the JustAct framework imposes some noteworthy constraints.

Firstly, \Cref{fig:framework} asserts that \textit{truth} is a static relation over (only) policies and facts.
This requires that any `fact~$f$ \textit{is true in} policy~$p$' is a proposition expressible and evaluable without ambiguity, contradiction, or nondeterminism.
For example, the truths of the same policy cannot change over time.
Recall that this is necessary for auditors to always agree on permission (\Cref{lem:prospection}). 
Consider the problems that arise from framing \textit{Prolog} as a policy language and adopting Prolog's usual notion of \textit{truth} as the result of evaluating Boolean queries.
At first glance, it is a natural choice, because Prolog programs seem to predicate the truth of queries, suggesting a relational semantics.
However, Prolog's semantics is generally not well-founded: not all Prolog queries terminate, leaving some truths unspecified.
Unfortunately, agents cannot generally decide if given queries are terminating~\cite{DBLP:journals/iandc/AptP93,DBLP:journals/umcs/Sasak06}.
The undecidability of truth would impact definitions dependent on truth, \eg, the effects of actions.
Clearly, some adjustments are needed before Prolog would serve as a policy language.
For example, one approach is to restrict agents to a terminating fragment of Prolog.

Secondly, the chosen policy language must yield definitions of \textit{truth} and \textit{valid} that afford definition the framework operators: agents rely on key propositions being decidable, and key sets being enumerable.
For instance, auditors must decide permission (\Cref{par:dec_enacted_permitted}).
Our approach relies on validity being decidable (applying \Cref{lem:decide_via}).
Moreover, affectors must be able to enumerate the effects of any enacted action they observe (\Cref{par:enum_effect_of}).
Our approach is to encode effects as truths, and ensuring that each policy's truths are enumerable.

\subsection{Rocq Specification versus Rust Implementation of \slick}

Prior to this work, \slick was semi-formally specified in \cite{chris_thesis} and implemented in Rust~\cite{esterhuyse_2026_21240014}.
In the rest of this section, we present our novel formal specification~\cite{esterhuyse_2026_21240014} of \slick in terms of the JustAct framework, \ie, framing it as a policy language.
As in \Cref{sec:framework}, definitions in this section mirror our definitions in our Rocq artefact.
These are formulated to mirror the corresponding type- and function-definitions in Rust.
For example, the \slick abstract syntax (to follow) has essentially identical definitions in Rocq and Rust as inductive data types.

However, there are two facets of the Rust implementation which we omit in Rocq:
\begin{enumerate}
    \item the \slick concrete syntax and the function parsing policies from strings, and

    \item the definition of $\mathit{eval}$, underlying the semantic $\mathit{truth}$ relation over policies and facts.
\end{enumerate}

\noindent 
We have left these facets of \slick out of the scope of our formal definitions.
Nevertheless, the reader can rest assured that our definitions and theorems are self-contained, necessarily extending to any future refinement of the Rocq specification to include these details.

In the future, we indeed want to formalise these as an effort to unify our specification, implementation, and case studies into one cohesive, executable, and verified Rocq codebase.

\Cref{lst:slick} displays the API of our \slick interpreter, implemented as a Rust library, which includes methods for parsing policies from strings, and for computing their truths and validity (their `denotations').
This library is a dependency of the runtime system detailed in \Cref{sec:impl}.

In \Cref{lst:slick} and henceforth, (only) listings in the Rust language have this blue colour.

\begin{figure}[hbt]
{\setlength{\columnsep}{2pt}
\begin{lstlisting}[language=Rust, backgroundcolor=\color{rustBackgroundColor}, multicols=2]
enum Fact {Leaf(String), Node(Vec<Fact>)}
enum Atom {
  Literal(String),
  Variable(String),
  AtomNode(Vec<Atom>),
}
enum CondKind { Diff, Same }
struct Cond {
  atoms: Vec<Atom>,
  kind: CondKind,
}
type SignedCond = (bool, Cond);
struct Rule {
  consequents: Vec<Atom>,
  pos_antecedents: Vec<Atom>,
  neg_antecedents: Vec<Atom>,
  conditions: Vec<SignedCond>
}
struct Policy { rules: Vec<Rule> }

impl Policy {
  fn parse(text: &str) -> Option<Self>;
  fn denotes(self) -> Option<Denotation>;
}

struct Denotation {
  trues:    Vec<Fact>,
  unknowns: Vec<Fact>,
}  
impl Denotation {
  fn is_valid(&self) -> bool;
}
\end{lstlisting}}
\refstepcounter{lstlisting}\label{lst:slick}
\addcontentsline{lol}{lstlisting}{\protect\numberline{\thelstlisting}Simplified API of the Slick parser and interpreter}
\makeatletter
\@makecaption{\lstlistingname~\thelstlisting}{Simplified API of the \slick parser and interpreter implementation.}
\makeatother
\end{figure}

\subsection{Abstract Syntax}

The \slick language is a logic programming language related to existing languages including Datalog, Prolog, Clingo, and Souffl\'e.
Readers accustomed to these languages will probably find \slick's abstract syntax to be familiar.

\Cref{def:facts_and_atoms} fixes the abstract syntax of \textit{facts} as an inductively-defined data type:
trees whose leaves are \textit{string literals}.
Precisely, they are \textit{rosetrees}, whose nodes have any number of branches, expressed as lists of nested rosetrees.
In this, \slick makes a minor generalization of what is typical in untyped logic programming languages such as Clingo: \slick requires no literal at node level, and it does not distinguish between \textit{predicate symbols} (literals on the left of any root node) and \textit{function symbols} (literals on the left of any non-root node).

We define the \textit{atoms} as the facts whose literals (strings) may alternatively be \textit{variables} (also strings), which act as abstractions for arbitrary nested facts.
Semantically, atoms are used to \textit{match} facts.
Syntactically, policies are constructed from atoms.

For brevity, we omit atom and list constructors from some atoms shown in the article, to make them easier to read (and to resemble the \slick concrete syntax we give in \Cref{sec:concrete_syntax}).
For example, we abbreviate the atom $\mathit{Node}([x, \mathit{Leaf}(\mlst{"reads"}), y])$ with the notation $x~\mlst{"reads"}~y$.

\begin{defi}[fact and atom abstract syntax]
\label{def:facts_and_atoms}
\begin{align*}
\mathit{rosetree}(t) &\coloneqq \mathit{Leaf}(t) \mid \mathit{Node}(\mathit{list}(\mathit{rosetree}(t)))
&&&
\mathit{fact} &\coloneqq \mathit{rosetree}(\textsf{string})
\\
\mathit{constant} &\coloneqq \mathit{Var}(\textsf{string}) \mid \mathit{Lit}(\textsf{string})
&&&
\mathit{atom} &\coloneqq \mathit{rosetree}(\mathit{constant})
\end{align*}
\end{defi}

\noindent
\Cref{def:rule} fixes the \textit{unrestricted} abstract syntax of \slick \textit{inference rule schemas}, which we call `rules' for short.
Intuitively, each of these rules expresses the set of `concrete rules' resulting from consistently substituting each variable $\mathit{Var}(v)$ in the rule with any fact.

\begin{defi}[unrestricted rule syntax]
\label{def:rule}
$\mathit{rule} \coloneqq (\mathit{atom}, \mathit{list}(\mathit{sign} \times \mathit{cond}))$\normalfont{, where}
\begin{align*}
    \mathit{head}((h, b): \mathit{rule}) &\coloneqq h
    &&&
    \mathit{sign} &\coloneqq \mathit{Pos} \mid \mathit{Neg}
    \\
    \mathit{body}((h, b): \mathit{rule}) &\coloneqq b
    &&&
    \mathit{cond} &\coloneqq \mathit{True}(\mathit{atom}) \mid \mathit{Same}(\mathit{list}(\mathit{atom})) \mid \mathit{Diff}(\mathit{list}(\mathit{atom}))
\end{align*}
\end{defi}

\noindent
Finally, \Cref{def:policy} fixes the \textit{policy} as lists of \slick rules.
However, policy rules must satisfy \textit{safety}: every variable inside its head is also inside its body, where `inside' is formalised as $\mathit{has}\text{-}\mathit{var}$ in \Cref{def:has_var}.
Safety is a prevalent property in logic programming languages with \textit{bottom-up} inference semantics, \eg, \cite{DBLP:journals/aicom/GebserKKOSS11} adopts its definition in \cite{abiteboul1995foundations} for Clingo.
Safety has a mostly practical motivation: it affords a class of implementations that use rules to guide the incremental enumeration of truths as combinations of prior truths.

\Cref{def:policy} formalises policies as collections of safe rules.
This encoding models our implementation in Rust: unrestricted rules are parsed first, programmatically checked for safety second (\Cref{thm:safety_is_decidable}), but then no informative proof object is stored at runtime.\footnote{
    $\{ r : \mathit{rule} \mid \mathit{safe}(r)\}$ is the more general and natural representation of `a rule that is somehow proven safe' in Rocq.
    \Cref{def:policy} can be understood to fix \textit{how} safety is proven as a pure function of the rule, such that the proof object that witnesses safety (if it exists) is not informative, requiring no storage.
}

\begin{defi}[atom has variable]
\label{def:has_var}
$
\dfrac
{}
{\mathit{has}\text{-}\mathit{var}(v, \mathit{Var}(v))}
\quad 
\dfrac
{\mathit{has}\text{-}\mathit{var}(v, a)\hfill\phantom{x}}
{\mathit{has}\text{-}\mathit{var}(v, \mathit{Node}([\dots, a, \dots]))}
$
\end{defi}
\begin{defi}
$\mathit{safe}((h,b)) \coloneqq \forall v, \mathit{has}\text{-}\mathit{var}(v,h) \ \rightarrow \ \exists (\mathit{Pos}, \mathit{True}(a)) \in b, \ \mathit{has}\text{-}\mathit{var}(v,a)$.
\end{defi}

\begin{thm}[safety is decidable]
\label{thm:safety_is_decidable}
$\mathit{dec}\text{-}\mathit{safe}: \forall (r: \mathit{rule}), \ \mathit{safe}(r)?$.
\end{thm}
\begin{defi}
\label{def:policy}
$\mathit{policy} \coloneqq list(\{ r: \mathit{rule} \mid \mathit{dec}\text{-}\mathit{safe}(r) = \top \})$.
\end{defi}

\noindent
Below, we define \textit{fact rules}.
These characterise a subset of \slick rules which is tersely defined and easily understood, but whose unique properties afford unique usages.
The name is suggestive; $\mathit{fact\text{-}rule}(f)$ asserts the unconditional (semantic) truth of (syntactic) fact~$f$.
And \Cref{lem:rules_without_conds_are_safe} 
shows that may always be included in policies, because they are trivially safe, intuitively, because they never have variables.
Later our definition of $\mathit{payload}$ (in \Cref{def:action}) uses \Cref{def:safe_fact} to `inject' the assertions of derived facts into policies.

\begin{defi}
\label{def:fact_rule}
$\mathit{fact}\text{-}\mathit{rule}(f: \mathit{fact}) \coloneqq \mathit{Rule}(f, [\ ])$.
\end{defi}
\begin{lem}
\label{lem:rules_without_conds_are_safe}
$\mathit{fact\text{-}rule\text{-}\mathit{safe}}:
\forall (f:\mathit{fact}), \ \mathit{safe}(\mathit{fact}\text{-}\mathit{rule}(f)) = \top
$.
\end{lem}
\begin{defi}
\label{def:safe_fact}
$\mathit{safe\text{-}fact}(f: \mathit{fact}): \mathit{safe\text{-}rule} \coloneqq (\mathit{fact}\text{-}\mathit{rule}(f), \mathit{fact}\text{-}\mathit{rule\text{-}safe}(f))$.
\end{defi}

\subsection{Concrete Syntax}
\label{sec:concrete_syntax}

The \slick concrete syntax is simple, and adopts many conventions of the logic programming community.
An obvious difference is that \slick prefers alphabetic tokens \mlst{if}, \mlst{and}, \mlst{same}, and \mlst{diff} in place of the traditional tokens (\mlst{:-}), (\mlst{,}), (\mlst{=}), and (\lstinline|!=|), respectively.
For example, \mlst{wet if rained and not sunny} is a \slick rule.
For convenience, \slick's concrete syntax admits a generalization, where rule heads are conjunctive lists, just like rule bodies.
For example, \mlst{a and b if x and y} abbreviates the two rules \mlst{a if x and y. b if x and y}.

Because \slick atom-nodes require no literal strings, parentheses surround whole atom-nodes only as needed to separate them from outer nodes, rather than the convention of using parenthesis to separate each node's leftmost literal from its other arguments.
For example, the Clingo atom \lstinline{f(X, g(Y,Z))} is represented by \mlst{f X (g Y Z)} in \slick, \eg, just like many functional programming languages such as Haskell and Rocq.
This uniformity in node arguments affords variables in the leftmost position.
Consequently, our policies in \Cref{sec:case_study} frequently adopt the word order of natural language, as in \mlst{error if Anyone reads Data and bob controls Data}.

Our Rust implementation goes beyond our specification of \slick's abstract syntax, by parsing \slick concrete syntax from text.
\Cref{lst:parser} gives a glimpse of the parser implementation.
Naturally, this snippet reflects details beyond the focus of this article; \eg, atoms require two mutually-recursive parsers to avoid runway recursive descent at runtime.

\begin{figure}[hbt]
{\setlength{\columnsep}{2pt}
\begin{lstlisting}[language=Rust, backgroundcolor=\color{rustBackgroundColor}, multicols=2]
// e.g.: "bob reads (data 1)"
fn atom(s: In) -> Res<In, Atom> {
  let tuple = nmap(many_m_n(2,
    usize::MAX, subatom), Atom::Tuple);
  first_match((tuple, subatom))(s)
}
fn subatom(s: In) -> Res<In, Atom> {
  let parenthesized = delimited(
    nchar('('), atom, nchar(')'));
  let v = nmap(variable, Atom::Variable);
  let c = nmap(constant, Atom::Constant);
  let w = nmap(wildcard,
    |_| Atom::Wildcard);
  first_match((parenthesized,v,c,w))(s)
}
// atom | "not" atom | atom "<" atom ...
fn antecedent(s: In) -> Res<In, Ante> {
    let p = nommap(atom, Ante::Pos);
    let n = nommap(
      preceded(neg,atom), Ante::Neg);
    let c = nommap(check, Ante::Compare);
    first_match((c, p, n))(s)
}
// e.g., "error if bob reads (data 1)."
fn rule(s: In) -> Res<In, Rule> {
    let h = many0(
        preceded(option(sep), atom));
    let right = many0(
      preceded(opt(sep), antecedent));
    let b = nommap(
      opt(preceded(ifsep,right)),
      Option::unwrap_or_default);
    nommap(terminated(pair(h,b),rulesep),
      |(head,body)| Rule {head, body})(s)
}

fn policy(s: In) -> Res<In, Policy> {
    nmap(many0(rule), Policy)(s)
}
\end{lstlisting}}
\refstepcounter{lstlisting}\label{lst:parser}
\addcontentsline{lol}{lstlisting}{\protect\numberline{\thelstlisting}Simplified excerpt of the Slick parser}
\makeatletter
\@makecaption{\lstlistingname~\thelstlisting}{A simplified excerpt of the Rust implementation of the \slick (concrete syntax) parser atop the popular \texttt{nom} parser combinator library.}
\makeatother
\end{figure}

\subsection{Semantics: Truth}
\label{sec:slick_semantics}

\Cref{def:truth} the fixes the \textit{truth} relation: they are the elements of $\mathit{eval}(p)$, which is a finite enumeration of all the truths in any policy~$p$, by definition.

\begin{parry}[enumerator of policy truths]
This is the only parameter of our Rocq formalisation which we never define.
Our definitions and theorems generalise over its value.
\label{par:eval}
\[\mathit{eval}: \mathit{policy} \rightarrow \mathit{list}(\mathit{fact})\]
\end{parry}
\begin{defi}
\label{def:truth}
$\mathit{truth}(p : \mathit{policy},f : \mathit{fact}) \coloneqq f \in \mathit{eval}(p)$.
\end{defi}

\noindent 
It is not difficult to create a useful formal definition of $\mathit{eval}$.
But it is more difficult to formalise precisely what is already implemented in Rust.
To follow, we give an account of the \slick semantics that suffices to understand the policies in our case study.

\subsubsection{Applying Rules}

Essentially, $\mathit{eval}$ is at the heart of the semantics, realising the reasoning procedure that characterises bottom-up logic programming in languages including Souffl\'e and Clingo.
Reasoning builds up the enumeration of \textit{true} facts from an initially empty set, and adds new truths, one at a time, to a fixed point, \ie, until no new truths are possible.
At each step, some program rule is selected and concretised (its variables are consistently substituted with variable-free atoms) such that each condition comprising its body are \textit{satisfied}.
Afterwards, the rule's head is true.
Rule conditions encode logical formulae over the truths.
For example, \slick rule $\mlst{error if bob reads X and ready}$ is applicable once concretised with $\{ \mlst{X} \mapsto \mlst{data 1}\}$ if $\mlst{bob reads (data 1)}$ and $\mlst{ready}$ are already true, making $\mlst{error}$ true.
Rules are satisfied when their conditions are all satisfied, and this depends on the case of the condition;
$\mathit{True}(a)$ is satisfied iff $a$ is true (iff it is not preceded by $\mathit{Neg}$), and $\mathit{Same}(l)$ and $\mathit{Diff}(l)$ are satisfied iff the atoms in~$l$ are all pairwise syntactically the same and different, respectively.\footnote{Note that $(\mathit{Neg},\mathit{Same}([X,Y,Z]))$ has a different meaning to $(\mathit{Pos},\mathit{Diff}([X,Y,Z]))$ when the list elements are neither all the same nor all different, \eg, after substitution $\{X \mapsto \mlst{amy}, Y \mapsto \mlst{amy}, Z \mapsto \mlst{bob}\}$.}

\subsubsection{Negation}

Logic programming languages differ in their treatment of \textit{negated} rule conditions, striking different compromises.
\slick adopts Van Gelder's \textit{well-founded semantics}~\cite{van_gelder_well-founded_1991,przymusinski_well-founded_1990}, and the Rust implementation uses the \textit{alternating fixpoint} algorithm for this semantics~\cite{van_gelder_alternating_1989}.
The choice of this semantics over alternatives is detailed in \cite{chris_thesis}.
In short, this semantics exhibits a desirable robustness, because it attributes meaning to programs expressing logical contradictions.
We expect such contradictions to arise in our context, when programs express composite policies, whose rules were created by different authors, with limited understanding of their peers' contributions~\cite{esterhuyse2024cooperative}.
For example, consider the program to which Amy contributes $\mlst{amy leads if bob leads}$, Bob contributes $\mlst{bob leads if not amy leads}$, and Dan contributes $\mlst{dan leads}$.
Does Amy lead or not?
Yes and no are both unsatisfactory.
As such, the popular \textit{stable model semantics}~\cite{DBLP:conf/iclp/GelfondL88} (which underlies answer-set solvers like Clingo) gives this program a trivial evaluation, leaving all queries unanswered.
In comparison, the well-founded semantics isolates contradictory facts by giving them a third \textit{unknown} value.
The other facts are evaluated to \textit{true} or \textit{false}, as usual, reflecting a consistent application of the rules.
In this example, the \slick interpreter warns that whether Amy and Bob lead is unknown, but $\mlst{dan leads}$ is certainly true.

For our purposes, it suffices to conflate false and unknown: they are not true.

\subsubsection{Termination}

If left unrestricted, some \slick policies (like Prolog programs) have no meaning, because their evaluation never terminates.
For example, \mlst{f X if X. x} induces an endless chain of inference steps, inferring the truth of infinite facts matching \mlst{f (f (f}~$\dots$~\mlst{f x}~$\dots$~\mlst{)))}.
Unfortunately, this problem is endemic in sufficiently expressive logic programming languages, and has been explored for decades~\cite{DBLP:journals/jlp/SchreyeD94}.
For our purposes, where agents have the luxury to decide which policies they wish to express and communicate, a na\"ive solution inspired by that in \cite{DBLP:journals/jlp/SchreyeD94} is sufficient:
the number of inference steps is limited to a statically fixed \textit{bound} (which we fix at 30,000 in our case study).
Whenever $\mathit{eval}(p)$ exceeds the bound, $\mathit{eval}(p) = [\mlst{error}]$.
Next, we explain how these policies are \textit{invalid}, hence safely ignored.

\subsection{Semantics: Validity}

\Cref{def:valid} completes the policy language.

\begin{defi}
\label{def:valid}
$\mathit{valid}(p) \coloneqq \neg \ \mathit{truth}(p,\mlst{error})$.
\end{defi}

\noindent 
As in \cite{esterhuyse2024cooperative}, this simple definition suffices to let agents specify subtle requirements on their peer's behaviour.
Intuitively, ordinary \slick rules with \mlst{error} at the head recognise conditions for invalidity.
In \Cref{sec:case_study}, we repeatedly demonstrate this pattern, and because \Cref{sec:impl} defines \textit{extract} and \textit{payload} to force policies to reflect their message- and action-contexts, these rules can express powerful constraints on enacted actions and effects.
Consider \Cref{eg:policies_facts_truth} once again, and note the rules constraining validity;
consider how injecting \mlst{error if bob reads data1} into any policy makes it useless whenever \mlst{bob reads data1} must be true.

Because $\mathit{eval}$ enumerates truths, \Cref{def:valid_dec} is easy to prove: validity is decidable.

\begin{lem}
\label{def:valid_dec}
$\mathit{dec}\text{-}\mathit{valid}: \forall (p: \mathit{policy}), \ \mathit{valid}(p)?$.
\end{lem}

\subsection{Reflecting Authorship in Rules}
\label{sec:reflecting_authorship_in_rules}
$\mathit{extract}: \mathit{message} \rightarrow \mathit{policy}$ is one of the two connections between policy language and communication statics.
To minimize the coupling between \Cref{sec:slick,sec:impl}, \Cref{def:reflect_author} presently fixes an intermediate definition: $\mathit{reflect}\text{-}\mathit{author}(r,a)$ is the transformation of a single syntactic \slick rule~$r$ to reflect its authorship by agent~$a$.
Put simply, $\mathit{reflect}\text{-}\mathit{author}$ pushes the head~$h$ of the given rule inside atom $a$~\mlst{says}~$h$, where $a$ is the author.
Readers may recognise the inspirations from typical modal and epistemic logics, \eg, as presented in \cite{fagin2004reasoning}, where the formula $K_A\phi$ proposes that  agent~$A$ \textit{knows}~$\phi$.

\Cref{lem:reflect_preserves_safety} proves that this transformation preserves the safety of the given \slick rule, \ie, the transformed rules may contribute to policies.
Finally, \Cref{lem:safe_reflect_author} lifts the $\mathit{reflect\text{-}author}$ function from unrestricted rules to safe rules.
This is what we use later.

\begin{defi}
\label{def:reflect_author}
$\mathit{reflect}\text{-}\mathit{author}(\mathit{Rule}(h, b), \ a) \coloneqq \mathit{Rule}(a~\mlst{"says"}~h,\ b)$.
\end{defi}
\begin{lem}
\label{lem:reflect_preserves_safety}
$
\mathit{reflect\text{-}safe} : \forall (r,a),\ \mathit{safe}(r) \ \rightarrow \ \mathit{safe}(\mathit{reflect}\text{-}\mathit{author}(r,a))? = \top$.
\end{lem}
\begin{defi}
\label{lem:safe_reflect_author}
$
\mathit{safe\text{-}reflect\text{-}author}((r,s), a)
\coloneqq
(\mathit{reflect}\text{-}\mathit{author}(r,a), \mathit{reflect\text{-}safe}(r,a,s))$.
\end{defi}

\noindent 
For \Cref{def:reflect_author} to encode agents in policies, \Cref{def:agent} identifies agents as facts.
For simplicity, we choose to identify agents only as strings, the simplest facts.
In \Cref{sec:impl} we discuss, and in \Cref{sec:case_study} we repeatedly demonstrate, how this connection between the worlds inside and outside policies lets them serve agents as meaningful \textit{meta}-policies: \slick rules specify which rules combine to comprise valid policies.  

\begin{defi}
\label{def:agent}
$\mathit{agent} \coloneqq \mathit{fact}$.
\end{defi}

\section{Generic Data Exchange Runtime System}
\label{sec:impl}

\lstset{
    language={slick}
}

This section instantiates the JustAct framework's communication statics and dynamics, shown in \Cref{fig:framework}.
In both our Rust and Rocq encodings~\cite{esterhuyse_2026_21240014}, the loose coupling with the policy language is expressed as abstraction, \eg, in Rust as functions generic over any type \mlst{P} implementing the \mlst{Policy} trait.
Dependencies on \slick in particular manifest as (only) \Cref{def:reflect_actor,def:extract,def:affects} using definitions from \Cref{sec:slick}.
Put together, \Cref{sec:slick,sec:impl} complete the system which we apply to particular usage scenarios in \Cref{sec:case_study}.

\begin{figure}[tbh]
    \centering
    \includegraphics[width=0.75\columnwidth]{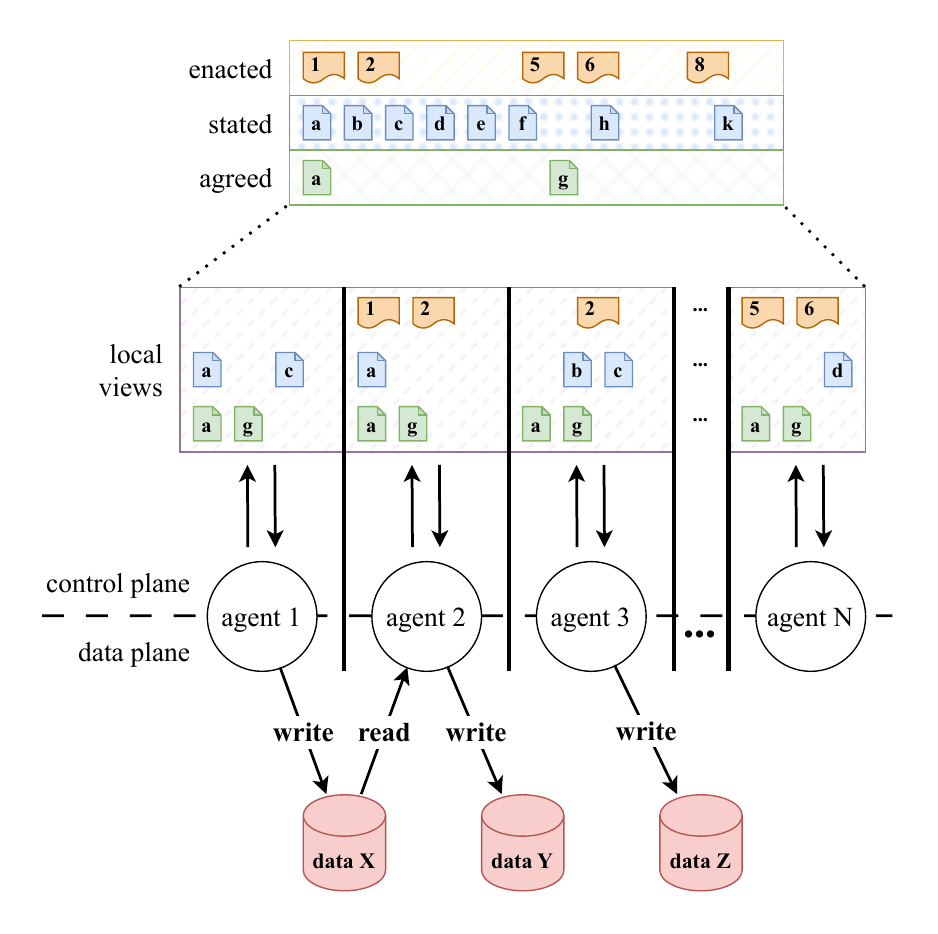}
    \vspace{-10pt}
    \caption{
    Example snapshot of our multi-agent system at runtime.
    In the control plane (above), agents interact via their partial views of their shared system configuration, \eg, which statements exist.
    Agents realise the effects of actions enacted in the control plane by reading and writing the assets shared in the data plane (below).
    But agents maintain consist views of the agreements.
    Here, $\{1, 2, 3, ... , 8\}$ and $\{a,b,c, ..., k\}$ identify distinguishable actions and messages, respectively.
    }
    \label{fig:system_impl}
\end{figure}

The runtime system is designed for its application to \textit{data exchange} use cases: it maintains store of \textit{asset data} or \textit{assets} that is shared by the agents, and it mediates agents' communications and access to assets.
Conceptually, and in the implementation, the system separates these concerns into two layers, as is visualised in \Cref{fig:system_impl}:
\begin{enumerate}
    \item \textbf{The Data Plane}: Agents are motivated to create and share \textit{data assets} with their peers. This is achieved by agents reading and writing assets in a shared asset store.
    \item \textbf{The Control Plane}: Agents create and share metadata in the control plane, which ultimately regulates the activities in the data plane.
    This metadata is partly centralised (agreements) but otherwise decentralised and communicated via explicit control messages.
\end{enumerate}

\noindent 
The concepts JustAct primarily concern the control plane.
Now \textit{messages}, \textit{actions}, and \textit{policies} introduced in \Cref{sec:framework} are vehicles for agent coordination and control.
And now \textit{effects} connect the control and data planes:
\textit{effects} \textit{enacted} in the control plane are externalised 
by the affectors (agents) as asset-reading and -writing events in the data plane.

\subsection{Design Considerations}
\label{sec:impl_design_considerations}

Data exchange systems are subject to many design considerations before they can be implemented in detail.
Here, we explore particular design considerations, and explain the design decisions reflected in our contributions.

Firstly, \textbf{how is the (meta)data managed}?
The software architecture and physical infrastructure underlying the data exchange runtime system has significant consequences for which agent behaviour is realizable, which dynamic (meta)data is accessible, and which dynamic system properties are provable, verifiable, and decidable.
Our Rocq and Rust implementations strike the same compromise:
agents have synchronized views of the agreements, but must explicitly forward statements to their peers.
In other words, our implementation adopts a distributed software architecture, but a centralised physical infrastructure.
This approach lays the groundwork for a fully distributed implementation, but postpones the effort and minutia that practical distributed implementations require.

Secondly, \textbf{how is agents' access to data specified}?
The expression and understanding of how agents should behave is aligned with the definition of \textit{well-behavedness} provided by the JustAct framework:
actors may only enact permitted actions, and affectors only externalize enacted effects.
This has two notable consequences.
Firstly, the data exchange runtime system adopts the view that actors are burdened to prove their well-behavedness, as actors alone define actions, whose contents suffice for an observer to check permission.
Consequently, the result of the audit is independent of the identity of the auditor.
Secondly, the specification of agent behaviour is expressed via the policy language, and dynamically (re)configured with changes to the dynamic configuration, \eg, as authors create new statements at runtime.

\begin{figure}[bth]
    \centering
    \includegraphics[width=1\columnwidth]{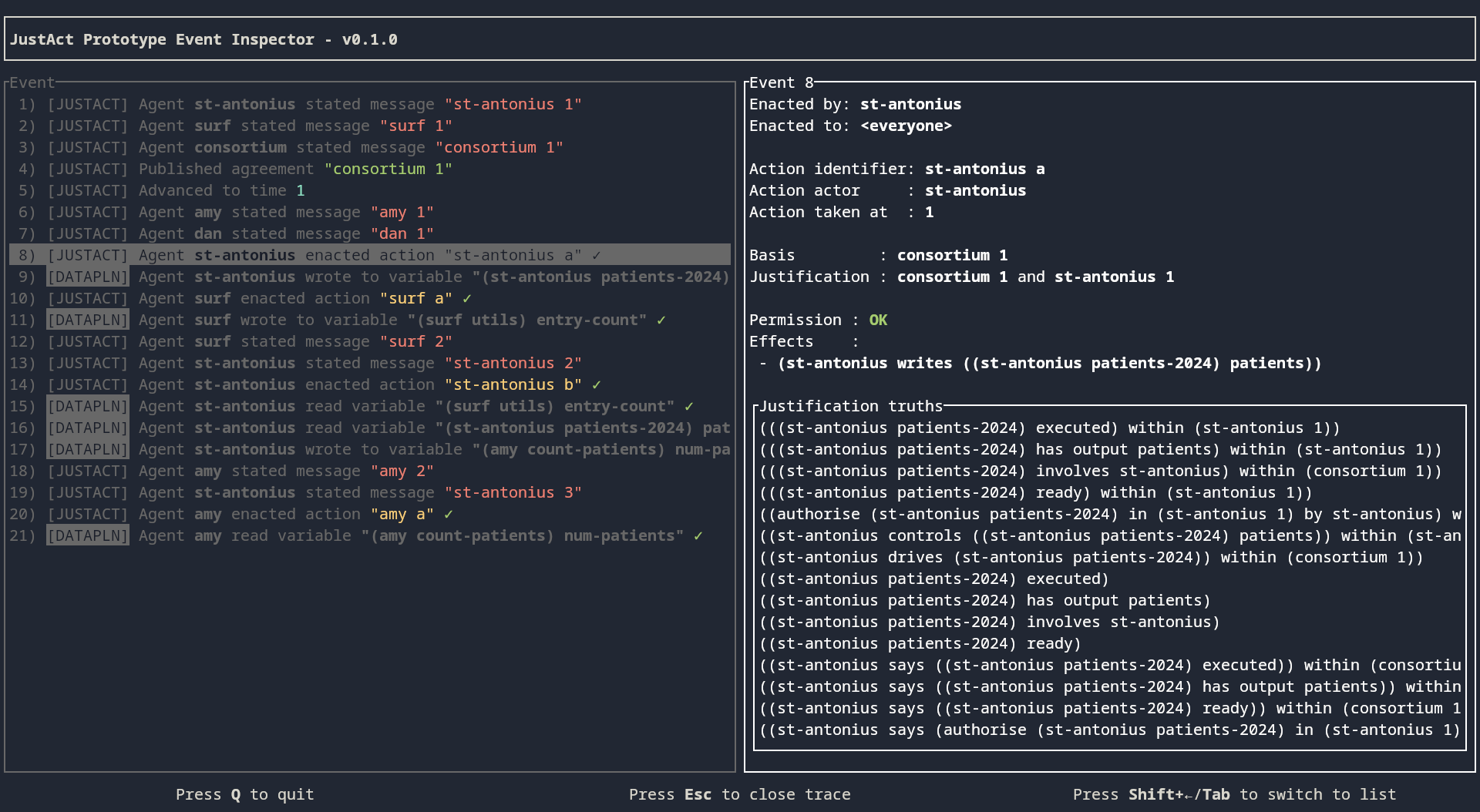}
    \vspace{-10pt}
    \caption{Screenshot from the interactive trace inspector, included in the Rust implementation~\cite{esterhuyse_2026_21240014}.
    The left window shows a all data- and control-plane activity so far.
    The right window scrutinises a selected enacted action, showing its actor, its payload, a breakdown of the permission decision, and an enumeration of its effects.
    Here, the trace produced by the first case study scenario (\Cref{sec:scenario:isolated_execution}) is shown.
    }
    \label{fig:prototype-tui}
\end{figure}

Finally, \textbf{how are agents' access to data controlled}?
Recall that the value of JustAct is that it fixes an unambiguous notion of well-behavedness which affords auditing, motivated by the need to enforce well-behavedness.
However, JustAct leaves the details of enforcement unspecified.
Many approaches to enforcement are available, and a wealth of existing literature explores their complex relationships to agent incentives and computation burdens on system components.
For example, when misbehaviour is \textit{prevented} (ex-ante enforcement), agents can reason in absolutes about (im)possibility via (im)permissibility, \eg, affording their assessment of the risks of harm incurred by misbehaviour. 
When misbehaviour is instead monitored and detected after the fact (ex-post enforcement), new actions may arise in \textit{compensation} (\eg, punishing the bad actor), and permitted actions may be further constrained in the future.
Ex-post enforcement is also noteworthy for leaving room for separating the enforcement from the runtime system implementation.
We leave these considerations out of scope of this article, and implement an enforcement mechanism that suffices to support our experiments and emphasise the unified view of permission:
the Rust implementation lets agents behave arbitrarily, but logs a trace of all events so far, to enable perfect (external) audits after the fact.
The logs suffice to recognise and diagnose all misbehaviour.
Our Rust implementation includes a supplementary tool for performing and visualising these diagnoses~\cite{esterhuyse_2026_21240014}.
An example of the visualisation is provided in \Cref{fig:prototype-tui}.
The details of enforcement in the data- and control-planes are discussed in \Cref{sec:impl_effects,sec:agent_knowledge_and_control_updates}, respectively.

\subsection{Effects in the Data Plane}
\label{sec:impl_effects}

Agents are ultimately motivated to read and write assets in the data plane.
We connect these to (the JustAct concepts in) the control plane with \Cref{def:affects}, which fixes the static \textit{affects} relation over $\mathit{agent} \times \mathit{fact}$, such that there are \textit{read} and \textit{write} effects, each identifies its subject~$a$ (the \textit{affector}) and the asset in the data plane, as identified by an arbitrary \slick fact~$f$.
This definition is practical, because it makes it easy for affectors to enumerate the effects of a given fact; there is one at most.
Because \slick is defined such that a policy's truths are enumerable, it is then a small step to filling in  \Cref{par:enum_effect_of}: enumerate the effects of a given fact.

\begin{defi}
\label{def:affects}
\raisebox{1.2ex}{$
\dfrac
{}
{\mathit{affects}(a, \ a~\mlst{"reads"}~f)} \quad \dfrac
{}
{\mathit{affects}(a, \ a~\mlst{"writes"}~f)}$}
\end{defi}

\noindent
Because our Rust implementation manages a centralised store of asset data, the agents necessarily have consistent views of the assets.

At runtime, agents input a stream of \textit{access requests} to access data.
With~$(v,d)$, an agent~$\alpha$ requests to access ($v=\mlst{"read"}$ or $v=\mlst{"write"}$) the asset in the data plane identified by~$f$, which is mapped to the effect~$e \coloneqq (\alpha, \alpha~v~f)$.
The runtime system's reaction is fixed in implementation:
the request is realized only if~$e$ is \textit{enacted} in the control plane.
But the burden of proof is on affector~$\alpha$:
the request is additionally parametrised by $\alpha$'s choice of action~$a$, and the system only verifies that $a$ is enacted, and that $(\alpha, f)$ is an effect of~$a$.

After the fact, auditors can inspect the system's logs to verify that the correspondence between access events in the data plane and effects in the control plane were preserved.

In the remainder of the section, we detail the control plane, by mapping the runtime system state to configurations, how agents reason about and update these configurations.

\subsection{Control Messages and Actions}
\label{sec:impl_statement_gossip}

The static set of \textit{messages} is defined as a simple record data type, whose fields identify the message \textit{author} (an agent), and the message \textit{contents}, an arbitrary policy, chosen by the author.
\Cref{def:extract} defines the policy \textit{extracted} from a message as a minor transformation of its contents:
for every given message~$m$ with given (safe) rule~$r$, another rule is injected, a version of $r$ that reflects the authorship of~$a$ (\Cref{lem:safe_reflect_author}).
Effectively, the image of $\mathit{extract}$ is a subset of the syntactic policies.
As only extracted policies are ultimately relevant to action permission and effects, agents can only reach -- and only have to consider -- this subset of policies.
So authors cannot assert any chosen (un)conditional truth without their authorship also being evident.
Consider how the perception of \mlst{dan is trusted} may depend on whether \mlst{dan says (dan is trusted)} is true!

\begin{defi}
\label{def:message}
$\mathit{message} \coloneqq \mathit{agent} \times \mathit{policy}$, with projections
\begin{align*}
    \mathit{author}((a,p)): \mathit{agent} &\coloneqq a
    &
    \\
    \mathit{contents}((a,p)): \mathit{policy} &\coloneqq p
\end{align*}
\end{defi}

\begin{defi}
\label{def:extract}
$\mathit{extract}(m) \coloneqq [r, \mathit{safe\text{-}reflect\text{-}author}(r,\mathit{author}(m)) \mid \forall r \in \mathit{contents}(m)]$.
\end{defi}

\noindent
The static set of \textit{actions} is defined similarly, as a record data type, whose fields identify the \textit{actor} (an agent), the \textit{basis} (a message), and the \textit{extra} messages (a message list).
In both Rocq and Rust, we use the standard polymorphic \textit{list} data type to define the \textit{message lists} in terms of messages, in the usual manner.
\Cref{lst:message_and_action} gives a glimpse of how (operators on) messages and actions are implemented in Rust; they generalize over the policy type~\mlst{P}. 

\begin{defi}
\label{def:reflect_actor}
$\mathit{reflect}\text{-}\mathit{actor}(a: \mathit{agent}): \mathit{message} \coloneqq (a, [\mathit{safe\text{-}fact}(\mlst{actor}~a)])$.
\end{defi}
\begin{defi}
\label{def:action}
$\mathit{action} \coloneqq \mathit{agent} \times \mathit{message} \times \mathit{list}(\mathit{message})$\normalfont{, with projections}
\[
\begin{array}{@{}l@{\qquad}l@{}}
    \mathit{actor}((A,B,E)) \coloneqq A
    &
    \mathit{basis}((A,B,E)) \coloneqq B
    \\
    \mathit{extra}((A,B,E)) \coloneqq E
    &
    \mathit{payload}(a) \coloneqq [\mathit{basis(a)}, \mathit{reflect}\text{-}\mathit{actor}(\mathit{actor}(a))] \dplus \mathit{extra}(a)
\end{array}
\]
\end{defi}

\begin{figure}[tbh]
{\setlength{\columnsep}{2pt}
\begin{lstlisting}[language=Rust, backgroundcolor=\color{rustBackgroundColor}, multicols=2]
#[derive(Clone, Debug, Eq, Hash,
    PartialEq, Deserialize, Serialize)]
pub struct Message<P: ?Sized + ToOwned> {
    pub author: String,
    pub payload: P::Owned,
}

impl<P: ?Sized + PolicySerialize
      + ToOwned> Message<P> {
    pub fn serialize(&self)
    -> Message<str> {
        Message { author_id:
            self.author_id.clone(),
        payload: self.payload
            .borrow().serialize() } }
}

#[derive(Clone, Debug, Eq, Hash,
    PartialEq, Deserialize, Serialize)]
pub struct Action<P: ?Sized + ToOwned> {
    pub actor: String,
    pub basis: Arc<Message<P>>,
    pub extra: justact::MessageSet<
        Arc<Message<P>>>,
}
\end{lstlisting}}
\refstepcounter{lstlisting}\label{lst:message_and_action}
\addcontentsline{lol}{lstlisting}{\protect\numberline{\thelstlisting}Rust encoding of the message and action datatypes}
\makeatletter
\@makecaption{\lstlistingname~\thelstlisting}{Rust encoding of the message and action datatypes, along with (some of) their operators, \eg, serialising messages to byte streams, and checking the equality of messages.}
\makeatother
\end{figure}

\noindent
\Cref{def:action} leads to \Cref{lem:payloads_contain_bases}: action's payloads necessarily include their bases.
This is convenient, as actors cannot help but implicitly satisfy this first conjunct of the \textit{based} criterion of permission (\Cref{def:based}).
But the second conjunct still depends on the dynamic configuration: is the basis an \textit{agreement} (in the current configuration)?
\begin{lem}
\label{lem:payloads_contain_bases}
$\mathit{payloads}\text{-}\mathit{contain}\text{-}\mathit{bases}: \forall (a: \mathit{action}), \ \mathit{basis}(a) \in \mathit{payload}(a)$.
\end{lem}

\noindent 
Similarly to how \textit{extract} is defined in terms of the message \textit{content} to reflect its author, \textit{payload} transforms the actor's chosen \textit{extra} messages to reflect the actor: $\mathit{payload}$ supplements the extra messages with $\mathit{reflect}\text{-}\mathit{actor}(a)$, an unconditional (safe) \slick rule asserting that $a$ is the actor.
Actors necessarily identify themselves in the policies extracted from their actions.
Agents can rely on this when choosing their message contents.
For example, statements including rule \mlst{error if actor dan} necessarily never contribute to Dan's permitted actions.

\subsection{Dynamic Configurations and Computability}

\Cref{def:config} uses conventional mathematical notation to capture the essence of (our Rocq formalisation of) our system implementation in Rust:
each system configuration predicates the \textit{enacted} actions, the \textit{stated} messages, and the \textit{agreed} messages.
However, we express the \textit{agreements} uniquely, as a \textit{list} which is computable from the configuration.
This makes clear that the \textit{agreed} messages are enumerable, and that checking a message for agreement is decidable, in any configuration.

\begin{defi}
\label{def:config}
$\mathit{config} \coloneqq (\mathit{action} \rightarrow \Prop) \times (\mathit{message} \rightarrow \Prop) \times \mathit{list}(\mathit{message})$,\\with projections
\begin{align*}
    \mathit{enacted}((E,X,A),a) &\coloneqq E(a)
    \\
    \mathit{stated}((E,X,A),m)&\coloneqq X(m) \ \vee \ \exists a, \ \mathit{enacted}((E,X,A),a) \wedge m \in \mathit{payload}(a)
    \\
    \mathit{agreed}((E,X,A),m) &\coloneqq m \in A
\end{align*}
\end{defi}
\begin{thm}
\label{thm:dec_agreed}
$\mathit{dec}\text{-}\mathit{agreed}: \forall(c:\mathit{config}, m: \mathit{message}), \ \mathit{agreed}(c,m)?$.
\end{thm}

\noindent 
In contrast, configurations arbitrarily predicate the \textit{stated} messages and \textit{enacted} actions.
This is how we express that these sets are generally not decidable or enumerable in a given configuration.
This is intentional.
Our model reflects the harsh realities of data exchange systems:
agents should not synchronise all dynamic information all the time, because
\begin{itemize}
    \item it would trivialise the distribution of the system over physical networks, and
    \item agents could never hide sensitive metadata (\eg, statement contents) from their peers.
\end{itemize}
But we include \Cref{thm:in_enacted_payload_impl_stated} as an important caveat.
By the definition of \textit{stated}, proof that a message is in the payload of some enacted message suffices to prove that it is stated.
Consequently, \Cref{thm:enacted_actions_are_sourced}: actions already enacted are necessarily sourced.

\begin{lem}
\label{thm:in_enacted_payload_impl_stated}
$\mathit{enacted}\text{-}\mathit{stated}: \forall(c,a), \ \mathit{enacted}(c,a) \ \rightarrow \ \forall m \in \mathit{payload}(a), \ \mathit{stated}(c,m)$.
\end{lem}
\begin{thm}
\label{thm:enacted_actions_are_sourced}
$\mathit{enacted}\text{-}\mathit{sourced}: \forall(c,a), \ \mathit{enacted}(c,a) \ \rightarrow \ \mathit{sourced}(c,a)$.
\end{thm}

\subsection{Agent Views and Control Updates}
\label{sec:agent_knowledge_and_control_updates}

To model the dynamics of our runtime system, we restrict our attention to the \textit{reconfigurations}: the updates to configurations the runtime system conceivably supports.
At any step, either a message is stated, an action is enacted, or the agreements are replaced.
But this does not (yet) capture extra constraints imposed by:
\begin{itemize}
    \item agents' limited knowledge of the existing statements and enacted actions, and
    \item the runtime system's enforcement of \Cref{thm:in_enacted_payload_impl_stated}, which implies that every newly enacted action has a payload consisting of messages that are already stated.
\end{itemize}
Precisely, the universe of reconfigurations are identified by \Cref{def:reconfiguration}, and then \Cref{def:reconfigure} gives \textit{reconfigure}, a reconfiguration-labelled total function over configurations.

\begin{defi}
\label{def:reconfiguration}
$\mathit{reconfiguration} \coloneqq \mathit{State}(\mathit{message})\mid \mathit{Enact}(\mathit{action}) \mid \mathit{Agree}(\mathit{list}(\mathit{message}))$.
\end{defi}

\begin{defi}
\label{def:reconfigure}
Let $\mathit{reconfigure}: \mathit{config} \times \mathit{reconfiguration} \rightarrow \mathit{config}$ be defined
\begin{align*}
    \mathit{reconfigure}((E,X,A), \mathit{State}(m))
    &\coloneqq \mathit{(E,X[m \mapsto \top], A)}
    \\
    \mathit{reconfigure}((E,X,A), \mathit{Enact}(a))
    &\coloneqq \mathit{(E[a \mapsto \top],X, A)}
    \\
    \mathit{reconfigure}((E,X,A), \mathit{Agree}(A'))
    &\coloneqq \mathit{(E,X, A')}
\end{align*}
\end{defi}

\noindent 
We use the reconfiguration-transition system to model an `upper bound' of what configurations can be traced by the system at runtime.
It is useful, because even at this level of abstraction, we observe \Cref{def:reconfigs_grow}: reconfigurations are \textit{growing} (\Cref{def:growing}), preserving everything already stated and enacted.
Recall from \Cref{def:dynamic_configurations} that this implies is the premise of important generic results such as \Cref{thm:enacted_effect_pres_while_growing}: enacted effects are always preserved.

\begin{lem}
\label{def:reconfigs_grow}
$\mathit{reconfigs}\text{-}\mathit{grow}: \forall (c: \mathit{config},r: \mathit{reconfiguration}), \mathit{growing}(c, \mathit{reconfigure}(c,r))$.
\end{lem}

\noindent
Finally, we precisely characterise the dynamics of our runtime system with \Cref{def:update_step}, which induces the \textit{update} transition system over the configurations, realising the full granularity of execution, \eg, as we have implemented in Rust.
Updates range over \textit{systems}, which supplement each configuration~$c$ with the agents' \textit{views} on~$c$, representing their partial knowledge of~$c$.
For simplicity, we focus on agents views of \textit{statements} (which are needed for the primary operation of acting) and we omit the agents' views of \textit{enacted actions} (which are managed similarly, but which are needed only for auditors, to discover actions to audit).

The `main' updates (on the left) correspond to the reconfigurations: agents can state a message, enact an action, or replace the agreements.
However, these updates additionally formalise the dependency of actors (who $\mathit{Enact}(a)$) on authors (who $\mathit{State}(m)$ for each $m \in \mathit{payload}(a)$):
actors must prove that their actions are sourced, where knowledge of each statement originates at its author, and must be communicated to the actor.
The other updates (on the right) preserve the configuration, but let agents manipulate their views.
First, agents expand their views via \textit{gossip}: at their own pace, for their own reasons, agents disseminate statements.
Finally, agents can choose to forget statements.
The freedom to forget is practical, \eg, to let agents avoid exhausting their storage in long-running systems.

\Cref{lst:gossip_update} samples the Rust implementation, showing how the `gossip' update is implemented: gossip copies a stated message from one view into another.

\begin{defi}
$\mathit{views} \coloneqq \mathit{agent} \rightarrow \mathit{list}(\mathit{message})$.
\end{defi}

\begin{defi}
\label{def:system}
$\mathit{system} \coloneqq \mathit{config \times \mathit{views}}$.
\end{defi}

\begin{defi}[update]
\label{def:update_step}
We inductively define the updates as $(\rightsquigarrow) \subseteq \mathit{system} \times \mathit{system} \coloneqq$
\begin{align*}
\text{(state)}~&
\mathmakebox[3em][l]{
    \dfrac
    {
    m \notin v(\alpha) \quad \wedge \quad \alpha=\mathit{author}(m)
    }
    {
    (c,v) \rightsquigarrow (\mathit{reconfigure}(c,\mathit{State}(m)), v[\alpha \mapsto [m] \dplus v(\alpha)])}
}
\\
\text{(enact)}~&
\dfrac
{
\mathit{payload}(a) \subseteq v(\alpha) 
}
{
(c,v) \rightsquigarrow (\mathit{reconfigure}(c,\mathit{Enact}(a)), v)}
&&&
\dfrac
{
m \in v(\alpha_\mathit{sender})
}
{
(c,v) \rightsquigarrow (c, v[\alpha \mapsto [m] \dplus v(\alpha)])}&~\text{(gossip)}
\\
\text{(agree)}~&
\dfrac
{
}
{
(c,v) \rightsquigarrow (\mathit{reconfigure}(c,\mathit{Agree}(M)), v)}
&&&
\dfrac
{
m \in v(\alpha)
}
{
(c,v) \rightsquigarrow (c, v[\alpha \mapsto v(\alpha)\setminus m])}&~\text{(forget)}
\end{align*}
\end{defi}

\begin{defi}[update chain]
\label{def:update_chain}
$
\dfrac{}{s \rightsquigarrow^* s}~\text{(0~updates)}
\quad
\dfrac{s \rightsquigarrow ^* s', \quad s' \rightsquigarrow s''}{s \rightsquigarrow^* s''}~\text{(+1~updates)}
$.
\end{defi}

\begin{figure}[htb]

\centering
\begin{minipage}[t]{.52\textwidth}
\begin{lstlisting}[language=Rust, backgroundcolor=\color{rustBackgroundColor},  caption={Rust implementation of the `gossip' update, forwarding a statement. Note how this if-condition reflects the condition $m \in v(\alpha_\mathit{sender})$ of the `gossip' rule shown in \Cref{def:update_step}: the sender must already view the statement!}, label={lst:gossip_update}]
impl<I: ?Sized + ToOwned, A, S, E>
View<I, A, S, E> {
    fn gossip<MS>(&mut self,
        to: Recipient<&I>, message: MS)
    -> Result<(), Error> where
        I::Owned: Clone,
        S: SetAsync<I, MS>
    {
        if !self.stated.contains(&message){
            return Err(Error::BadGossip);
        }
        self.stated.add(to, message)
            .map_err(Error::Set)
}   }
\end{lstlisting}
\end{minipage}%
\hspace{1pt}%
\begin{minipage}[t]{.47\textwidth}
\begin{lstlisting}[language=Rust, backgroundcolor=\color{rustBackgroundColor}, caption={Rust implementation of the \textit{system}, which stores configuration data (\eg, statements) alongside agents' views of the configuration. Note how agents have no (distinct) views of agreements.
}, label={lst:system}]
#[derive(Default)]
struct SetAsync<T> {
    views: HashMap<String, HashSet<T>>,
}
#[derive(Default)]
struct System<P: ?Sized + ToOwned> {
    stated:  SetAsync<Arc<Message<P>>>,
    enacted: SetAsync<Action<P>>,
    agreed:  HashSet<Arc<Message<P>>>,
}
impl<P: ?Sized + ToOwned> System<P> {
    fn initial() -> Self {
        Self::default() } }
}
\end{lstlisting}
\end{minipage}
\end{figure}

\noindent
We generally want agents' views to always be \textit{accurate} with respect to the configuration (\Cref{def:accurate_knowledge}).
Of course, accuracy is trivial in the \textit{initial} system, where (everyone knows) there are no statements.
\Cref{thm:reachable_implies_accurate} follows from induction on
\Cref{def:update_step}: all systems reachable from the initial system via updates are accurate.
The Rust implementation enforces an even stronger result: all conceivable systems are accurate, because views \textit{alias} the same ground-truth collection of statements;
\Cref{lst:system} shows how agent views alias any messages in memory by sharing and copying messages stored behind \textit{atomic reference-counted} pointers (Rust's polymorphic \mlst{Arc} type).  
The Rocq formalisation models this inherent accuracy via refinements to \Cref{def:update_step} beyond what we show here, such that views on~$c$ are stores of pairs of type $(\Sigma m: \mathit{message} \mid \mathit{stated}(c,m))$, whose proofs depend on~$c$ and~$m$.
Consequently, the Rocq user cannot define systems without proving the accuracy of their views.
But the five cases of \Cref{def:update_step} give users transformers of views and proofs of accuracy in lockstep.

\begin{defi}
\label{def:accurate_knowledge}
$\mathit{accurate}((c: \mathit{config}, \ v: \mathit{views})) \coloneqq \forall \alpha, \ \forall m\in v(\alpha),  \ \mathit{stated}(c,m).$
\end{defi}
\begin{defi}
\label{def:initial}
    $\mathit{initial}: \mathit{system} \coloneqq (((\lambda a, \bot), (\lambda m, \bot), [\,]): \mathit{config}, \ (\lambda a, [\,]) : \mathit{views})$.
\end{defi}

\begin{thm}
\label{thm:reachable_implies_accurate}
$\mathit{reachable}\text{-}\mathit{implies}\text{-}\mathit{accurate}:
\forall s, \
    (\mathit{initial} \rightsquigarrow^* \mathit{s}) \  \rightarrow \ \mathit{accurate}(s)$.
\end{thm}

\noindent
In distributed implementations, where gossip must traverse physical networks connecting the agents, enforcing accuracy becomes a matter of \textit{consistency}.
Our framework and implementations were designed with this eventuality in mind.
The integrity of statements can be reduced to checking their \textit{cryptographic signature} by their author.
In the same way, agents can build and share other kinds of knowledge, \eg, of which actions have been enacted.
In this context, \Cref{thm:updates_grow} is especially significant: knowledge of statements is never stale, so it suffices to gossip over slow and unreliable networks, even as new updates occur.
For example, we imagine re-implementing the statements as a \textit{conflict-free replicated data type}~\cite{shapiro2011conflict}.
Namely, we imagine the statements as a \textit{grow-only set}, whose messages are added by distributed agents' \textit{state}-updates and merged by \textit{gossip}-updates.
In fact, in the distributed context, we expect the management of agreements to pose the bigger challenge.
As we demonstrate in \Cref{sec:case_study}, we expect agreements to change infrequently, so that costly synchronisation for each agreement is acceptable.
For example, we imagine agents maintaining a distributed ledger, recording and serialising their \textit{agree}-updates.

\Cref{thm:updates_grow} follows from \Cref{def:reconfigs_grow}, ensuring that every update applied at runtime grows the current configuration.
Consequently, auditors and actors more extensively reason about permission in the past and in the future, in terms of their present knowledge.

\begin{thm}
\label{thm:updates_grow}
$\mathit{updates}\text{-}\mathit{grow}:
\forall (c,k,c',k'), \ (c,k) \rightsquigarrow^* (c',k') \rightarrow \mathit{growing}(c,c')
$.
\end{thm}

\noindent 
In Rocq, we have verified that the above definitions make the permission of enacted actions decidable, enabling the decision of auditors, \ie, \Cref{par:dec_enacted_permitted}.
Because of its definition, decisions of permission are necessarily replicated by anyone at any time.

\subsection{Scripting Agent Behaviour}
\label{sec:agent_scripting}

For the purpose of the experiments in \Cref{sec:case_study}, the Rust implementation includes a final \textit{agent-scripting} layer, which closes the runtime system, and fixes each agent's interactions with the system in the data- and control-planes.
Each of the \textit{scenarios} in \Cref{sec:scenarios} fixes the initialisation of the agents with bespoke scripts.

\begin{figure}[tb]
{\setlength{\columnsep}{2pt}
\begin{lstlisting}[language=rustagentscript, backgroundcolor=\color{rustBackgroundColor}, multicols=2]
Agent::new("amy".into()).program()
.state_on_truth( // `amy 1`
  ground_atom!((surf utils) executed),
  slick::parse::program(
    include_str!("./slick/amy_1.slick")),
)
.state_on_truth( // `amy 2` 
  ground_atom!
    ((amy count_patients) executed),
  slick::parse::program(
    include_str!("./slick/amy_2.slick")))
.enact_on_truths([
  // from `amy 1`
  ground_atom!((amy count_patients)
    has output num_patients),
  // from `amy 2`
  ground_atom!((amy end) executed),
  // from `st antonius 1`
  ground_atom!((st_antonius
    patients_2024) executed),
  // from `st antonius 2`
  ground_atom!
    ((amy count_patients) executed),
  // from `st antonius 3`
  ground_atom!(authorise read of
    ((amy count_patients) num_patients)
    for (amy end) by amy),
  // from `surf 1`
  ground_atom!((surf utils)
    has output entry_count)
  ]);
\end{lstlisting}}
\refstepcounter{lstlisting}\label{lst:eg_agent_script}
\addcontentsline{lol}{lstlisting}{\protect\numberline{\thelstlisting}Agent script specifying Amy's behaviour}
\makeatletter
\@makecaption{\lstlistingname~\thelstlisting}{This script specifies the behaviour of agent Amy in \Cref{sec:scenario:isolated_execution}:
Amy creates two statements and then enacts an action with the payload of statements whose contents extract to policies with the given truths.}
\makeatother
\end{figure}

The agent scripts realise a small, domain-specific language, embedded in Rust.
Each script specifies one agent's behaviour as a collection of reactions to events.
Examples of events include initialisation, synchronised changes to the agreements, and incoming gossip of a statement meeting a specified condition.
Examples of reactions include creating and gossiping statements to specified peers.
During the application of this runtime system in the medical data processing scenarios of \Cref{sec:case_study}, our personification of the agents' goals and intentions are encoded via these agent scripts.
For example, \Cref{lst:eg_agent_script} shows how agent Amy is specified to behave in the first scenario, shown in \Cref{sec:scenario:isolated_execution}: Amy awaits a statement where \lstinline{(surf utils) ready} is true, and reacts by creating and broadcasting
a statement encoding a computational workflow Amy wishes to execute, to access its results.

We expect realistic systems to automate some agent behaviour via scripts, \eg, for tasks requiring agents to react very quickly.
By adopting the JustAct framework, these kinds of runtime systems place the same fundamental standard of permission and well-behavedness on all of the agents, regardless whether they play the roles of actor, author,
or auditor, and regardless of whether they represent human users or automated services.

\section{Case Study: Processing Distributed Medical Data}
\label{sec:case_study}

In this section, we demonstrate and evaluate the JustAct framework (\Cref{sec:framework}), instantiated with the \slick policy language (\Cref{sec:slick}) as a data exchange runtime system (\Cref{sec:impl}), and then applied to policy-regulated medical data processing across institutional boundaries.
\Cref{sec:the_case} overviews the subject of our case study: the \brane component of the EPI framework.
\Cref{sec:init_impl_for_brane} presents \brane through the lens of the JustAct framework, and initialises our runtime system.
\Cref{sec:scenarios} plays out five representative usage scenarios.
Finally, \Cref{sec:discussion_of_the_case_study} reflects on the usage scenarios, evaluating our specific instantiation of JustAct, and how it exhibits the characteristics of  JustAct in general.

The results in this section were produced by our Rust implementation.
~\cite{esterhuyse_2026_21240014} includes the instructions, Rust source, and agent scripts required to reproduce our results. 

\subsection{The Case: The \brane Component of the EPI Framework} 
\label{sec:the_case}

The \textit{Enabling Personalised Interventions} (EPI)\footnote{The EPI project website is hosted at \url{https://enablingpersonalizedinterventions.nl}.} project aimed to create digital twins in healthcare to improve patient treatment prediction~\cite{DBLP:series/oasics/KassemAAKMTBBGHGLK24}.
Research on EPI investigated the challenges of safely coordinating the federated processing of medical data across organisational boundaries within the European Union, \eg, between hospitals, universities, and research centres.

An overarching technical contribution of the EPI project is the \textit{EPI Framework}, which is conceptualised as a modular suite of software components that work together to control and automate facets of data-processing pipelines~\cite{DBLP:journals/access/KassemLTG20,KASSEM2024107550}.
The most complete and recent overview of the EPI framework is given in~\cite{KASSEM2024107550}.
Various articles report on the development of its specialised components.
For example, \cite{DBLP:conf/eScience/KassemVBG21} (re)routes network messages between containerised functions to the satisfaction of security constraints.

In this work, we focus on \brane, the component responsible for unfolding users' processing workflows by planning, orchestrating, and coordinating the execution of workflow tasks.
Workflow tasks are executed by \textit{worker} services and have the effect of computing new data from existing data.
Workers and data-related assets are distributed over physical machines and (technical approximations of) \textit{organisational domains}~\cite{DBLP:conf/eScience/ValkeringCB21}.
For example, a hospital is a domain that provides and controls some medical datasets.

\cite{DBLP:conf/eScience/EsterhuyseMBB22} focuses on the regulatory facet of \brane: each domain is represented by a \textit{checker} service that reasons about and enforces the domain's \textit{policies}, asserting control over the usage of the assets provided by the domain.
For example, via the checker, a human policy expert defines the hospital's policy, which expresses constraints on the permitted processing of the hospital's medical records in order to protect the privacy of patients.
In this context, an extra complication is that domain-local policies express decisions in terms of \textit{meta} data that may also be sensitive, \eg, `the processing of data-asset \lstinline{foot-fracture-x-rays1.png} requires the consent of Amy' reveals sensitive information about Amy.
\cite{DBLP:conf/eScience/EsterhuyseMBB22} addresses this complication by prescribing a policy-worker interface that gives domains maximal control over the extent to which private policy information is publicised via their policy decisions.
Precisely, the execution of planned execution steps await concrete \textit{authorisations} from the involved domains, just in time.
Hence, domains control the usage of their data, and they also decide how to balance policy privacy against system productivity.

Below, we clarify the usage of \brane by enumerating the kinds of its constituent services and overviewing their roles and interactions.
They are listed in an order that suggests the progression of a successful workflow execution, from top to bottom:
\begin{enumerate}
    \item A \textbf{driver} defines computational workflows: directed acyclic diagrams expressing the functional dependencies.
    Each result (data) depends on a containerised function (data) and arguments (data).
    The driver reads their workflow's results once they are available.
    \item The \textbf{planner} plans workflows: it assigns workflow tasks to workers for execution.
    \item A \textbf{checker} \textit{authorises} workflow plans whose execution (and the necessary reading and writing of data by workers) conforms to the checker's local policies. 
    \item A \textbf{worker} executes their assigned workflow tasks which have sufficient authorisations.
    The worker reads the function and argument data, then computes and writes the result.
\end{enumerate}
Of the above services, only the planner is generally centralised, ultimately orchestrating the other services' tasks by planning the assignment of tasks to workers.
Each organisational domain hosts drivers, checkers, and workers, which participate in data processing and regulation.
In principle, any subset of services may be automated, but generally, \brane drivers and checkers are automated, interfacing with human data scientists (defining work to be executed) and policy experts (regulating data access via workflow execution).

\subsection{Initialising our Runtime System for \brane Scenarios}
\label{sec:init_impl_for_brane}

\begin{figure}[tbh]
    \centering
    \includegraphics[width=0.74\columnwidth]{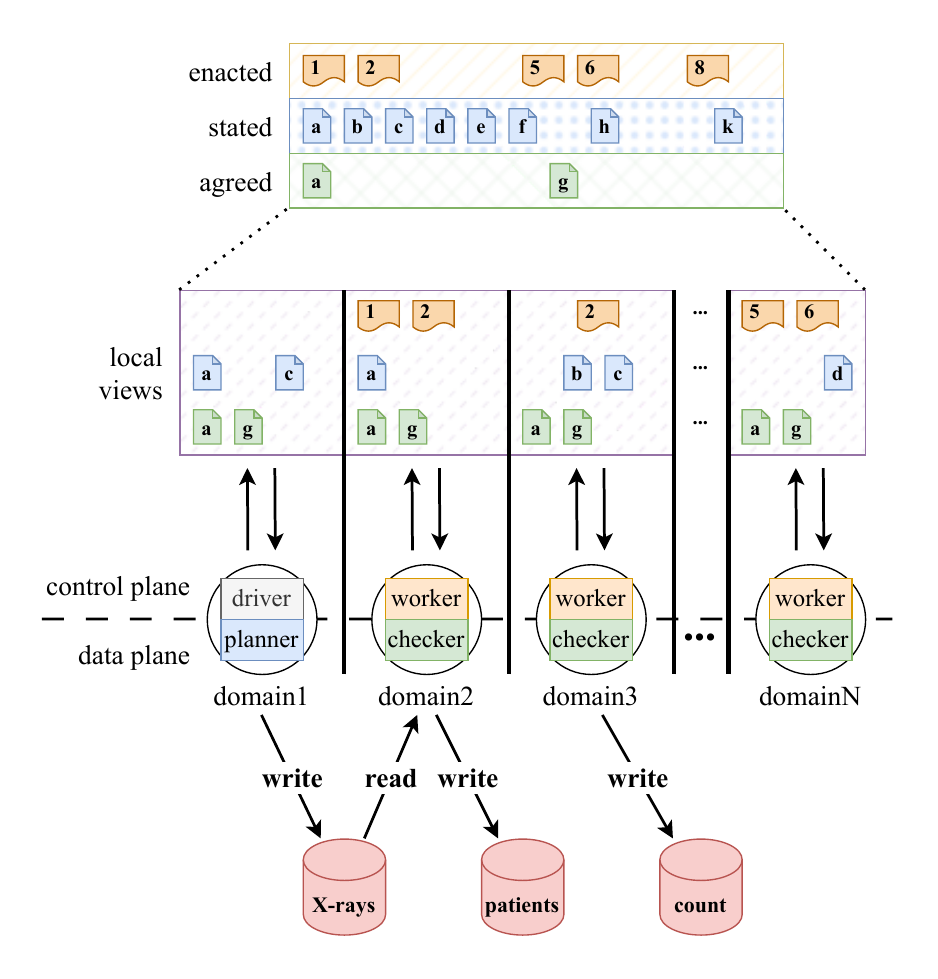}
    \vspace{-10pt}
    \caption{Graphical depiction of our runtime system instantiated for application to \brane usage scenarios.
    Hence, this figure refines \Cref{fig:system_impl}: \brane services map to our agents, and (the distinct kinds of) \brane control meta data are mapped to meta data in our control plane.
    \brane's medical data is stored in our data plane.
    }
    \label{fig:system_brane}
\end{figure}

\Cref{fig:system_brane} visualises the refinement of our runtime system such that \brane concepts are mapped to elements of our runtime system.
First and foremost, we map \brane's notion of (organisational) \textit{domain} to our notion of \textit{agent}, such that distinct domains are distinct agents.
This determines the granularity at which control is distributed, and at which the autonomous entities are distinguished.
Hence, each agent can \textit{play the role of} several of \brane's services.
For simplicity, this abstracts away the internal organisation of each domain.
For example, we represent the organisational domain of the St.\ Antonius hospital as agent \lstinline{st-antonius}, and abstract all its internal resources, human users, and automated services behind this agent.
Throughout this section, we use St.\ Antonius (a Dutch hospital) and SURF (a Dutch research institute) as examples of particular domains.
These are real contributors to the EPI project, but our accounts of their behaviour and their users (Amy, Bob, and Dan) are fictitious.

In \Cref{sec:brane_data_plane,sec:brane_control_plane}, to follow, we complete the mapping of \brane concepts to facets isolated to the data and control planes, respectively.

\subsubsection{The \brane Data Plane}
\label{sec:brane_data_plane}

Our runtime system and \brane rely on a similar abstraction of a shared (initially empty) data store, whose data is read and written by the agents.

Agents' access to data in the data plane is regulated via the connection to enacted effects, as specified previously in \Cref{sec:brane_data_plane}.
Any agent~\lstinline{Agent} can introduce fresh data~\lstinline{Data} using the existing mechanism of executing workflow tasks: it suffices for \lstinline{Agent} to define a fresh workflow (playing the role of driver) with no inputs and the output~\lstinline{Data}, and executing the task (playing the role of worker).
In the next section, we design the policies such \lstinline{Agent} is always permitted to enact the above, and it has the desired effect: \lstinline{Agent writes Data}.

\subsubsection{The \brane Control Plane \& The Initial Agreement}
\label{sec:brane_control_plane}

In the original \brane, the roles and interactions between (distributed) automated services and between users are documented in publications such as \cite{KASSEM2024107550}.
However, naturally, there is a gap between these concepts and their encoding in the \brane implementation, which is what ultimately determines the capabilities of services and users at runtime.
For example, \cite{KASSEM2024107550} uses precise wording and logical notation to define the obligation of workers to observe authorisations from involved checkers.
In our version of \brane, we aim to narrow the gap between our conceptualisation and implementation of these same agent roles and interactions by encoding them in \slick policies and including these in our runtime system configuration as statements and agreements.
Namely, we intend for our reasoning about \slick rules in this section to closely align with the reasoning performed by the agents themselves at runtime.

In the remainder of this subsection, we discuss the features of \brane and encode them as \slick rules in message \mlst{consortium 1} which we call the \textit{initial agreement}.
Each scenario in \Cref{sec:scenarios} begins with the system initialised without any statements or enacted actions, but with just this initial agreement in place.
The author of the initial agreement is \mlst{consortium}, which corresponds to no particular physical entity or human user at runtime.
Instead, it represents the consensus of the other agents;
its only purpose is to author agreements.

\Cref{fig:initial_agreement} gives an ontological view of the initial agreement.
This view is intended to guide the reader's interpretation of the agreement itself;
\eg, the figure emphasises the need for each \textit{task} to have exactly one \textit{driver} agent.
Ontological sets are encoded in the agreement itself as facts matching structural patterns, \eg, \textit{inputs} are encoded as facts matching \lstinline{Task has input Variable}.
Ontological functions are encoded either structurally, \eg, a sub-fact of each \textit{input} is a \textit{task}, or they are encoded in truth via rules, \eg, $\mathit{ready} \Leftarrow \mathit{executed}$ is encoded as \lstinline{error if Task executed and not Task ready}.
The initial agreement layers additional meaning atop the ontology.
Notably, via rules conditioning invalidity, it specifies the inter-agent power dynamics.
For example, only the driver of a task may define its inputs.

\begin{figure}
\centering
\begin{framed}
\begin{tikzcd}[cramped,column sep=6em,row sep=1.7em,math mode=false,cells={font=\normalsize\itshape},labels={inner sep=2pt,font=\footnotesize,pos=0.5}]
inputs
    \arrow[d]
    \arrow[dr]
&
&authorisation
    \arrow[ll,"of"]
    \arrow[d,shift left =3pt,"authoriser"]
    \arrow[d,shift right=3pt,swap,"worker"]
    \arrow[dl,swap,"for"]
\\
variables
    \arrow[]{r}[pos=0.3]{output of}
&tasks
    \arrow[r,"driver"]
&agents
\\
&reads,writes
    \arrow[ul]
    \arrow[u,"for"]
    \arrow[ur]
&drives,involves,controls
    \arrow[u]
    \arrow[ul]
\end{tikzcd}
\end{framed}
\caption{
Graphical ontology of the initial agreement of our reproduced \brane system.
Sets (italicised) are related by pure, total functions ($\rightarrow$) from domain to co-domain.
}
\label{fig:initial_agreement}
\end{figure}

The following snippet shows the first part of the initial agreement, which prevents agents from mimicking the transformations to \slick rules applied during \textit{extract} by \textit{reflect-author} (\Cref{def:reflect_author}) and \textit{reflect-actor} (\Cref{def:reflect_author}).
Consequently, valid policies have truths arising \textit{only} from rules injected by these transformations.
The agents can rely on conditions matching \mlst{actor Agent} really identifying the actor, and \mlst{Sayer says Fact} really witnessing that the truth \mlst{Fact} arose from a rule authored by \mlst{Sayer}.
The topmost rule is included to create a convenient indirection: because \mlst{error Reason} implies \mlst{error}, recognisers of invalidity can supplement \mlst{error} with a term that documents the reason for invalidity.
In the article, this helps us document the rules, but it is also very useful at runtime, because the \mlst{Reason} provides a frictionless way to diagnose invalidity.
Note how the other two rules use this indirection.

Here and henceforth, (only) listings of \slick policies in statements have this yellow colour.
\begin{lstlisting}
// Statement (consortium 1) ...
error if error Reason.
error (illegal (Sayer says (Agent says Fact))) if Sayer says (Agent says Fact).
error (two actors X also Y) if actor X and actor Y and diff { X Y }.
\end{lstlisting}

The remaining parts of the initial agreement concern more the concrete facets of \brane.

The next part characterises \textit{drivers} as agents that
\begin{enumerate*}
    \item define the input-output dependencies of \textit{tasks} on data, and
    \item mark tasks as being \textit{ready} for execution.
\end{enumerate*}
For example, a user defines a workflow via a typical scripting language (\eg, BraneScript, typically used with \brane~\cite{DBLP:conf/eScience/ValkeringCB21}), including the assignment statement $y \coloneqq f(x)$.
The corresponding driver encodes this as a fresh task~\lstinline{t} by stating \lstinline{t has input (t f).} \lstinline{t has input (t x).} \lstinline{t has output y.}

\Cref{fig:initial_agreement} visualises the need for each task to have a unique driver as the function arrow $\mathit{tasks} \funarrow{driver} \mathit{agents}$.
We encode this mapping in the structure of the task identifier itself: task $(\alpha, n)$ identifies $\alpha$ as its driver, and $n$ as its $\alpha$-local task \textit{name}.
This uses agents as task-namespaces.
Hence, we reduce the (complex) problem of distributed drivers managing a global task namespace to the (simple) problem of each driver managing their own local task namespace.
For example, $(\alpha, n)$ is the global identifier for a task named~$n$ by driver~$\alpha$.

The following rules enforce that only $\alpha$ may perform the activities of a driver of task $(\alpha, n)$: other agents are prohibited from defining its input-output dependencies with data, and marking the task ready for execution.
\begin{lstlisting}
// ... continuing statement (consortium 1) ...
error (Sayer defined input Variable for (Driver Name) illegally)
    if Sayer says ((Driver Name) has input Variable) and diff { Sayer Driver }.
error (Sayer defined output Label for (Driver Name) illegally)
    if Sayer says ((Driver Name) has output Label) and diff { Sayer Driver }.
\end{lstlisting}

The next part prevents the premature execution of tasks.
Precisely, tasks cannot be executed until they are ready, and the tasks producing their inputs are also executed.
\begin{lstlisting}
// ... continuing statement (consortium 1) ...
error (illegal (Task2 executed) when input (Task1 Label) aint executed yet)
    if Task2 executed and Task2 has input (Task1 Label) and not Task1 executed.
\end{lstlisting}

The next part defines the \textit{effects} of enacting the execution of tasks.

As with tasks and their drivers, we fix the \textit{output of} function from variables to tasks in  \Cref{fig:initial_agreement} via the encoding of the variables themselves:
each variable is a task-\textit{label} pair, where labels discriminate the variables of the same task.
By including the labels, we leave room for tasks which have multiple outputs, and distinguish them by labels.

The first two rules infer the \textit{enacted effects} of actions enacting the execution of a task:
the acting worker reads each input, and writes each output as truths matching \lstinline{Worker reads Variable} and \lstinline{Worker writes Variable}.
The other rules prevent these effects from arising in any other way.
Recall from \Cref{sec:impl_effects} that the runtime system uses these effects to regulate agents' access to asset data;
agents can only read or write asset data when these are the effects of their (permitted and taken) actions.
The scenarios will demonstrate several examples of agents reasoning backward from the effects they desire to the necessary actions and statements.
\begin{lstlisting}
// ... continuing statement (consortium 1) ...
Worker reads Variable
    if Task has input Variable and actor Worker and Worker says (Task executed).
Worker writes (Task Label)
    if Task has output Label and actor Worker and Worker says (Task executed).
error (illegal (Worker Verb Variable) mimicking effect) 
    if Sayer says (Worker Verb Variable) and diff { consortium Sayer }
    and not diff { Verb writes reads }.
\end{lstlisting}

The final part of the initial agreement encodes the role of checkers in \brane: a worker's execution of a task is conditioned on proof that agent~$\alpha$ \textit{authorises} each of input~$d$, whenever $\alpha$ \textit{involves}~$d$, in which case \cite{KASSEM2024107550} calls $\alpha$ a \textit{checker} of~$d$.
We reproduce the key characteristic of the \textit{involves} relation from \cite{KASSEM2024107550}.
In the base case, \textit{involves} is \textit{controls}, where the latter is used in \brane on a case-by-case basis, capturing the (shared) ownership of agents' data, \eg, a steward of medical data controls a sensitive medical dataset.
And in the inductive case, involvement of $\alpha$ with data~$d$ propagates to data derived from~$d$, \ie, from task inputs to their outputs.
This mechanism gives agents the power to veto the execution of any task that would process data under their control, and all of its (transitive) derivatives.
The agent's own motivations for creating particular authorisations can be hidden from the \slick policy, \eg, hiding the underlying reasoning about sensitive local policies~\cite{DBLP:conf/eScience/EsterhuyseMBB22}.
For example, a hospital authorises a particular trusted worker to process their sensitive medical records, but vetos the result being further processed elsewhere by withholding any further authorisations.

\begin{lstlisting}
// ... completing statement (consortium 1).
error (illegal read by Worker of Variable without Checker authorisation)
    if Worker reads Variable and Worker says (Task executed)
    and Task has input Variable and Variable involves Checker
    and not Checker says (authorise read of Variable for Task by Worker).

Variable involves Checker if Checker controls Variable.
(Task Label) involves Checker if Variable involves Checker
    and Task has input Variable and Task has output Label.
\end{lstlisting}

As we have done in the article, we expect uses to supplement their formal agreements with informal summaries and motivations.
However, the real value in treating the formal agreement as the ground truth is that it can ultimately stand alone.
End users can scrutinise candidate agreements systematically, relying on the precision of the policy semantics to convey unambiguous meaning.
For example, our \slick formalisation of \brane is more specific about how authorisations are represented and related to worker tasks than in the original article in natural language~\cite{KASSEM2024107550}.
Then users can rely on (trusted) automated tools to reason about hypothetical cases in detail, beyond what humans can reliably infer from interactions specified in the agreement, and beyond what its authors originally considered.
For example, in this manner, readers of this article could check corner cases of our initial agreement by interacting with out \slick interpreter (although we intend this to be unnecessary).
In general, we expect this systematic approach to developing and evaluating agreements to afford productive, transparent, and reproducible negotiations and compromises between~users.

\subsection{Usage Scenarios}
\label{sec:scenarios}

Here we walk through a sequence of scenarios in which we apply our initialised runtime system to new and existing usage scenarios of \brane.
Note that our discussion of the first scenario is the most detailed, because most of its details are novel.

Each scenario begins the same way, immediately after, from the initial (empty) configuration, the initial agreement is stated and agreed.
Each agent has trivially accurate knowledge: no other messages are stated and no actions are enacted.

\subsubsection{Existing Usage: Isolated Authorisation and Execution}
\label{sec:scenario:isolated_execution}

We demonstrate the application of our system to a representative usage of the original \brane system: several agents cooperate to drive the execution of data-processing workflows, as permitted by agent reasoning in isolation about their local policies.
In our scenario, this isolation manifests as truths of the form \lstinline{authorise Task in Msg by Worker} being stated only as facts, i.e., inferred by variable- and condition-free rules.
Effectively, agents entirely externalise their reasoning about local policies, and reveal only the resulting authorisations, on a case-by-case basis.

The scenario begins with the \lstinline{surf} agent creating fresh data by executing a zero-input task in the statement \lstinline{surf 1}.
The data is bound to  variable \lstinline{(surf utils) entry-count}.
\Cref{fig:task_graph} depicts these, and all (relations between) tasks and variables used in \Cref{sec:scenarios}.
\begin{lstlisting}
// Statement (surf 1).
(surf utils) has output entry-count.
(surf utils) executed.
\end{lstlisting}

\begin{figure}[t]
\centering
\begin{framed}
\begin{tikzcd}[cramped,column sep=0.56em,row sep=0.4em,math mode=false,cells={font=\scriptsize},labels={inner sep=2pt,font=\footnotesize,pos=0.5}]
\fbox{$\overset{\text{surf}}{\text{utils}}$}
    \arrow[r,dashed]
&[-5pt]entry-count
    \arrow[rrr]
    \arrow[ddrrr]
&[+10pt]
&[+10pt]
&[-0pt]\fbox{$\overset{\text{amy}}{\text{count-patients}}$}
    \arrow[r,dashed]
&[+5pt]num-patients
    \arrow[r]
&[+16pt]\fbox{$\overset{\text{amy}}{\text{end}}$}
\\[-5pt]
\fbox{$\overset{\text{st-antonius}}{\text{patients-2024}}$}
    \arrow[r,dashed]
    \arrow[dash,dotted,ultra thick,shift right=15pt,start anchor=west,end anchor=east]{rrrrrr}
&patients
    \arrow[urrr,crossing over]
    \arrow[dr]
&&&&&
\phantom{xxxxx}
\\[+5pt]
\fbox{$\overset{\text{bob}}{\text{step1}}$}
    \arrow[r,dashed]
&filter-consented
    \arrow[r]
&\fbox{$\overset{\text{bob}}{\text{step2}}$}
    \arrow[r,dashed]
&consented
    \arrow[r]
&\fbox{$\overset{\text{bob}}{\text{step3}}$}
    \arrow[r,dashed]
&num-consented
    \arrow[r]
&\fbox{$\overset{\text{bob}}{\text{step4}}$}
\end{tikzcd}
\end{framed}
\caption{
Bipartite graph of workflow tasks (task-\textit{names} overset by -\textit{drivers} in boxed vertices) and asset variables (variable-\textit{labels} in unboxed vertices), in scenarios of \Cref{sec:scenarios}, related by data-dependencies in the direction of data flow; output (dashed arrows) relates tasks to variables, and input (solid arrows) relates variables to tasks.
The dashed line separates tasks defined in Scenarios~1 (above) and~2~(below).
}
\label{fig:task_graph}
\end{figure}

We let `Analyst' Amy drive the completion of a one-task medical data workflow.
Amy is a (human) data scientist working to collect basic statistics of European hospitals, starting with the St.\ Antonius hospital.
Amy defines a fresh task with the suggestive name \lstinline{count-patients}, and marks it ready for execution.
Intuitively, it encodes an assignment that would be rendered as $\mathit{numPatients} \coloneqq \mathit{entryCount}(\mathit{patients})$ in typical imperative workflow languages.
Input variables with labels \lstinline{entry-count} and \lstinline{patients} are outputs of tasks driven by \lstinline{surf} and \lstinline{st-antonius}, respectively.
Amy publicises this statement to her peers.

On its own, \lstinline{amy 1} defines a task, but the task is not executable yet, because this (first) requires the execution of \lstinline{st-antonius patients-2024}, to produce the needed \mlst{patients} asset.
\begin{lstlisting}
// Statement (amy 1).
(amy count-patients) has input ((surf utils) entry-count).
(amy count-patients) has input ((st-antonius patients-2024) patients).
(amy count-patients) has output num-patients.
\end{lstlisting}

After some time, Amy acquires the following statement by St.\ Antonius, and propagates it to the workers in the system.
Whether this statement was made before, or in reaction to Amy' statement is not important; it suffices that all its observers agree that it was authored by St.\ Antonius.
Note that it includes an assertion that St.\ Antonius controls the patients data, so tasks cannot process it (or its derivatives) without the authorisation of St.\ Antonius.
This authorisation is given by St.\ Antonius to themselves for task \lstinline{st-antonius patients-2024}, but not yet for tasks processing the outputs of \lstinline{st-antonius patients-2024}, \eg, task \lstinline{amy count-patients}.
\begin{lstlisting}
// Statement (st-antonius 1).
(st-antonius patients-2024) has output patients.
st-antonius controls ((st-antonius patients-2024) patients).

(st-antonius patients-2024) executed.
\end{lstlisting}

`Disruptor' Dan observes these statements.
Inspired by \lstinline{st-antonius 1}, Dan attempts to establish control over \lstinline{patients} by stating \lstinline{dan 1}.
All observers agree that this statement is valid, and, as no action is taken, Dan remains well-behaved.
Indeed, it is possible that, after some more statements, actions may be justified using \lstinline{dan 1}.
However, Dan has misunderstood his role in the system.
The crucial difference between Dan and St.\ Antonius is the incentives they provide for agents to accept their conditions of control over \lstinline{patients}.
Only St.\ Antonius places this condition within the same statement as other useful rules, namely, the definition and execution of task \lstinline{st-antonius patients-2024}.
In contrast, agents have no incentive to include \lstinline{dan 1} in the payloads of their actions, so the statement goes unused.
Agents who expect this outcome can exercise their right to forget \lstinline{dan 1} altogether.
There also exists no statement (\eg,~by~Dan) that alters anyone's perception that \mlst{st-antonius} drives task \lstinline{st-antonius patients-2024}, because the identity of St.\ Antonius is encoded in the task name, and recall (from \Cref{sec:reflecting_authorship_in_rules}) that the runtime system prevents anyone (Dan) from creating messages that reflect the authorship of anyone else (St.\ Antonius).

\begin{lstlisting}
// Statement (dan 1)
dan controls ((st-antonius patients-2024) patients).
\end{lstlisting}

After some external reasoning, St.\ Antonius reasons that they prefer to execute Amy's task themselves, as it requires little work, yet it handles extremely sensitive data.
They state \lstinline{st-antonius 2}.
Next,
St.\ Antonius enacts $(\mlst{st-antonius}, \mlst{consortium 1}, [\mlst{surf 1}, \mlst{amy 1}, \mlst{st-antonius 1},$ $\mlst{st-antonius 2}])$, \ie, using all statements but \mlst{dan 1}.
This action is permitted (\Cref{def:permitted}):
\begin{enumerate}
    \item The action is \textit{sourced}: each message in its payload is \textit{stated}.
    \item The action is \textit{based} (on \lstinline{consortium 1}), which is an agreement.
    \item The policy extracted from the payload of the action is \textit{valid}: its truths exclude \lstinline{error}.
\end{enumerate}
Every observer of this action necessarily reproduces this decision, and so, agrees that St.\ Antonius remains well-behaved.
Moreover, they agree on the effects: St.\ Antonius reads inputs \mlst{entry-count} and \mlst{patients}, and writes output \mlst{num-patients}.
Soon afterward, St.\ Antonius externalises these effects in the data plane by reading and writing the corresponding assets.

\begin{lstlisting}
// Statement (st-antonius 2)
authorise read of ((st-antonius patients-2024) patients)
    for (amy count-patients) by st-antonius.
(amy count-patients) executed.
\end{lstlisting}

Amy works toward \textit{justifying} effect \lstinline{amy reads ((amy count-patients) num-patients)}, \ie, enacting it while preserving well-behavedness.
After learning \lstinline{st-antonius 2}, Amy states \lstinline{amy 2}.
In this case, Amy's reasoning is straightforward; no existing task has the desired effect, and this task is minimal, having only this effect.
But in any case, the reasoning and motivations that underlie what agents choose to state, enact, or agree have no consequence to actions and their effects or permission, so agents generally keep these details to themselves.

Amy has the physical capability to enact this statement already, but not without observers agreeing that the action is not permitted!
Amy must include \lstinline{st-antonius 2} in the justification (to witness the execution of the input), but then Amy must include \lstinline{st-antonius 1} (for the same reason), but then Amy must provide evidence that St.\ Antonius authorises Amy executing the new task.
No existing statement states this, and Amy cannot state it herself (while preserving validity).
Amy chooses to remain well-behaved by not acting just yet.

\begin{lstlisting}
// Statement (amy 2)
(amy end) has input ((amy count-patients) num-patients).
(amy end) executed.
\end{lstlisting}

Upon observing Amy's new statement, St.\ Antonius is driven to the obvious conclusion; Amy awaits the authorisation to act.
St.\ Antonius reasons that stating this authorisation would enable Amy enacting the execution of the task, with the effect of Amy reading count, which is a derivative of patients.
However, St.\ Antonius concludes that this derivative of the sensitive data it not sensitive itself, and so there is no harm in Amy reading it.
All of this reasoning is hidden from external observers, but its result is publicised as \lstinline{st-antonius 3}.
\begin{lstlisting}
// Statement (st-antonius 3)
authorise read of ((amy count-patients) num-patients) for (amy end) by amy.
\end{lstlisting}

Finally, Amy enacts the previously stated \lstinline{amy 2}, as justified by all statements so far, except for \lstinline{dan 1}.
Once again, all observers independently come to the same decision: Amy's action is permitted, and so it preserves the good behaviour of Amy.
All observers agree that this action has the (intended) effect of Amy reading the variable labelled \mlst{num-patients}.
Amy externalises this effect by reading the asset data identified by this variable.

\subsubsection{Existing Usage: Distributed Workflow Execution}
\label{sec:scenario:distributed_execution}

Like the original \brane system, our system affords the execution of multi-step workflows over multiple domains.

In this scenario, `Bookkeeper' Bob wants to know how many patients at St.\ Antonius have consented to the processing of their patient data being read.
To achieve this goal, Bob plays the role of another data scientist; in the new statement \lstinline{bob 1}, Bob defines a workflow of sequential tasks, which Bob names Step~1 to Step~4.
Bob also plays the role of a software provider in Step~1 by providing the \lstinline{filter-consented} function to be used in Step~2.

In \lstinline{bob 1}, Bob supplements the usual task definition rules with rules about how the tasks are authorised and executed.
This demonstrates how all agents (not just \lstinline{consortium}) enjoy the expressive power of conditional rules to impose complex conditions on how their statements are used.
Specifically, first (in \mlst{authorise read}~$...$), Bob provides a blanket authorisation for any \mlst{Worker} agent to execute tasks; these authorisations in fact apply to \textit{any} task globally, but recall that authorisations are useless (and harmless) for tasks in which the authoriser is not involved.
Finally (in $...$~\mlst{executed}), Bob couples the definitions of Steps~1 and~4 with the stipulation that Bob executes these himself.
Consequently, agents using \lstinline{bob 1} to justify any action are obligated to acquire authorisation on behalf of Bob.
Informally, observers can understand this to signal Bob's intention to acquire authorisation for these steps.

\begin{lstlisting}
// Statement (bob 1).
(bob step1) has output filter-consented.

(bob step2) has input ((bob step1) filter-consented).
(bob step2) has input ((st-antonius patients-2024) patients).
(bob step2) has output consented.

(bob step3) has input ((surf utils) entry-count).
(bob step3) has input ((bob step2) consented).
(bob step3) has output num-consented.

(bob step4) has input ((bob step3) num-consented).

authorise read of Variable for Task by Worker
    if Worker says (Task executed) and Task has input Variable.

(bob step1) executed.
(bob step4) executed.
\end{lstlisting}

St.\ Antonius and SURF observe and reason about Bob's statement.
Externally to the system, they exchange communications, negotiating their possible roles in facilitating the execution.
For example, SURF observes that Bob's statement implies Bob's authorisation to any executed task, but this is harmless, as it remains useless for tasks not involving Bob.
While St.\ Antonius is willing to execute Step~3, they are unwilling to execute Step~2, as this requires them to execute Bob's untrusted \lstinline{filter-consent} with the hospital infrastructure.
Fortunately, St.\ Antonius is willing to let SURF execute Step~2 instead;
St.\ Antonius trusts SURF to read the sensitive patient data, while SURF accepts the risk of executing Bob's mysterious function. 
None of this reasoning is revealed to the other agents, but its results are revealed as the following two statements by SURF and St.\ Antonius, respectively.
\begin{lstlisting}
// Statement (surf 2).
(bob step2) executed.
\end{lstlisting}
\begin{lstlisting}
// Statement (st-antonius 4).
(bob step3) executed.
authorise read of ((st-antonius patients-2024) patients) for (bob step2) by surf.
authorise read of ((bob step2) consented) for (bob step3) by st-antonius.
authorise read of ((bob step3) num-consented) for (bob step4) by bob.
\end{lstlisting}

At this point, the execution of all of Bob's tasks are permissible.
Independently, Bob, SURF, and St.\ Antonius can build payloads and act.
As they see fit, agents can cooperate with their peers.
For example, if Bob acts first, he can gossip the action to SURF, guiding them to action.
As always, everyone will agree on the permission and effects of these actions.

While their data-dependencies are \textit{modelled} by their statements as inputs, outputs, reads, and writes, they are not enforced by policy;
for example, the above statements suffice to let St.\ Antonius enact Step~3, even before SURF enacts Step~2.
However, of course, these actions are also constrained by their data-dependencies in the data plane, so St.\ Antonius cannot execute Step~3 until the effects of executing Step~2 are externalised.
However, this do not concern validity or permission, so they are not enforced via policies in the control plane.

\subsubsection{Existing Usage: Concurrent Driving, Execution, and Authorisation}
\label{sec:scenario:concurrency}

In this section, we briefly remark on the observation that our system faithfully reproduces the isolation of independent workflows, \ie, the definition, authorisation, and execution of different workflows proceed concurrently.
Concurrent tasks are clearly visible in \Cref{fig:task_graph} as each task-pair where neither depends on the other, \eg, \lstinline{amy end} and \lstinline{bob step2}.
But moreover, unrelated statements are created concurrently also.
For example, all the activities of Scenarios 1 and 2 (in \Cref{sec:scenario:isolated_execution,sec:scenario:distributed_execution}, respectively) occur concurrently after \lstinline{patients} is available.
For example, if Amy crashes immediately after initialisation, and then stays entirely unresponsive,
Scenario~1 cannot complete, but Scenario~2 will still complete successfully.

\subsubsection{New Usage: Inter-Domain Policies}
\label{sec:scenario:inter-domain-policies}

Thus far, our scenarios have faithfully reproduced a characteristic of the original \brane; domain-local policies are entirely private: hidden from other agents, exposed only in the form of particular authorisations, publicised only when needed, on a case-by-case basis.
Indeed, this is an important feature of \brane, which was intentionally developed to handle the eventuality that domain-local policies often reflect private information, thus becoming private themselves~\cite{DBLP:conf/eScience/EsterhuyseMBB22}.
However, the original \brane offers no way for domains to share policy information.
For example, while St.\ Antonius can send information to SURF via channels outside of \brane, it is disconnected from SURF's authorisations.
In contrast, this scenario demonstrates how our system lets agents cooperate in defining composite \slick policies, leading to authorisation.
In general, statements are useful for unifying the informative and regulatory qualities of policy.
In the case of \brane, statements can publish non-sensitive parts of domain-local policies.
Recipients of these statements benefit from having insight into the reasoning of the author, while the author benefits from being decoupled from (the work of) justifying other agents' actions.

From \Cref{sec:scenario:isolated_execution}, it is clear that St.\ Antonius already models and reasons about the trustworthiness of some agents above others, which ultimately informs which tasks they choose to authorise.
But with the following statement, St.\ Antonius formalises some of this reasoning as conditional authorisation.
Consequently, any observer can predict the authorisation of St.\ Antonius, and under the right conditions, apply it on their behalf.
In other words, by creating this statement, St.\ Antonius \textit{delegates} some power of authorisation to highly trusted peers.
Moreover, the statement formalises the \mlst{is trusted} and \mlst{is highly trusted} relations in a form that other agents can reuse for their own purposes in the future.
\begin{lstlisting}
// Statement (st-antonius 5).
st-antonius is highly trusted. surf is highly trusted.

Agent is trusted if Sayer says (Agent is highly trusted)
          and st-antonius says (Sayer is highly trusted).
          
authorise Task in Msg by Worker if (Task ready) within Msg
    and st-antonius says (Worker is trusted).
\end{lstlisting}

Given \lstinline{st-antonius 5}, SURF defines and executes the following task, which reads the patient data.
Most interestingly, although this task is defined to \lstinline{involve} St.\ Antonius,
their participation in the action is not required at any point.
In this sense, \lstinline{st-antonius 5} delegated a limited power to authorise on the behalf of St.\ Antonius to SURF, despite this requiring SURF to reason about a publicised part of the local policy of St.\ Antonius.
\begin{lstlisting}
// Statement (surf 3).
(surf read-patients) has input ((st-antonius patients-2024) patients).
(surf read-patients) executed.
\end{lstlisting}

Next St.\ Antonius makes the following statement.
Like \lstinline{st-antonius 5}, this statement formalises a facet of their local policy, to the benefit of internalising their reasoning about tasks, with the benefit of letting others systematically reason on their behalf.
In this case, it captures the link between (simplified) consent of patients to the processing of their data, and the authorisation by St.\ Antonius of tasks processing patient data.
The statement enables a fruitful cooperation with SURF, because SURF is highly trusted by St.\ Antonius.

However, unlike \lstinline{st-antonius 5}, this statement captures sensitive policy information: Billy, Wally, and Berty are patients of the St.\ Antonius hospital!
St.\ Antonius does not wish to disclose this to \textit{un}trusted agents!
Ideally, statements could be used to regulate such communications via justification and permission, as usual.
Unfortunately, JustAct does not (yet) offer a means for agents to internalise this.
For example, if SURF forwards \lstinline{st-antonius 6} to Dan, SURF still satisfies the definition of well-behaved.

\begin{lstlisting}
// Statement (st-antonius 6).
billy is a patient. wally is a patient. berty is a patient.

trusted consent for Task in Msg if st-antonius says (Sayer is trusted)
    and Sayer says (Patient consents to Task in Msg).

Task lacks trusted consent if authorise Task in Msg by Worker via consent
    and Patient is a patient and not trusted consent for Task in Msg.
    
authorise Task in Msg by Worker if Patient consents to Task in Msg
    and Worker says (Task executed) and not Task lacks trusted consent.
\end{lstlisting}

\subsubsection{New Usage: Dynamic Amendment of the Agreement}
\label{sec:scenario:dynamic_agreement}

So far, all actions have been controlled via the concepts defined in the initial agreement: \mlst{consortium 1}.
In this final scenario, we demonstrate the more fundamental changes agents can make to alter which actions are justifiable in the future, by changing the agreements.

After some negotiations, the agents reach consensus: the initial agreement has some flaws.
Namely,
\begin{enumerate*}
    \item SURF lacks control over \lstinline{entry-count}; SURF regrets providing the asset in \lstinline{surf 1} unconditionally, and
    \item St.\ Antonius finds it inconvenient that they cannot express their involvement in the sensitive \lstinline{patients} without also being involved in \lstinline{num-patients}.
\end{enumerate*}

Working together (externally), the agents formulate their changes as
a new message \lstinline{consortium 2}, which differs from \lstinline{consortium 1} in only the following, small details:
\begin{enumerate}
    \item the new agreement includes a new rule:
\begin{lstlisting}
(surf utils) involves surf.
\end{lstlisting}
    \item the existing rules concerning involvement are each given a new condition that the variable in question is not stated to be \textit{insensitive}, which is a novel concept.
    Precisely, this change is realised as a replacement of those rules with the following:
\begin{lstlisting}
Variable involves Checker
    if Checker controls Variable and not Checker says (Variable is insensitive).
(Task Label) involves Checker if Variable involves Checker
    and Task has input Variable and Task has output Label
    and not Checker says ((Task Label) is insensitive).
\end{lstlisting}
\end{enumerate}

\noindent 
The changes take effect by theagents updating the \textit{agreements} from $\!\![\mlst{consortium 1}]\!\!$ to $\!\![\mlst{consortium 2}]\!$.
The prior actions are entirely unaffected, because their permission was based on \mlst{consortium 1}.
However, actions in the future based on \mlst{consortium 1} are prohibited.

St.\ Antonius takes advantage of the new notion of insensitive variables by making the following statement.
Agents can then justify processing (derivatives of) \lstinline{num-patients} without the authorisation of St.\ Antonius, but authorisation is required to process \lstinline{patients} as before.
SURF is also pleased, because they have regained control of \lstinline{entry-count}, letting them judge each execution before it happens, \eg, to avoid it being used for any nefarious purposes.
\begin{lstlisting}
// Statement (st-antonius 7).
((amy count-patients) num-patients) is insensitive.
\end{lstlisting}

Some time later, an administrator of the system becomes suspicious that sensitive asset data has somehow leaked to `Disruptor' Dan, via their exploitation of a vulnerability of the external asset store.
The administrator exploits their power to `break the glass', seizing emergency powers to regain control of the situation; all actions must be paused to prevent more reading and writing events, until the administrator completes their diagnosis.
Precisely, the administrator updates (everyone's synchronised view of) the agreements to $[\; ]$.
Until the agreements are updated again, well-behaved agents cannot act!

After performing some analyses, the administrator determines that it was a false alarm;
Dan did not do anything wrong (yet).
The productivity of the system is restored by updating the agreements to $[\mlst{consortium 2}]$ agreeing that \lstinline{consortium 2} again, permitting new actions.

\subsection{Evaluation of the Case Study}
\label{sec:discussion_of_the_case_study}
We conclude the case study by evaluating the JustAct framework in general via its implementation as a data exchange system with \slick policies applied to \brane's usage scenarios in particular.

We structure the discussion around the \textit{desirable characteristics} of \brane-like systems, distinguishing those already (E)xisting in \brane from those (N)ew to \brane, \ie, those present only in our own system.
In the discussion to follow, we explicitly connect discussion points to particular desirable characteristics, along with whether each point contributes positively~($+$) or negatively~($-$):
\begin{enumerate}[label=\fbox{$E_{\arabic*}$}] 
    \item Data scientists can define arbitrary workflows and read the results of their execution.
    \item Software engineers can introduce persistent assets for use as inputs to workflows.
    \item The usage of assets is regulated by domain-local policies.
    \item Checkers control the exposure of their domain-local policies to other agents.
    \item Agents work autonomously, unimpeded by other agents (not) doing unrelated work.
    \item Agents agree on how they are permitted to act.
\end{enumerate}
\vspace{0.3em}
\begin{enumerate}[label=\fbox{$N_{\arabic*}$}]
    \item Checkers can express inter-domain policies to share and delegate control over processing.
    \item Inter-agent power dynamics are enforced via user-facing domain-local policies.
    \item Agent violations of their peers' domain-local policies are reliably detected by auditors.
    \item Agents can robustly reason about and prove the policy-permissibility of their actions.
    \item (Inter-)domain policies are easily and effectively refined and changed at runtime.
\end{enumerate}

\noindent 
\Cref{sec:scenario:isolated_execution,sec:scenario:concurrency,sec:scenario:distributed_execution} focused on the reproduction of existing \brane functionality.
\Cref{sec:scenario:isolated_execution} demonstrated defining tasks let `Analyst' Amy, St.\ Antonius, and SURF each define computational workflow tasks in their own statements.
This lets Amy play the role of a workflow provider, ultimately reading the result of executing the workflow~($+E_1$).
It also lets SURF and St.\ Antonius play the role of asset providers, as their task outputs were inputs to Amy's workflow~($+E_2$).
In the case of St.\ Antonius, patient data was provided along with an assertion of control over the processing of the asset and its derivatives; hence, workers could not justify executing Amy's workflow tasks without authorisation from St.\ Antonius~($+E_3$).
Actions interfaced with the domain-local policies of their controllers via the latters' explicit authorisations, which publicised only their policy-decisions themselves~($+E_4$).
In fact, as in \cite{DBLP:conf/eScience/EsterhuyseMBB22}, domains could intentionally postpone their authorisations, or mask the lack of authorisation as inactivity, \eg, to mask their reasoning process~($+E_4$). 
\Cref{sec:scenario:concurrency} demonstrated how agents' work went unimpeded by the unrelated (in)activity of their peers, \eg, the processing of Amy's workflow by St.\ Antonius was unaffected by SURF offering only an ignored task-execution plan~($+E_5$).
Unfortunately, to afford agents playing their \brane roles, the runtime system had to store many statements, and the agents had to reason about their many complex interactions~($-E_6$).
Fortunately, this generalised several distinct processes of the original \brane system, resulting in an implementation that is more concise by re-using this general solution~($+E_6$).
Moreover, as the JustAct framework is based on concepts that stakeholders in \brane are expected to understand, our hardcoded portion of \brane is smaller and simpler, and thus easier to reason about and maintain~(+$E_6$).
Hence, complex \brane-specific notions (\eg, workers, domains, task execution, and asset providers) can be understood via JustAct concepts, which are common and well-understood in computer science (\eg, agents, messages, signatures, rules, and first-order logic) and in legal regulations (\eg, auditing, actions, agreements, qualification rules, and burdens of proof)~(+$E_6$).

Our system confers the \brane system with new desirable characteristics.
Via the complex \slick rules expressible in statements, domain-local policy providers can choose to reveal facets of their policies into the statements themselves~($+N_1$).
For example, in \Cref{sec:scenario:inter-domain-policies}, St.\ Antonius used this feature in two cases.
In Case~1, it enabled the explicit communication of a non-sensitive facet of the St.\ Antonius-local policy to the other agents.
SURF benefitted from having more information to drive its local reasoning~($+E_6$), and moreover, it effectively delegated the power to authorise some tasks from St.\ Antonius to SURF~($+N_1$).
By decoupling St.\ Antonius from some of its work, the system became more fault-tolerant, as new St.\ Antonius-authorisations could be created even after the St.\ Antonius agent or network crashed~($+E_5$).
In Case~2, St.\ Antonius internalised a sensitive facet of its local policy~($+N_1$).
The JustAct framework enabled just the agents trusted by St.\ Antonius to observe this information, however, this control was external, and was not enforced using the usual action-justification mechanism~($-E_3$).
In either case, this delegated reasoning work from domains to other agents, potentially complicating their roles~($-E_6$).
But fortunately, agents retained the power to prefer inactivity, \eg, to forget and ignore messages, to protect themselves from being overworked~($+E_5$).
Throughout the scenarios, external effects at the domain-level (\eg, the reading and writing of assets) and at the policy-level (\eg, the proof that actions were justified) are connected via explicitly created, signed, communicated, and collected policy metadata.
On the one hand, the correspondence between user input and runtime behaviour is systematically driven, enforced, logged, and explained to the users concerned with correctness~($+N_2$) and to the agents responsible for the actions~($+N_4$).
On the other hand, these activities are programmable via manipulations of the user inputs~($+N_5$).
\Cref{sec:scenario:dynamic_agreement} demonstrated that these manipulations can include the fundamentals of the inter-agent power dynamics which were hard-coded in the original \brane, such as arbitrarily re-defining the notion of \textit{involvement} that relates domain-local policies to the tasks whose execution they regulate~($+N_5$).
Fortunately, via the composition control features of \slick~\cite{esterhuyse2024cooperative}, many changes are expressible as controlled policy-refinements in statements as usual, thus not needing agents to synchronise. 
For example, in \Cref{sec:scenario:distributed_execution}, the statement \lstinline{bob 1} enforces the obligation to plan Bob's four workflow tasks together before any are executable, and in \Cref{sec:scenario:dynamic_agreement}, \lstinline{st-antonius 6} lets St.\ Antonius opt out of involvement in a particular asset, after judging that its contents are not sensitive information, creating new ways for workers to justify their processing of this asset~($+N_5$).
Throughout, agents preserved their autonomy over their access to data: each agent has the final word on which actions (and the consequent read and write effects) they enact~($E_6$).
But any such prohibited action is systematically logged, and certainly identified by any auditors afterwards~($+N_3$).

In summary, our system exhibited all the desirable characteristics to some extent.
These improvements (and their limitations) resulted from the more unified and dynamic approach JustAct provides to regulating actions with policies.
Our version of \brane moves some of the burden from the static \brane implementation to the agents at runtime, thus enabling greater flexibility than in the original \brane, but requiring care to shield agents from the potential complexity of their interacting policies.
The case study also exposed cases where the dynamism of JustAct still does not yet go far enough; \eg, in \Cref{sec:scenario:inter-domain-policies} St.\ Antonius lacked a way to control how SURF shared a sensitive statement.
In all scenarios, our version of \brane benefits from its greater emphasis on accountability.
More than ever before, it is clearer how to define, implement, and audit the policy-compliant execution of users' medical workflows.
This functionality was achieved with acceptable runtime performance.

These findings motivate further experiments to integrate the features of our experimental system into the real \brane.
The next steps include distributing the agents in our control plane, connecting to \brane's existing data plane, and re-implementing (or interfacing with) the services that \brane has already automated, \eg, planner and workers.

\section{Discussion of the JustAct Framework}
\label{sec:discussion}

This section evaluates the framework in general by reflecting on the processes and results of the implementations and experiments from \Cref{sec:impl,sec:slick,sec:case_study}.

\subsection{Strengths of the Framework}
\label{sec:strengths}

\subsubsection{Highly Dynamic and Extensible}
The framework is highly abstract, making it applicable to many policy languages and runtime systems.
Notably, the framework is parametric to any chosen policy language, as long as it satisfies the requirements.
This leaves significant room for systems instantiating the framework to adopt various notions of policy, with various representations and semantics.
For example, the framework's notion of policy affords the $n$-ary relations and logical constraints underlying Bell-LaPadua security policies and various Rule-Based Access control policies, which are summarised and compared in \cite{DBLP:conf/wetice/ZhaoC08}.

Moreover, the relations between agents that ultimately control actions are highly dynamically configurable, by agents making statements.
Different design decisions confer different characteristics on the system, \eg, allowing for dynamic specialisation for various use cases in reaction to runtime information.
We recognise two noteworthy spectra on which particular system configurations fall.
Together, these help to clarify the ways systems can change their characteristics at runtime.
Firstly, systems can be centralised (where inter-agent consistency is high, \eg, by agents extensively updating agreements) or decentralised (where many agents can act independently).
Secondly, systems can be highly static (where permission is static and systematically prevented, resulting in predictability and efficiency) or highly dynamic (where permission is often updated but loosely enforced, resulting in flexibility and autonomy).

\subsubsection{Formal Inter-Agent Power Dynamics}
The framework is useful if agents strive to stay well-behavedness: acting only as permitted.
Thus, policies with complex conditions for permission create complex inter-agent power dynamics, \eg, modelling various useful normative concepts.
For example, in \Cref{sec:scenario:inter-domain-policies}, by stating \lstinline{st-antonius 5}, the St.\ Antonius hospital formalises who they trust in the \lstinline{X trusts Y} relation, and delegates their own power of authorisation to trusted peers.
Despite its complexity, \lstinline{st-antonius 5} is unambiguous.

Well-behavedness is also robust: an agent that violates their own well-behavedness by taking a non-permitted action preserves the well-behavedness of their peers.
This affords meaningful cooperation between well-behaved peers, even in the presence of misbehaving peers.
The framework lays the foundation for agents to recognise and punish these bad actors, \eg, by cutting them out of future cooperations, ignoring their statements, or expediting their punishment by participating in future audits of their actions.

\subsubsection{Autonomy and Parallelism}
Agents synchronise to change agreements.
All other communication can be asynchronous, delayed, and lossy.
Agents are very autonomous, as they are never fundamentally compelled to act or make statements.
Hence, agents are robust to unreliable peers.
The framework affords realistic inter-agent enforcement of well-behavedness: agents monitor actions, and actors must prove that their actions are permitted.
This comes at the cost of burdening the actors with reasoning and bookkeeping work.
However, in cases where this burden is too great, agents are free to remain inactive instead;
inactivity does not require justification.
The framework is also robust to agents arbitrarily forgetting what they know about messages and statements.

\subsubsection{Consistent Permission despite Privacy}
\label{sec:strength_permission_despite_privacy}

Implementations of the framework ensure that the permission of observed enacted actions is decidable, despite permission being dependent on policy information that agents create at runtime.
This apparent contradiction is resolved by agents being able to decide when their knowledge suffices to decide permission.
Moreover, permission is objective; agents are certain that other agents (\eg, future auditors) agree that their actions were permitted, without involving them at all.
For example, in \Cref{sec:scenario:inter-domain-policies}, the SURF agent can justify actions using \lstinline{st-antonius 6}, despite this statement being kept private between SURF and St.\ Antonius.
This sensitive statement must only be observed at the moment the permission of this action is audited.
For example, it can be audited by a third party trusted by St.\ Antonius not to reveal \lstinline{st-antonius 6}, and trusted by the other agents to accurately judge the permission of the action.

\subsection{Limitations of the Framework}
\label{sec:limitations}

\subsubsection{Costly Justification Search}
\label{sec:limitation_costly_justifications}
Well-behaved agents cannot act until they have found an action that \textit{justifies} a desired effect (\Cref{def:justifies}): it is permitted, and enacts the desired effects.
But we expect the justification process to generally be difficult, as it requires agents to explore a vast combinatorial search space:
which combinations of agreements, actions, and statements by various actors and authors lead to the goal?
This cost depends on the chosen policy language, with complex policies potentially requiring complex reasoning to compute the permission and effects of given actions via their extracted policies.
This dependency means we cannot meaningfully quantify the cost of justification in abstract.
And we have not included an analysis of this cost for our case study in the scope of this article.

However, by the same rationale, users have many ways to reduce the cost on agents to reason about policies.
Firstly, this cost should be a factor in selecting the policy language.
For example, \slick was sufficiently complex to capture the normative concepts required to reproduce the \brane system in \Cref{sec:case_study}, despite \slick being able to express only enumerable relations, which JustAct does not generally require.
The same goes for other implementation decisions.
For example, where such is possible, implementations can make statements more predictable by restricting what agents can do.
For example, in our case study, we could have fixed Amy, Bob, and Dan as the only workers, such that (everyone knows that) no one else executes workflow tasks.
Secondly, agreements should be carefully designed to restrict the statements agents must consider.
For example, static analysis of the initial agreement in \Cref{sec:brane_control_plane} reveals that any action executing any task in the namespace of~$a$ requires at least one statement from~$a$, because rules must define the task's dependencies, and they must be authored by~$a$.
Thirdly, agents can choose to avoid communication and reasoning effort that does not contribute to reaching their own goals.
Amy demonstrates a simple example; she contributed no further to the scenario after reading the desired \mlst{count-patients} dataset.
Bob demonstrates a more subtle example;
in \mlst{bob 1}, he authorised everybody to read everything, because Bob was neither an auditor (who guards authorisations), nor involved in any sensitive data (whose authorisations are meaningful).
In general, whenever reasoning about policies becomes to costly, agents can fall back on inaction to remain well-behaved.

Future work can design particular policy languages and agreements to simplify these search problems.
We are also interested in leveraging existing literature and tools to improve how efficiently agents solve them.
We consider answer set solving (\eg, with Clingo~\cite{DBLP:journals/aicom/GebserKKOSS11}) or model-checking rewrite systems modulo theories (\eg, with Maude~\cite{clavel2003maude}).

\subsubsection{No Obligations to Act}
Recall that JustAct provides no mechanism to let agents compel one another to act before a deadline.
On the one hand, the benefits of this approach were shown in the case study.
For example, \Cref{sec:scenario:inter-domain-policies} presented a scenario where data processors were free to ignore and discard incoming control messages, because doing so never threatened their own well-behavedness and never impeded the work of auditors.

On the other hand, JustAct policies cannot express arbitrary \textit{obligations to act}, because only actions can be prohibited (and punished), but the absence of timely action cannot.
This hinders the practical applicability of JustAct.
For example, JustAct is unsuitable to \textit{real-time systems}, where the timely coordinated actions of agents are of paramount importance. 

However, \Cref{sec:case_study} demonstrated how certain cases of obligation are expressible in policies as conditional permission.
These sufficed to implement the functionality of \brane in \Cref{sec:case_study};
\eg, data processors were (indirectly) obligated to acquire consent to process data from data subjects in order to process the subject's data while staying well-behaved.

Currently, JustAct does not fundamentally prohibit the expression and enforcement of obligations to act outside of the JustAct abstraction.
However, we expect that agents may have difficulty compelling their peers to act before specific deadlines, as a consequence of the point in \Cref{sec:limitation_costly_justifications}: agents cannot efficiently consider the possible policies and actions.

Work on a variant of JustAct is underway to support the desired notion of obligation.
This requires two changes.
Firstly, the ontology must be expanded to capture \textit{obligation} as a dynamic relation over agents, whose elements are mutated by actions.
Secondly, mechanisms must be added so that agents are not punished for obligations beyond their knowledge.

\subsubsection{Specification of Communication}
\label{sec:limitations_specification_of_communication}
A key feature of the JustAct framework is that agents do not need complete knowledge of the enacted actions and stated messages.
The intention is that agents decide how they share this information via communication.
However, while policies specify how agents act, policies do not specify how agents communicate.
\Cref{sec:scenario:inter-domain-policies} demonstrated how this leaves room for agents to hide sensitive policy information from their peers by withholding their messages.
The St.\ Antonius hospital prevented Dan from observing the statement of~$m$ by sharing it with SURF but not Dan, but St.\ Antonius has no recourse if SURF then forwards it to Dan.
As far as JustAct is concerned, SURF would remain well-behaved, and no alteration of $m$ would give St.\ Antonius more control.

In future work, we want to supplement the current notion of agreements and permissions with a similar mechanism for the communication of statement and action information, which regulates the communications from one agent to another, and is violated when information is sent without permission.
But it remains unclear how to achieve this without burdening agents too much with reasoning about the permission of communications.

\subsubsection{No Privacy from the Actor}
\label{sec:no_privacy_from_the_actor}
Recall from \Cref{sec:strength_permission_despite_privacy} that a key strength of JustAct is that agents can cooperate to realise effects based on sensitive data without widely revealing the sensitive information.
However, auditors of an action must generally inspect the payload policies of enacted actions to decide if they are permitted (\Cref{par:dec_enacted_permitted}).
And in practice, actors striving to stay well-behaved must verify this permission, \ie, pre-empting the decision of the auditor.
As such, authors of statements must expect prospective actors to observe any sensitive contents.
For example, in \Cref{sec:scenario:inter-domain-policies}, to let SURF act on statement \lstinline{st-antonius 6}, St.\ Antonius had to reveal to SURF that Billy, Wally, and Berty are patients at the hospital, so that SURF could be sure that sufficient proof of consent was included.

What is needed is robust methods for agents to compute with actions while hiding parts of their policy payloads.
We have left any systematic developments out of scope.
But here we briefly present three approaches that we see as promising avenues for future work.

The first approach simply abstracts sensitive details out of policies.
For example, in \Cref{sec:scenario:inter-domain-policies}, instead of the St.\ Antonius encoding consent per patient, `sufficient consent' could have been one atomic fact.
Then St.\ Antonius could manage the individual patients' consent outside the system.
This is comparable to how the original \brane system hides authorisation information~\cite{DBLP:conf/eScience/EsterhuyseMBB22}.
An advantage of this approach is that it is always available in some form.
Authors must already decide the granularity of their policies.
But the downside is that it disconnects the sensitive information from the auditing process.

The second approach is to let (only) a trusted third party process the sensitive information.
For example, St.\ Antonius could propose how SURF should act, identifying the sensitive statements but withholding their contents.
SURF can then act, trusting (but not knowing) that they remain well-behaved.
However, auditors would still need the sensitive data.
The trustworthiness of the third parties could be improved by automating and encapsulating them.
For example, a containerised process on trusted hardware could accept a self-contained action (with sensitive data) as input, but output only the permission decision.

The third approach is applying the privacy preserving computation methods overviewed in \Cref{sec:related_privacy_preserving_computation}: decouple the proof of permission from the underlying sensitive data.
In a sense, this uses cryptography to force anybody to stay trustworthy enough to audit permission.

\section{Related work}
\label{sec:related_work}

In this section, we discuss some noteworthy related work, and remark on the ways it can complement the implementation of JustAct, or replaces (parts of) JustAct.
We cluster the works according to the problem that they solve which is most relevant in our context.
\begin{enumerate}
    \item How can agents' views of the distributed configuration be synchronised?
    \item How can agents check action permission and effects while hiding the underlying policies?
    \item Which languages can capture the necessary normative relations between the agents?
\end{enumerate}

\subsection{Many Ways to Synchronise Agent-Local Data}

Distributed ledgers are an abstraction of consistent state over decentralised processes.
Blockchains afford robust, probabilistic implementation of distributed ledgers, but they differ in their details; \eg, Fabric emphasises scalability~\cite{DBLP:conf/eurosys/AndroulakiBBCCC18}, while Ouroboros emphasises provable security~\cite{DBLP:conf/crypto/KiayiasRDO17}.

SmartAccess \cite{DBLP:journals/access/OliveiraRVMO22} uses distributed ledgers to store policies and (meta)data, enabling decentralised implementations of the access-control model whilst affording accountability regarding authorisation.
However, this approach imposes a burden on agents to maintain a high level of synchrony in the policy information they create and collect.
This imposes unnecessary overhead whenever policy information is irrelevant.
For example, the concurrency demonstrated in \Cref{sec:scenario:concurrency} would be impossible; agents would be forced to synchronise their statements just in case they affect the permissibility of their actions.
Moreover, the synchronisation of policies would make it impossible to support cases where policy information captures private information, and thus, should not be broadly publicised.
This is demonstrated in \Cref{sec:scenario:inter-domain-policies}, where the St.\ Antonius hospital sends the (trusted) SURF agent some private information to enable new permitted actions.

Other ledger-based systems allow a heterogeneous view on the policy state.
For example, Canton \cite{canton-better} (whitepaper) replaces the (sequential) block\textit{chain} with a (hierarchical) block\textit{tree}.
Agents must only synchronise the relevant sub-trees with their neighbours.
This lays the groundwork for private policies.
This idea is promising and requires further investigation.

The \textit{FastSet} protocol guarantees eventual consistency despite each action only being synchronised by a \textit{quorum} (roughly: a majority) of agents~\cite{chen2025fastsetparallelclaimsettlement}.
The approach is fruitful when the changes to the shared data \textit{commute} in some sense that arises in many financial transactions.
For example, it does not matter in which order Amy and Dan send their money to Bob.
We see value in implementing protocols such as FastSet in particular systems also implementing JustAct.
For example, would FastSet offer a practical way for agents to usually act in parallel, but at the same time, let policies express mutual exclusion between actions: where the permission of one actions precludes the prior enactment of another.

\subsection{(Policy) Privacy Preserving Computation}
\label{sec:related_privacy_preserving_computation}

In general, there is a tension between keeping data private while publicising its derivatives.
\Cref{sec:no_privacy_from_the_actor} discussed how this tension manifests in JustAct, but it is curtailed by letting agents decide how policy information is shared.
Specifically, agents can limit sensitive information (in statements) to the agents that directly depend on them (in actions).
\Cref{sec:scenario:inter-domain-policies} demonstrated; the sensitive policy of the St.\ Antonius hospital (that names of patients Billy, Wally, and Berty) was revealed only to SURF, who verified permission (by checking that each patient had consented).

Various lines of research develop robust solutions to this kind of problem through the careful application of cryptography.
In \textit{homomorphic encryption} (HE), the result of function application $f(x)$ is computed despite $x$ remaining encrypted. 
For example, thus auditors can decide whether actions are permitted without observing the underlying policies in unencrypted form.
Since foundational works on HE such as~\cite{gentry2009fully}, there have been incremental improvements to its efficiency~\cite{brakerski2014leveled} and applicability~\cite{cheon2017homomorphic}; \cite{acar2018survey} provides a survey of HE.
\textit{Zero-knowledge proof} (ZKP) protocols are related; a \textit{prover} reveals some artefact $p'$ that convinces an observer that some proposition~$P$ is witnessed by some proof $p$, which remains secret.
Obviously, the key is the relationship between $p$ and $p'$. 
For example, thus agents can propose actions dependent of policies that are never revealed, yet convince all observers that the action is permitted.
ZKP was surveyed in \cite{li2014survey} and more recently in \cite{DBLP:journals/corr/abs-2408-00243}, pointing out the pivotal improvements to efficiency as those in~\cite{ben2018scalable,parno2016pinocchio} that make ZKP feasible today. 
Note how HE and ZKP differently locate the secret and computation.
\textit{Multi-party computation} (MPC) generalises to groups of agents jointly computing some results, each hiding intermediate values from one another.
In MPC research,
\cite{damgaard2012multiparty} is a landmark paper, and \cite{zhao2019secure} is an illuminating survey.

All of these approaches have the advantage of systematically isolating what information agents reveal to one another, \eg, such that auditors can verify permission, actors can enumerate effects, but only the author knows underlying private policies.
The downside is that effort is required to encode the desired policies in the needed cryptographic framework.
Moreover, the added cryptographic operations add computational overhead to auditing, where it is least desirable.
Fortunately, active research continues to increase their applicability and reduce their cost.
We particularly note works studying \slick-style logic programming languages (\eg, Datalog) in ZKP protocols~\cite{wen2024practical}, MPC~\cite{pankova2020short}, and with HE~\cite{kouhounestani2022datalog}.

\subsection{Curie: A Data Sharing System with Policy Privacy Preserving Computation}
\textit{Curie}~\cite{DBLP:conf/codaspy/CelikAASMU19} is an existing distributed medical data sharing system.
Like JustAct, Curie encodes agents' permission to act in communicable policies.
And specifically like the original \brane and the version that instantiated JustAct, Curie is specialised for the sharing of medical data, whose sensitive nature motivates an emphasis on controlling how data is shared.
In fact, Curie goes further, surpassing what has been developed for JustAct, as is discussed in \Cref{sec:no_privacy_from_the_actor}: in Curie, the kind of cryptographic methods in \Cref{sec:related_privacy_preserving_computation} decouple the checking of permission from the observation of the sensitive policy information on which permission depends.
So our case study system cannot reproduce every Curie usage scenario.
For example, in Curie, a hospital's sharing policy can be conditioned on user data surpassing a threshold of differential privacy \cite{DBLP:conf/sp/MohasselZ17}, which depends on the data itself, but the policy decision only requires processing the data in homomorphically encrypted form \cite{DBLP:journals/tjs/DoanMGD23}.

However, upon closer inspection, Curie and JustAct make largely orthogonal contributions.
Curie also cannot reproduce our case study system.
Curie policies are statically distributed, and specify local requirements.
In contrast, JustAct generalises how agents represent and disseminate policies, affording ad hoc system reconfiguration via agreements and statements, as we demonstrated in our case study, \eg, where hospitals delegate power on the fly.

Future work can explore the combination of the best features of JustAct and Curie.
Can Curie can be reframed as a JustAct implementation?
Which systems support policies that express multi-agent power dynamics that are dependent, but do not reveal the the underlying policies?
Which policy languages are rich enough to express these complex relationships, yet constrained enough to afford the application of the necessary cryptographic protocols?
What would such policy languages look like and which features can they inherit from \slick or the Curie policy language?
In which cases is the increased computational burden on auditors acceptable?
Which problems can these very expressive policies help to solve?

\subsection{Trust Management: Logical Expression and Proof of Access to Resources}
Traditional methods of \textit{access control} offer languages and tools for specifying and checking a requester's permission to access data.
\textit{Trust management} reifies the role of the accessor as a \textit{certificate}, enabling access control in a decentralised environment, where the identities of particular requesters are not known ahead of time~\cite{DBLP:conf/sp/BlazeFL96}.
Much literature dates to the 1980's and 1990's, investigating policy languages suited to defining certificates and inferring them at request time from context.
Many of these are extensions of Datalog, \eg, adding non-monotonicity~\cite{DBLP:journals/tissec/LiGF03},  constraints~\cite{DBLP:conf/padl/LiM03}, and weights~\cite{DBLP:conf/IEEEares/BistarelliMS08}.

Like access control, trust management focuses agent reasoning on the access-request decision, whereas our framework emphasises the inter-relationship between agents and their actions via their synchronised agreements.
However, the bulk of trust management research complements our work, because it informs the selection of particular policy languages suited to particular purposes.
\cite{DBLP:conf/otm/Sacha11} overviews and compares (the complexity of) noteworthy trust management languages.
Which trust management languages are good candidates for policy languages in the JustAct framework?
We expect these to be even more desirable if the framework is extended to support the desirable features explained in \Cref{sec:limitations_specification_of_communication}, empowering policies to specify how agents can communicate statements and actions.

\section{Conclusion}
\label{sec:conclusion}

In this article, we define \textit{JustAct}, a framework for systems driven by automous agents whose actions on the system are regulated by the policy information that they define, communicate, and assemble.
Our work is motivated by the problem of exchanging sensitive (\eg, medical) data between autonomous agents, where it is crucial that \begin{enumerate*}
    \item agents can define adjustable policies that regulate their peers' uage of their data, and
    \item it is undesirable or infeasible for every agent to maintain their overview of all actions and the underlying policy information. 
\end{enumerate*}

Our framework is instantiated by multi-agent systems whose behaviour is unfolded at runtime as the autonomous agents communicate statements (carrying policy information) and act (whose effects are externalised by the agents, \eg, as reading a particular dataset).
To maximise the applicability of our work to existing literature and in existing systems, we do not fix the language agents use to express policies.
In general, it suffices for the language to meet our minimal specification.
Intuitively, each policy must be communicable in messages, and determine the \textit{permission} and \textit{effects} of a given action.
For implementation, these properties of policies must also be decidable, \eg, such that agents' agreement on an action guarantees their agreement on its meaning.
Actors rely on this guarantee to ensure that auditors in the future will agree that their actions were permitted at the time.
The agents are in control of the fundamental trade-off between predictability and reliability in which actions are permitted at any time, by selecting their shared \textit{agreements}.
Because of the key role of agreements in permission, agreements give system designers and administrators a means of controlling which actions are permissible.
And because permission is highly re-configurable by agents' asynchronous statements -- even when agreements are fixed -- agreements can change infrequently, amortising any associated synchronisation overhead.

We evaluated our framework in application to a representative case;
we apply a Rust implementation of JustAct in a runtime system and policy interpreter to usage scenarios of \brane, an existing medical workflow execution system.
Desirable features were reproduced successfully: data scientist users can define workflow tasks, workers can execute them, while domain-checkers regulate the processing of domain-controlled medical data.
Users enjoy the typical benefits of \brane: domain-policies are programmable, and data-independent tasks complete concurrently.
Moreover, our approach affords greater expressivity and flexibility in the regulation of workflow execution.
For example, we demonstrate agents dynamically delegating their power to authorise workflow tasks, and amending the roles of agents in workflow execution, while preserving their fundamental consensus on which actions (in the past and future) are permitted, and what are their effects on the medical data.

We discussed the generalisability of these findings to other systems, remarking on the strengths and limitations of our approach in general.
The strengths follow from systematising the connection between agents and programmable policies, laying the groundwork for unified approaches to various tasks that are difficult when they are separated;
for example, our agents use policies to \begin{enumerate*}
    \item formalise the connection between system events and external sources of regulatory norms such as the GDPR, 
    \item audit the permission of existing actions,
    \item plan actions that have desirable effects, and
    \item identify policy amendments and changes that afford desirable actions.
\end{enumerate*}
There are two kinds of limitation to our work.
Firstly, the additional flexibility and dynamism our framework demands of the system comes at the cost of computational complexity.
To some extent, this is unavoidable, arising as a necessity of systems being designed to exploit the novel flexibility we provide to the dynamics between agents.
However, the logical, systematic nature of our policies affords the application of existing language-design and system-analysis techniques to designing policy languages and policies to strike better compromises between flexibility and simplicity.
Secondly, we have identified extensions of this framework that push it yet further in the direction of linking policies to agent behaviour.
For example, we see promise in future work extending the notion of permission; we want statements to regulate how statements are communicated.


\section*{Acknowledgment}
\noindent 
This research is partially supported by the projects:
\begin{enumerate*}
    \item
    EPI (NWO grant 628.011.028),
    \item
    AMdEX-fieldlab (Kansen Voor West EFRO grant KVW00309),
    \item
    BNVw: budget-neutral sustainable home improvements (MOOI-224049), and
    \item
    the Dutch Metropolitan Innovations ecosystem for smart and sustainable cities made possible by the National Growthfund.
\end{enumerate*}

\bibliographystyle{alphaurl}
\bibliography{bib}

\end{document}